\documentclass[aps,twocolumn,amsfonts,nofootinbib,preprintnumbers,eqsecnum,superscriptaddress]{revtex4-1}

\usepackage{epsfig} 
\usepackage{psfrag}
\usepackage{amsmath}
\usepackage{amssymb}
\usepackage{color}
\usepackage{graphicx}
\usepackage{subfigure}
\usepackage{hyperref} 
\usepackage{rotating}
\usepackage{enumerate,mdwlist}

\newcommand{\reef}[1]{(\ref{#1})}
\def\be{\begin{equation}}
\def\ee{\end{equation}}

\def\bea{\begin{eqnarray}}
\def\eea{\end{eqnarray}}

\newcommand{\nb}[1]{\color{blue}}

\newcommand{\hl}[1]{\color{magenta}}

\newcommand\ov{\over}

\usepackage[normalem]{ulem}
\usepackage{amsfonts}
\usepackage{epsfig}
\usepackage{mathbbol}

\usepackage{amsfonts}

\def\Tr{\mathop{\rm Tr}}

\newcommand\half{{\ensuremath{\frac{1}{2}}}}
\newcommand\p{\ensuremath{\partial}}

\newcommand\field[1]{{\ensuremath{\mathbb{{#1}}}}}

\newcommand\vev[1]{{\ensuremath{\left\langle{#1}\right\rangle}}}

\newcommand\ket[1]{\ensuremath{\lvert{#1}\rangle}}

\newcommand\sP{{\ensuremath{{\mathcal P}}}}

\newcommand{\RR}{\field{R}}

\newcommand{\bi}{\begin{itemize}}
\newcommand{\ei}{\end{itemize}}
\newcommand{\ben}{\begin{enumerate}}
\newcommand{\een}{\end{enumerate}}
\newcommand{\bca}{\begin{cases}}
\newcommand{\eca}{\end{cases}}
\newcommand{\bln}{\begin{align}}
\newcommand{\eln}{\end{align}}
\newcommand{\bst}{\begin{split}}
\newcommand{\est}{\end{split}}
\def\ie{\begin{equation}\begin{aligned}}
\def\fe{\end{aligned}\end{equation}}

\newcommand\al{{\alpha}}

\newcommand\sig{\sigma}
\newcommand\Sig{\Sigma}
\newcommand\lam{\lambda}

\newcommand\om{\omega}
\newcommand\Om{\Omega}

\newcommand\de{{\ensuremath{{\delta}}}}
\newcommand\De{{\ensuremath{{\Delta}}}}

\def\th{{\theta}}

\newcommand\ha{{\half}}

\def\le{\left}
\def\ri{\right}

\newcommand\sE{{\ensuremath{{\mathcal E}}}}

\newcommand\sN{{\ensuremath{{\mathcal N}}}}
\newcommand\sO{{\ensuremath{{\mathcal O}}}}

\newcommand\bk{{\mathbf k}}

\newcommand\bv{{\bf v}}

\newcommand\bx{{\mathbf x}}

\newcommand{\eql}{\ell_{\rm eq}}
\newcommand{\Vee}{v_{E}}

\begin{document}

\preprint{MIT-CTP/5065}

\title{Holographic systems far from equilibrium: a review}
  \author{Hong Liu}
    \affiliation{Center for theoretical physics, Massachusetts Institute of Technology, Cambridge, MA 02139, U.S.A.}
 \author{Julian Sonner}
    \affiliation{Department of theoretical physics, Universit\'e de Gen\`eve, 24 quai Ernest-Ansermet, 1211 Gen\`eve 4, Switzerland}
\begin{abstract}
In this paper we give an overview of some recent progress in using holography to study various far-from-equilibrium condensed matter systems. Non-equilibrium problems are notoriously difficult to deal with, not to mention at strong coupling and including quantum effects. Remarkably, using holographic duality one can describe and follow the real time evolution of far-from-equilibrium systems, including those which are spatially inhomogeneous and anisotropic, by solving partial differential gravity equations (PDEs). We sample developments on two broad classes of questions which are of much recent interest in the condensed matter community: non-equilibrium steady states (NESS), and  quantum systems undergoing a global quench. Our discussion focuses on the main physical insights obtained from the gravity approache, rather than comprehensive treatments of each topic or detailed descriptions of gravity calculations.
The paper also includes an overview of current numerical techniques, as well as the holographic Schwinger-Keldysh approach to real-time correlation functions.
\end{abstract}

\date{\today}
\maketitle
\tableofcontents

\section{Introduction}

Holographic duality is an equivalence between a quantum gravity system in an asymptotic $(d+1)$-dimensional anti-de Sitter (AdS$_{d+1}$) spacetime and a $d$-dimensional quantum many-body system living on its boundary~\cite{Maldacena:1997re,Gubser:1998bc,Witten:1998qj}. This is reminiscent of an optical hologram, where a
three-dimensional object is encoded in terms of data living on a two-dimensional surface.
Below we will refer to the gravity and the many-body system in a duality relation as bulk and boundary systems respectively. 
A striking implication of the duality is that the classical gravity regime of the bulk theory turns out to 
correspond to the strong coupling and large number of degrees of freedom limit of the boundary system, thus 
providing new powerful tools for studying strongly correlated quantum matter using classical gravity techniques. 
Important insights into many aspects of quantum many-body systems have been obtained, see e.g.~\cite{casalderrey2014gauge,zaanen2015holographic,Son:2007vk,Hartnoll:2009sz,Herzog:2009xv,McGreevy:2009xe,Adams:2012th,DeWolfe:2013cua,Hartnoll:2016apf,Hartnoll:2016apf} for reviews. 

In this paper we give a short overview of some recent progress in using holography to study various far-from-equilibrium condensed matter systems. 
Non-equilibrium problems are notoriously difficult to deal with, not to mention at strong coupling and including quantum effects.
Remarkably, using the duality one is able to describe and
follow real time evolution of a far-from-equilibrium system, including those which are spatially inhomogeneous and anisotropic, by solving partial differential gravity equations (PDEs). While solving gravity PDEs is often a nontrivial 
task, it is much more manageable than explicitly following the time evolution of a wave function (or density matrix)
of  a system consisting of large numbers of constituents. 
Furthermore, {the} gravity description provides  new ways to organize physics of a system which could be particularly valuable in non-equilibrium contexts.  For example, the extra spatial dimension in the bulk, often referred to as the radial direction, can be considered as a geometrization of renormalization group flow of a boundary theory. By analyzing the gravity geometry at different  
radial location, one can learn about the boundary system at different scales. 

Using holographic duality to study non-equilibrium problems is by now a big subject with many 
directions. For example,  deep connections between Einstein's equations and hydrodynamics have been uncovered, which led to
new understandings of hydrodynamics for systems with quantum anomalies. 
Thermalization from various homogeneous and inhomogeneous non-equilibrium states have been extensively studied with the motivation for understanding the creation and thermalization of the quark-gluon plasma following heavy ion collisions 
at RHIC and LHC. Interesting new dynamical insights have been discovered. 
 For example it was found that local thermalization (i.e. non-conserved quantities become locally equilibrated) can happen 
 extremely fast for strongly coupled systems, at time scales of order $1/T$ with $T$ the final equilibrium temperature. 
 It was also found that  hydrodynamics can become valid much earlier than 
expected, before a system isotropizes. Extensive reviews~\cite{Hubeny:2010ry,DeWolfe:2013cua,casalderrey2014gauge} on these topics exist and will not be recounted here. 

In this review we focus on two broad classes of questions which are of much recent interest in condensed matter communities. 

The first is concerned with characterizations of various newly discovered non-equilibrium steady states (NESS). 
NESS arise for systems subject to external forcing, for example an applied electric field or a heat gradient, or a gradient of a chemical potential. If the system is able to adjust itself in such a way as to establish a balance between the resulting currents and the applied forcing it will settle into a stationary state. More generally, one can view such a steady state as an intermediate-time description of a system, on scales small compared to the ultimate equilibration time. If the latter can be made parametrically large then the steady-state persists.  A particularly interesting question concerning a NESS is whether there could exist some effective thermodynamic description. 
We will discuss one example each of the three main types of steady states, namely current driven, heat driven and momentum driven NESS.

The second class of questions concerns the dynamics after a system has undergone a quench, i.e. an un-adiabatic  change of  certain parameter(s) of the system. These could include  temperature, or energy density, or external fields such as a magnetic field, or some parameter in the Hamiltonian such as a mass or a coupling constant.  One is interested in the subsequent dynamics as well as the properties of the asymptotic state approached at late time. Quenches are of great interest as they provide simplest  ways, both theoretically and experimentally, to drive a system out of equilibrium, yet yielding rich dynamics and wide range of phenomena. 
We will discuss three topics: entanglement propagation after a global quench of energy density, sudden driving of the order parameter of a superfluid, and defect production across a critical point. 

Due to limitation in length, our discussion of these questions will focus on the main physical insights obtained from
gravity approaches, rather than comprehensive treatments of each topic or detailed descriptions of gravity calculations. 

We have also included several sections discussing techniques of gravity approaches. Again this is a vast subject with many reviews and books. Our choice is based on relevance for non-equilibrium problems and (un)availability in other reviews. 
In Sec.~\ref{sec:SK} we discuss how to compute real-time correlation functions on a Schwinger-Keldysh contour using gravity. 
In Sec.~\ref{sec:num} we give an overview of existing numerical methods. 
In Sec.~\ref{sec.AspectsOfDuality} we highlight some basic aspects of the duality which will be used in subsequent sections. 
Again these sections are not meant to be comprehensive, but rather emphasize key conceptual elements.

There are many other exciting topics which we will not be able to go into. One is holographic turbulence, which we will just briefly describe here.  Turbulent flows are ubiquitous in fluid motions. It is thus a natural question to ask what the gravity dual of a turbulent flow looks like. Such a turbulent gravity solution has been constructed numerically in~\cite{Adams:2013vsa} for a $(2+1)$-dimensional boundary system. As appropriate for $(2+1)$-dimensional, the authors observed an inverse energy cascade with energy being transferred from short to large distances through merging of vortices, and the $-{5 \ov 3}$ Kolmogorov scaling law of the energy spectrum.  Furthermore, they provided support for and argued more generally that the gravity geometry for a turbulent fluid is a black hole whose event horizon is fractal with fractal dimension $D = d+4/3$, where $d$ is the boundary spacetime dimension, and the extra $4/3$  is related to the $-{5 \ov 3}$ Kolmogorov scaling law. Other discussions of holographic turbulent fluids include~\cite{Green:2013zba}.

Another area that has seen significant progress in recent years is the theory of transport of inhomogeneous systems without quasiparticles\footnote{A more general review of transport in systems without quasiparticles is given in \cite{Hartnoll:2016apf}}. By this we mean (holographic) field theories whose preferred ground state has inhomogeneities or otherwise exhibits some form of momentum dissipation. Examples of such ground states involve spatial modulation of some kind, for examples stripes \cite{Donos:2011bh}, helical phases \cite{Donos:2012wi}, or checkerboards  \cite{Withers:2014sja}, as well as holographic systems with an explicit lattice deformation \cite{Horowitz:2012ky} (including Q-lattices \cite{Donos:2013eha}). Remarkably it was shown that one can determine the DC conductivities (and other transport coefficients) in terms of a simple hydrodynamic system of equations  \cite{Davison:2014lua,Lucas:2015vna,Donos:2015gia}, which is located at the dual black-hole horizon. In fact, intuition gleaned from transport of strongly-coupled transport in holography has already found fruitful application in the theory of bad metals, as described for example in \cite{Lucas:2015pxa,Lucas:2017idv} and reviewed in \cite{Hartnoll:2016apf}. Furthermore, an interesting perspective on holographic conductivity with momentum dissipation is afforded by considering massive gravity in the bulk, \cite{Vegh:2013sk}, which may morally be seen as an effectve theory of broken spatial translation invariance \cite{Blake:2013bqa,Blake:2013owa,Davison:2013jba}.

\section{Aspects of the duality}\label{sec.AspectsOfDuality}

We first quickly highlight certain aspects of the duality which will be central to subsequent sections of the review. 
We will take the boundary spacetime dimension to be $d$, with the bulk gravity spacetime being $d+1$-dimensional.
For more detailed expositions of the duality, see e.g.~\cite{Aharony:1999ti,DHoker:2002nbb,nuastase2015introduction,ammon2015gauge}. 
Readers who are already familiar with the basic ideas of holography may safely skip this section.

\subsection{Some basic dictionary} 

Holographic duality is an equivalence between two quantum systems. Clearly symmetries of the two systems must coincide. Furthermore, there should be a one-to-one correspondence between quantum states. On the bulk side, in the classical gravity regime, states are represented by solutions to equations of motion which satisfy appropriate boundary conditions. Thus each such bulk solution/geometry corresponds to some quantum state of the boundary system.  More explicitly, we can write the action for the bulk theory as 
\be \label{nmb}
S_{\rm bulk} =  S_{\rm grav} + S_{\rm matter}
\ee
with $S_{\rm grav}$ the gravitational action in AdS  
\be\label{eq.EHaction}
S_{\rm grav} ={1 \ov 16 \pi G_N}  \int d^{d+1}x \sqrt{-g} \left( R + \frac{d(d-1)}{\ell^2} \right)\,,
\ee 
and $S_{\rm matter}$ that for possible matter fields. 
In~\eqref{eq.EHaction}, $\ell$ is the AdS radius. Newton's 
constant $G_N$ is inverse proportional to the number of degrees of freedom $\sN$ of the 
boundary theory. Thus gravity is weak if $\sN$ is large. 
The  spectrum of matter fields and the specific form of $S_{\rm matter}$ vary with the 
specific dual boundary system, while the gravity action $S_{\rm grav}$ is universal to all systems with an Einstein gravity dual.

\begin{figure}
\begin{center}
\includegraphics[width=\columnwidth]{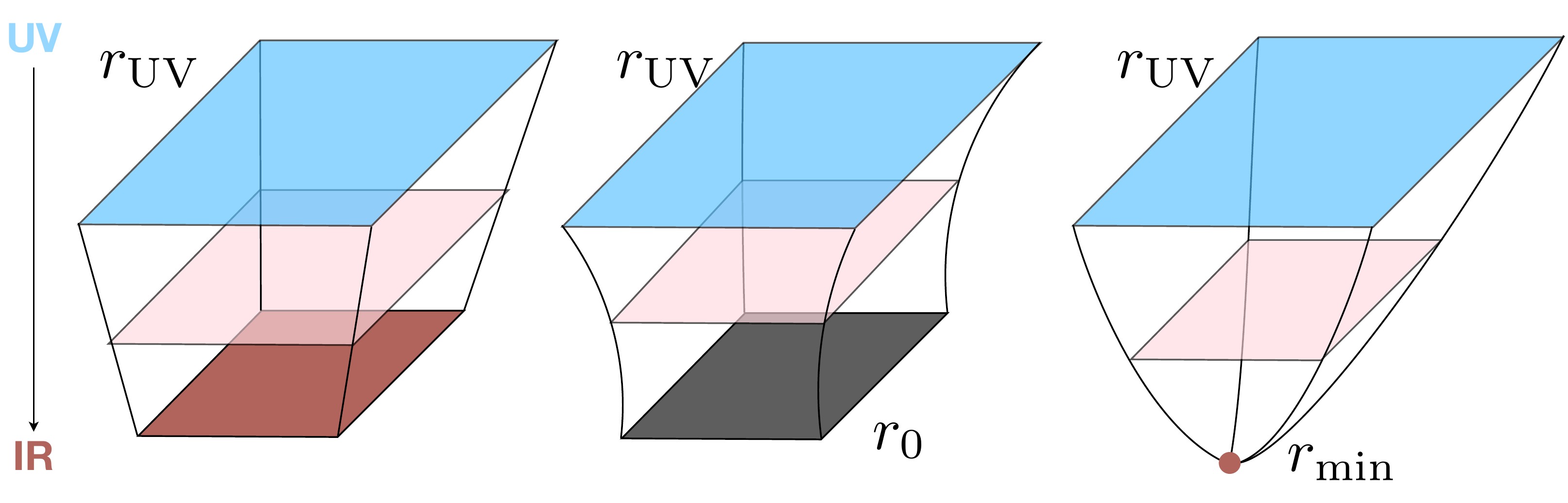}\\
\end{center}
\caption{AdS/CFT as an RG flow. The left panel represents an RG fixed point, so that the entire geometry is scale-invariant (empty AdS). The middle panel shows a thermal state, where the IR geometry is instead a black hole with horizon at $r_0$. The third panel represents an RG flow where the UV fixed point flows to gapped theory in the IR, ending smoothly at a minimum radius $r_{\rm min}$. Only the first geometry is fully scale invariant.}
 \label{fig:AdS}
\end{figure}

As an illustration let us look at some simple solutions to~\eqref{nmb} with no matter excited (which then reduce to solutions of~\eqref{eq.EHaction}). 
The simplest and most symmetric solution is the anti-de Sitter spacetime, 
\be\label{eq.poincareadsmetric}
ds^2 = \frac{\ell^2}{z^2} \left( -dt^2 + dz^2 + \sum_{i=1}^{d-1}dx_i^2 \right) \ .
\ee
It corresponds to the {\it vacuum} of a dual field theory which is conformally invariant, i.e. a CFT. 
Below we will use the notations $x^M  = (z, x^\mu)$ where $x^{\mu} = (t, x_i)$ run over the coordinates of the boundary theory. $z \in (0, +\infty)$ is the extra ``holographic'' coordinate, with the AdS boundary lying at $z=0$,\footnote{Notice that as $z \to 0$, the overall prefactor $1/z^2$ in~\eqref{eq.poincareadsmetric} blows up which is analogous to $r \to \infty$ limit of 
a flat Euclidean metric $ds^2 = r^2 d \Om^2$ in spherical coordinates.} and large value of $z$ can be considered as the ``interior'' of AdS (see Fig. \ref{fig:AdS}). 

 The metric~\eqref{eq.poincareadsmetric} has a large number of isometries (i.e. coordinate transformations which leave the metric invariant), which are in one-to-one correspondence with conformal transformations of  
the boundary system.

Among all the isometries of~\eqref{eq.poincareadsmetric} we would like to draw particular attention to the following scaling symmetry
\be
z \to \lambda z \qquad x^{\mu} \to \lambda x^{\mu} \ . \label{scalisom}
\ee
We see that as we scale the boundary coordinates $x^\mu$ we must accordingly scale the holographic coordinate $z$. This indicates that $z$ represents length scales of the boundary theory: we scale to short distances (UV) in $x^\mu$ as $z$ scales to $0$, and to long distances (IR) in $x^\mu$ as $z$ scales to $\infty$. In other words, going from the boundary $z=0$ to some large values of $z$ along the radial direction may be considered as going from UV to IR in the boundary system. This turns out to be a general feature of all bulk geometries, including those for which~\eqref{scalisom} is no longer an isometry. 
Recall that the central idea of renormalization group (RG) is to organize physics of a many-body system in terms of scales, thus 
the radial direction of AdS can be considered as a geometrization of renormalization group (RG) flow of a boundary theory! 

Another simple solution to~\eqref{eq.EHaction} is the Schwarzschild black hole
 \be \label{equi1}
 ds^2 = \frac{\ell^2}{z^2} \left( -f(z) dt^2 + \frac{dz^2}{f(z)} +dx_i^2  \right)\,,
 \ee
 where $f(z) = 1 - \left({  z \ov z_h}\right)^d$ and $z_h$ is a constant. Equation~\eqref{equi1} has an event horizon at $z = z_h$ with  topology\footnote{Thus strictly speaking one should call~\eqref{equi1} a black brane rather a black hole whose horizon has the topology of a sphere.} 
 $\RR^{d-1}$.  From the discoveries of Hawking and Bekenstein in the 1970s,  black holes are known to be thermodynamic objects. Thus it is natural to identify the solution~\eqref{equi1} with a thermal state of the boundary system,  
 with the Hawking temperature 
 \be \label{hawT}
 T_H = {d \ov 4 \pi  z_h}
 \ee
  identified with the boundary system temperature. 
 Note that as $z \to 0$, $ f(z) \to 1$, equation~\eqref{equi1} reduces to~\eqref{eq.poincareadsmetric}. This is consistent with the above discussion of $z$ as representing length scales: as $z \to 0$ we go to short distances and recover vacuum physics, while 
 the whole geometry~\eqref{equi1} tells us how the system flows from vacuum physics at short-distances (UV) to thermal physics at IR scales. The presence of event horizon at some finite value of $z=z_h$ can be considered as an IR ``cutoff'' representing the inverse temperature scale. In contrast, in~\eqref{eq.poincareadsmetric}  the values of 
 $z$ extend all the way to $+\infty$, reflecting that near the vacuum, there  there exist excitations of arbitrarily low energies.

Now let us turn to another crucial aspect of the duality dictionary: correspondence of operators. 
On the gravity side, the operators are fields living in AdS spacetime, each of which is mapped to an operator 
in the boundary system. 
Clearly the dual pair  must have the same quantum numbers under symmetries of the theory.
For example, a scalar boundary operator should be dual to some bulk scalar field.
Since different boundary theories have different operator spectra,  the precise dictionary depends on 
specific systems. Nevertheless,  there are some common elements universal to all theories: (i) 
the boundary stress tensor $T^{\mu \nu}$ is dual to spacetime metric  $ g_{MN}$; (ii) 
 a conserved boundary current $J^\mu$ is dual to a bulk 
gauge field $A_M$.

Solving equations of motion of a bulk field $\phi$ (say, dual to some boundary operator $\sO$), one finds that near the 
boundary a general solution can be written as a superposition of two independent powers of $z$ 
 \be \label{asymp}
\phi(z \to 0,x^{\mu}) = a(x) z^{\al_1} + b(x)z^{\al_2} , \quad \al_1< \al_2  
\ee
where $a(x), b (x)$ are ``integration constants'' and $\al_{1,2}$ are some constants which depend on the mass and spin of $\phi$. 
Here we have suppressed possible spacetime indices in $\phi, a, b$, which could be tensors or spinors.
As $z \to 0$, the first term in~\eqref{asymp} dominates and is often referred to as the non-normalizable term, while the second term as the normalizable term.   

Various quantities in~\eqref{asymp} turn out to have important physical interpretations 
in the boundary system:
\bi

\item $a(x)$ can be identified as the source for $\sO$. More explicitly, a nonzero $a(x)$ corresponds to deforming 
 the boundary CFT by a term 
\be 
S_{\rm CFT} \to S_{\rm CFT} + \int d^d x \, a(x) \sO (x) \ . \label{singtrace}
\ee 
In other words, the presence of a non-normalizable term modifies  the boundary theory itself. 

\item $b(x)$ can be identified with the expectation value of the operator $\sO(x)$ in the corresponding state described by the bulk geometry, i.e.
\be
\langle \sO(x) \rangle \propto b(x)\,, \label{vevdef}
\ee
where we have suppressed possible proportionality constant (which can depend on scaling dimension of $\sO$ and tensor structure). Thus the $b$'s for different $\phi$'s tell us important information about the state of the system.  

\item  
$\al_{1,2}$ are related to the conformal dimension $\De$ of  $\sO$
\be 
\al_1 = d - \De -n, \qquad \al_2 = \De -n 
\ee
where 
$n$ is the number of tensor indices of $\phi$. 
Thus from boundary behavior~\eqref{asymp} of a bulk field one could read the boundary conformal dimension of the corresponding operator. 
\item  For a scalar field of mass square $m^2$, by solving the associated bulk equations, one finds that $\al_1 = d - \De$ and $\al_2 = \De$ 
with 
\be
\Delta = \frac{d}{2} + \nu, \qquad \nu =
 \sqrt{\frac{d^2}{4} + m^2 \ell^2}. \label{UVdim}
\ee
Note that in an AdS spacetime $m^2$ can in fact  be negative as far as not too negative  that the square root in~\eqref{UVdim}  
becomes complex. 
For $m^2 < 0$, we have $\Delta < d$, i.e. the corresponding operator $\sO$ is relevant, while $m^2=0$ corresponds to a marginal operator, and  $m^2 >0$ to an irrelevant operator. For a scalar, the precise version of~\eqref{vevdef} is 
\be
\vev{\sO (x)} = 2 \nu b(x) \ .
\ee
An interesting subtlety arises when the bulk mass satisfies
\be
-\frac{d^2}{4} < m^2 < -\frac{d^2}{4}+1\,.
\ee
For an operator that is dual to a bulk field in this range, there exists a second allowed prescription, called `alternative quantization', where the role of source and expectation value above are exchanged. This is described in detail in \cite{Klebanov:1999tb}.

\item A conserved current $J^\mu$ is dual to a bulk gauge field $A_M$ whose action at quadratic level is simply the Maxwell action 
\be \label{Max}
S_{\rm Max} = -{1 \ov 4e^2} \int dz \, d^{d} x \, \sqrt{-g} \, F_{MN} F^{MN}  \ . 
\ee 
One finds $\al_1 = 0, \al_2 = d-2$, i.e. 
\be \label{bgr}
A_\mu (z \to 0, x) = a_\mu (x) + b_\mu (x) z^{d-2}  
\ee
which implies $\De = d-1$ and is indeed consistent with the scaling dimension of a conserved current. The proportional constant in~\eqref{vevdef} is
\be \label{jvev}
\vev{J_\mu} = - \frac{\ell^{d-3} }{e^2} (d-2) b_\mu \ .
\ee

\item For metric perturbations, one finds 
$\al_1 = -2, \al_2 = d-2$, i.e. 
\bea
\de g_{\mu \nu} (z \to 0, x) &=& a_{\mu\nu} (x) \left(z^{-2} + \cdots \right) \nonumber \\
  &&+ b_{\mu \nu} (x) \left(z^{d-2} + \cdots \right)  
\eea
which is consistent with $\De = d$ for the stress tensor, and~\eqref{vevdef} has the form
\be
\vev{T_{\mu \nu}}=  c b_{\mu \nu}  \,,
\ee
where, given the canonical normalization of the Einstein-Hilbert action, the constant $c$ takes on the value
\be
c = \frac{d\ell^{d-1}}{16 \pi G_N} \,.
\ee
We are  ignoring the subtleties of holographic renormalization, such as possible contribution from the Weyl anomaly in even dimensions (see \cite{deHaro:2000vlm} for details). A nonzero $a_{\mu \nu}$ implies that the boundary metric is deformed to $\eta_{\mu \nu} + a_{\mu \nu}$ with 
$\eta_{\mu \nu}$ the flat Minkowski metric. 

\item For any relevant operator, the backreaction of the corresponding non-normalizable term to the spacetime geometry 
always goes to zero as $z \to 0$. This can be seen, for example, from that for a scalar $\al_1 = d - \De > 0$, while for $A_M$ it becomes $z$-independent.

\ei
There are also systematic procedures for finding higher point functions of boundary operators. 
We describe how to compute retarded two-point functions in next subsection and general real-time multiple-point functions defined on a Schwinger-Keldysh contour in Sec.~\ref{sec:SK}.

Now consider a CFT deformed by a relevant operator $\sO$, i.e. by adding to the action a term $\lam \int \sO$. As we go to larger distances or lower energies, the system will move farther and farther away from the UV fixed point,  and eventually to some other IR fixed point or a gapped phase. 
On the gravity side, this amounts to finding the gravity solution in which the bulk field $\phi$ dual to $\sO$ satisfies $a(x) = \lam$, but the corresponding $a(x)$ for other fields must all vanish. Near the boundary, the geometry is close to~\eqref{eq.poincareadsmetric} (see the last item above), but as $z$ increases the deviation becomes larger and larger, and eventually transitions to 
the geometry representing the IR state (see Fig. \ref{fig:AdS}).

As a simple example, let us consider turning on a chemical potential for a conserved $U(1)$ charge, i.e. 
adding a term $\mu \int J^t$ to the boundary action, with $J^t$ the time component of a conserved current $J^\mu$. 
Since $J^\mu$ has dimension $\De = d-1$, this is a relevant perturbation.  
On the gravity side, in the simplest situation, only $A_M$ is excited and one needs to find a solution to $S_{\rm grav} + S_{\rm Max}$ (i.e. combining~\eqref{eq.EHaction} and~\eqref{Max}) which satisfies the boundary condition $A_t (z \to 0) = \mu$. 
The most general solution satisfying the boundary condition has the form~\eqref{equi1}
but with a different $f$ and a nonzero $A_t$ given by
 \be \label{bhga2}
 f = 1 + { Q^2 z^{2d-2}} - {M  z^d}, \quad A_t = \mu \le(1- {z^{d-2} \ov  z_0^{d-2}}\ri) \ ,
 \ee
where $Q, M, z_0$ are constants. This solution again has an event horizon located at $z=z_h$ where $z_h$ is the largest root of  $f(z_h) =0$.  The geometry~\eqref{bhga2} describes a CFT at finite chemical potential $\mu$, and a finite temperature which is again identified with the Hawking temperature of~\eqref{bhga2}.

\subsection{Linear responses and quasi-normal modes}  \label{sec:ret}

In preparation of our discussion of far-from-equilibrium systems, here we briefly mention some key aspects 
concerning near-equilibrium systems.

When a weak external field is applied to an equilibrium system, the resulting displacement from the equilibrium state is small, and at lowest order can be treated as linear in the external source.  For example, turning on a source $a(x)$ coupled to a Hermitian operator $\sO$ (whose expectation values are taken to be zero in equilibrium), we have 
\be \label{linr}
\vev{\sO} (\om, \bk) = G^R (\om, \bk) a (\om, \bk) \,, 
\ee
where $G^R (\om, \bk) $ is the retarded Green function
\be 
G^R (x) = i \th (x^0) \vev{[\sO (x), \sO (0)]}
\ee
in momentum space. 
The linear responses of a system under various external fields  are most commonly used experimental probes 
and contain a wealth of dynamical information: 

\ben 

\item The static susceptibility is obtained as 
\be 
\chi= \lim_{\bk \to 0} \lim_{\om \to 0}  G^R (\om, \bk) \ .
\ee

\item  When $\sO$ in~\eqref{linr} is the current for a conserved quantity, various transport coefficients can be obtained from 
the zero momentum and zero frequency limit. For example, the DC conductivity $\sigma$ along some direction $i$, is obtained by taking $\sO = J^i$ where 
$J^i$ is the $i$-th component of the conserved current: 
\be 
\sig = \lim_{\om \to 0} \lim_{\bk \to 0} {1 \ov i \om} G^R (\om, \bk) \ .
\ee

\item The imaginary part of $G^R$ gives the spectral function, 
\be 
\rho (\om, \bk) = {\rm Im} G^R\,,
\ee
which encodes the spectral weight of $\sO$.

\item For a general non-conserved operator $\sO$, $G^R$ generically has singularities such as poles or branch points
in the lower half $\om$-plane which are a finite distance away from the real $\om$-axis as $\bk \to 0$.  The nearest singularity controls the relaxation times of $\sO$. More explicitly, suppose  
$G^R$ has a pole at $\om_* (\bk) = \om_1 (\bk) - i \om_2 (\bk)$, the contribution of the pole to 
$\sO (t, \bk)$ is then given by 
\be \label{ghl}
\vev{\sO(t, \bk)} \propto e^{- i \om_* (\bk) t} \propto e^{- \om_2 (\bk) t - i \om_1 (\bk) t}  \ .
\ee
The late-time behavior is thus dominated by the pole with the smallest $\om_2$. 
Note that for a stable state, $\om_2$ of any pole must be positive. 
A negative $\omega_2$ leads to exponential growing behavior in~\eqref{ghl} and 
signals instability.

\item For $\sO$ given by a conserved quantity such as energy, momentum or charge densities, $G^R$  exhibits poles in the complex $\om$ plane which approach
the origin as $\bk \to 0$. These are hydrodynamical modes such as sound  and diffusion modes which reflect that conserved quantities relax much slower than typical time scales of microscopic interactions.

\een

The response function $G^R$ for an operator $\sO$ in a thermal equilibrium state 
can be obtained as follows~\cite{Son:2002sd}. 
One solves the linearized equation of motion for the bulk field $\phi$ corresponding to $\sO$ 
in the black hole geometry for the thermal state. In a classical black hole geometry, things can only fall into a black hole, and cannot come out. Otherwise causality is violated. Thus to obtain the retarded function, which is causal, one should choose the solution for which there is only the ingoing behavior at the horizon. For a second order differential equation, this fixes 
the asymptotic behavior~\eqref{asymp}, i.e. $b$ and $a$ up to an overall multiplicative factor. 
From~\eqref{linr} and the identifications of $b$ and $a$ respectively as the expectation value and the source, we then conclude that, for example for a scalar 
\be
G^R (\om, \bk) = 2\nu \frac{b(\om, \bk)}{a(\om, \bk)} \,,  \label{GRdef}
\ee
where $b(\om, \bk)$ and $a(\om, \bk)$ are the Fourier transform in $t$ and $x_i$ directions. 
This prescription can also be derived by analytic continuation from Euclidean signature or the more elaborate real-time formalism discussed in Sec.~\ref{sec:SK}.

As reviewed in the above, key information regarding $G^R$ is its pole structure in the complex $\om$-plane. 
From~\eqref{GRdef} we see that poles of $G^R$ correspond to\footnote{From the structure of the bulk differential equation one can show that $b$ cannot have poles.}
\be \label{eq.QNMnaive}
a (\omega,\mathbf{k}) = 0  \ .
\ee
Equation~\eqref{eq.QNMnaive} in turn implies that the non-normalizable piece in Equation~\eqref{asymp} vanishes, i.e. the solution should be normalizable. To summarize, the poles of $G^R$ correspond to bulk solutions which are in-falling at the horizon and 
normalizable at infinity. Since this involves two-sided boundary conditions, for a given $\bk$, the allowed values of $\om$ should be discrete, i.e.~\eqref{eq.QNMnaive} has a discrete spectrum $\om_n (\bk), n=1,2, \cdots$. 
 Furthermore, the spectral problem defined by studying modes with ingoing boundary conditions is not self-adjoint, and the corresponding eigenfrequencies are in general complex.  
 
The eigenmodes $\{\om_n (\bk)\}$  have long been studied in the general relativity community, and are known as quasinormal modes (often abbreviated QNM). QMNs play an important role in gravitational dynamics, as they describe how a small normalizable perturbation around a black hole evolves with time, and thus characterize the long-time behavior of dynamical black holes. A black hole formed from collapse, for example, after an initial non-equilibrium phase, displays a characteristic ring-down at late times. This ring-down is directly related to the quasinormal modes. Holography thus translates the question of the analytic structure of retarded correlation functions into the study of quasinormal modes of the dual black-hole geometry. 
Note that any perturbation will eventually fall into the black hole, which implies that, unless there is an instability, the imaginary parts of $\{\om_n (\bk)\}$ should be all negative, which is consistent with general boundary theory expectation discussed below~\eqref{ghl}.

\subsection{A simple example}

As an illustration of the discussion of the last subsection we work out the small $\om, \bk$ behavior of the 
thermal retarded two-point function of a conserved $U(1)$ current $J^\mu$ in (2+1)-dimension. The calculation recovers the expected diffusion behavior and computes explicitly the conductivity and diffusion constant for a strongly coupled conformal field theory with a gravity dual.

As mentioned earlier,  a conserved current operator $J^\mu$ is dual to a gauge field $A_M$, with action~\eqref{Max}. 
To compute thermal two-point functions of $J^\mu$, as in \cite{Policastro:2002se}, we consider equations of motion for $A_M$ around the black hole geometry~\eqref{equi1} (here we take $d=3$). It is convenient to choose the gauge $A_z =0$, and 
we will work in Fourier space
 \be\label{nh}
 A_\mu (z, x^\mu) = 
 e^{-i\omega t + i q y}A_\mu(z, \omega,q)\,,
 \ee
where without loss of generality we have oriented the spatial momentum along the $y$ direction and we distinguish the original variables from the Fourier transforms merely by their arguments.
The  Maxwell equations in the geometry~\eqref{equi1} then read (with $'$ denoting $z$ derivatives)
 \bea\label{eq.chargediffusion}
 \mathfrak{w} A_t' + \mathfrak{q} f A_y'&=&0\\
 A_t'' - \frac{1}{f} \left(  \mathfrak{q}^2 A_t + \mathfrak{w} \mathfrak{q} A_y\right) &=&0\\
 \label{axeq}
 A_x'' + \frac{f'}{f}A_x'- \frac{1}{f} \left( \mathfrak{q}^2 - \frac{\mathfrak{w}^2}{f} \right)A_x &=&0\\
 A_y'' + \frac{f'}{f}A_y' + \frac{\mathfrak{w}^2}{f^2}A_y + \frac{\mathfrak{w q}}{f^2}A_t&=&0\,.
 \eea
 We have rescaled the radial coordinate by $z\rightarrow z/z_h$ which means that we now have $f(z) = 1-z^3$ with a horizon at $z=1$, while $ \mathfrak{w} = \om z_h $ and  $\mathfrak{q} = q z_h $ are frequency and momentum measured in units of $1/z_h$ which is turn proportional to the Hawking temperature~\eqref{hawT}. 
 
 It can be directly verified that the solutions to these equations indeed have the asymptotic behavior~\eqref{bgr}, i.e. 
 \be\label{eq.asymptAmu}
A_\mu(z,\omega,\mathbf{k}) = a_\mu (\omega,\mathbf{k}) + b_\mu (\om,\bk) z + {\cal O}(z^2)  \ .
\ee
 Thus from~\eqref{jvev}, we find
 \be \label{fgh}
 \vev{J_\mu (\om, \bk)} = - \frac{1}{e^2} b_\mu (\om,\bk) = - \frac{1}{e^2}A_\mu' (z=0) \ .
 \ee
  Now notice that equation~\eqref{eq.chargediffusion} contains only first derivative in $z$, thus should be considered as a constraint equation for the $z$-evolution. Using~\eqref{fgh} we see that, when evaluated at the boundary, it is simply 
 the conservation equation $\p_\mu J^\mu =0$ in Fourier space. This is a general feature: conservation laws of the boundary correspond to constraint equations in the bulk. Note that the $A_x$ equation~\eqref{axeq} decouples from the rest; this is again expected on general ground, with $\bk$ aligned in the $y$ direction, $J_x$ decouples from the rest of the current in the conservation equation. 
Thus we will focus on the sector $\left\{ A_t,  A_y \right\}$. 
 
The equations for $A_t, A_y$  can be reduced to a single second order differential equation 
 \be\label{ezeq}
 E_z'' + \frac{f'}{f} E_z' + \frac{1}{f} \left( \frac{\mathfrak{w}^2}{f} - \mathfrak{q}^2 \right)E_z =0\,,
 \ee
with $E_z \equiv A_t'$ and 
 \be\label{eq.AyAtrelation}
  A_y = \frac{f}{\mathfrak{q w}} A_t'' - \frac{\mathfrak{q}}{\mathfrak{w}} A_t\,.
 \ee
Near the horizon $z =1$, $f \to 0$,  one finds~\eqref{ezeq} can be written as 
\be 
\p_{z_*}^2 E_z +  \mathfrak{w}^2 E_z = 0
\ee
where $z_* = - \int {dz \ov f} = -{1 \ov 3} \log (1-z) + \cdots$ (as $z \to 1$) is the 
 so-called tortoise coordinate. Thus $E_z$ 
behaves as a plane wave in $z_*$
\be \label{op}
E_z \sim e^{\pm i  \mathfrak{w} z_*} \sim (1-z)^{\pm {i \ov 3} \mathfrak{w}} , \quad z \to 1\ .
\ee
Note that as $z \to 1$, $z_* \to - \infty$. Including $t$-dependence~\eqref{nh}, the $+$ ($-$) sign  
in~\eqref{op} then describes a wave moving away from (going toward) the horizon. 
As mentioned around~\eqref{GRdef}, to find the retarded Green function we need to take the solution which goes into 
the horizon, i.e. $-$ sign in~\eqref{op}. 

Now our task is to find the solution to~\eqref{ezeq} which behaves at the horizon with the $-$ sign in~\eqref{op}, 
expand the solution at the boundary $z \to 0$, and then read from~\eqref{eq.asymptAmu}--\eqref{fgh} the explicit expression for $\vev{J_\mu}$ with any given $a_\mu$. The components of the retarded Green functions can be read from
\be
\vev{J_\mu(\om, \bk)} = G^R_{\mu \nu} (\om, \bk)  a_\nu (\om, \bk) \ .
\ee
While~\eqref{ezeq} cannot be solved analytically for general $\om, q$, it can be solved analytically 
for small $\om, q$ which is enough to extract transport behavior. Expanding in small $\mathfrak{w}, \mathfrak{q}$  one finds that 
\be 
\vev{J_t} = \frac{1}{e^2}  \frac{\mathfrak{q}^2 a_t  + \mathfrak{w q} a_y } { i \mathfrak{w}-\mathfrak{q}^2}\,
\ee
with $e^2$ the dimensionless bulk gauge coupling.
 \bea
 G^R_{tt}  = \langle J_t(\omega,q)J_t(-\omega,-q)\rangle &=& \frac{1}{e^2}\frac{q^2}{i\omega -  D q^2} \nonumber \\
 \\
 G^R_{ty}  = \langle J_t(\omega,q)J_y(-\omega,-q)\rangle &=&  \frac{1}{e^2}\frac{q\omega }{i\omega  - D q^2} \nonumber\\ 
 \eea
where $D = \frac{3}{4\pi T}$. As expected, the above expression exhibits a pole at 
$\om = - i D q^2$ , corresponding to the physics of charge diffusion with diffusion constant $D$. Note that $D \propto {1 \ov T}$ as expected for a scale invariant theory.  Similarly, one finds that 
\be
 G^R_{yy}  = \langle J_y(\omega,q)J_y(-\omega,-q)\rangle  = \frac{1}{e^2}\frac{\omega^2}{D q^2-i \omega}
 \ee
from which we can extract the conductivity
\be
\sig =  \lim_{\om \to 0, q\to 0} {1 \ov i \om} G_{yy}^R (\om, q) =  \frac{1}{e^2}\,,
\ee
which is a constant in $2+1$ dimensions \cite{Herzog:2007ij}.

\subsection{Entanglement entropy}\label{sec:RTsurface}

So far we have been discussing local operators. A theory can also contain nonlocal observables. 
For example, in a gauge theory one can have Wilson loops. When the Hilbert space of a quantum system has 
a tensor product structure, one can also define entanglement entropy associated with a subset of degrees of freedom. 
Here we briefly review the prescription for computing entanglement entropy associated with a region, also called geometric entropy, using gravity. 

Consider a spatial subregion $A$ of a boundary system. Imagine a UV regularization (say putting the system on a lattice) such that the Hilbert space factorizes into ${\cal H} = {\cal H}_A \otimes {\cal H}_{\overline A}$, where $\overline A$ denotes the complement of $A$.  The entanglement entropy of subregion $A$, $S(A)$, is then defined as 
 the von-Neumann entropy of the reduced density matrix $\rho_A$ 
\be
\rho_A = {\rm tr}_{{\cal H}_{\overline A}} \rho\qquad \Rightarrow \qquad S(A) = - {\rm tr} \rho_A \log \rho_A\,.
\ee
In a many-body system (even including non-interacting systems!), computing $S(A)$ is a very difficult task. One typically proceeds by computing first the R\`enyi entropy
\be
S_n := \frac{1}{1-n}{\rm log}\, {\rm tr}\rho_A^n\qquad \Rightarrow \qquad S(A) = \lim_{n\rightarrow 1} S_n\,,
\ee
where $S_n$ for $n\in \mathbb{Z}$ is computed using the so-called replica trick \cite{Holzhey:1994we,Calabrese:2004eu}, and the limit $n\rightarrow 1$ is taken formally by analytically continuing the result away from integer values.

\begin{figure}
\begin{center}
\includegraphics[width=\columnwidth]{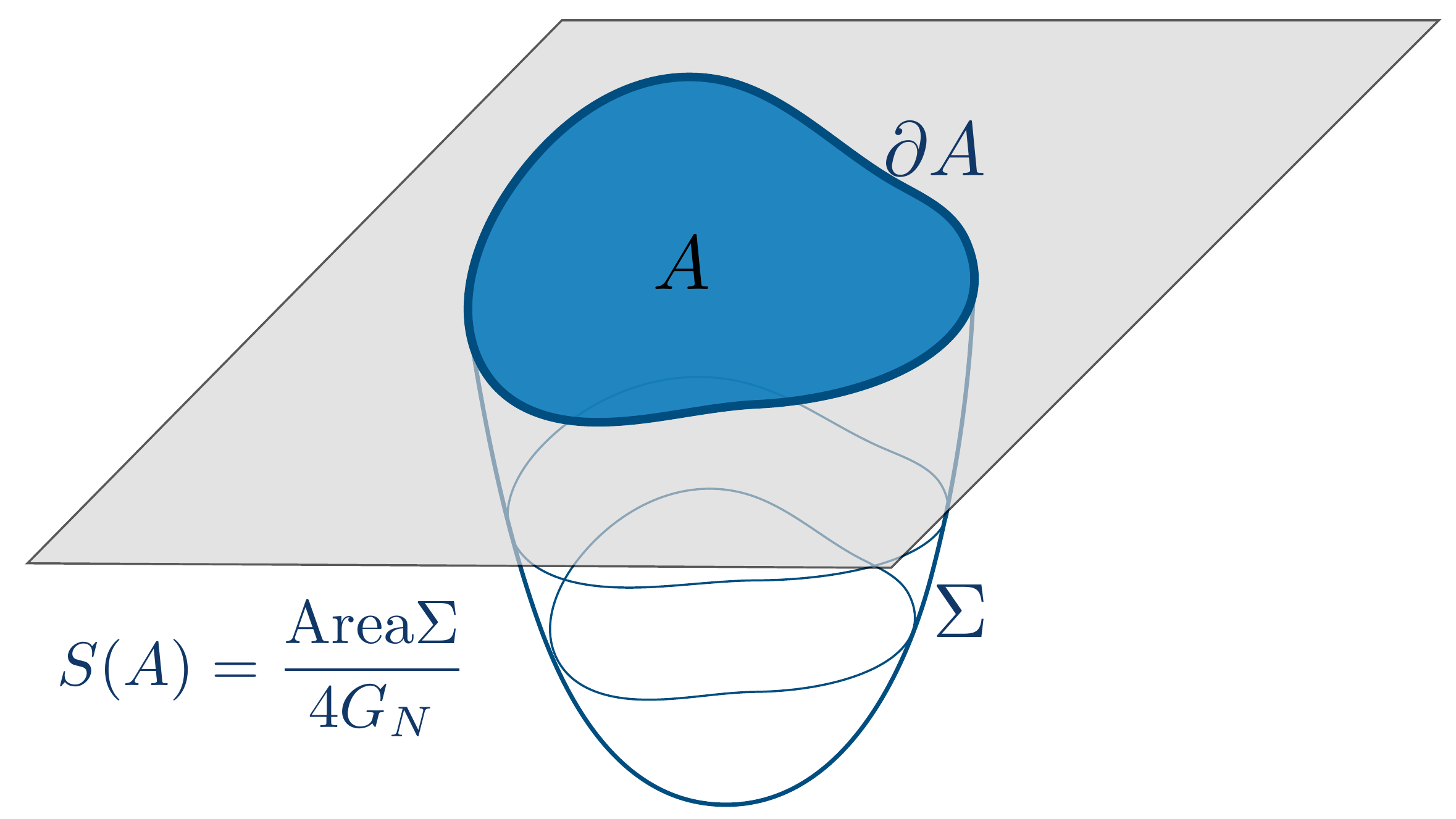}\\
\end{center}
\caption{Ryu-Takayangi prescription for the calculation of the entropy of entanglement of the subregion $A$ with respect to its complement $\overline{A}$. The bulk minimal surface $\Sigma$ shares the boundary of the subregion $A$, i.e. $\partial A = \partial \Sigma$ and the entropy $S(A)$ is then given by the area of $\Sigma$ in the way indicated.}
 \label{fig:RT}
\end{figure}

In holographic duality $S(A)$ can be directly obtained without using the replica trick. It involves a beautiful geometric formula first proposed by Ryu and Takayanagi \cite{Ryu:2006bv} and later generalized to time dependent situations in~\cite{Hubeny:2007xt}. 
More explicitly, $S(A)$ for the system in a given state is obtained by 
\be\label{trf}
S(A) = \frac{{\rm area} \, \Sigma}{4 G_N}\,.
\ee 
In the above equation $\Sig$ is an extremal surface homologous to $A$ in the corresponding bulk geometry for the state, with the boundary of $\Sig$ ending on the boundary of region $A$ (see Fig.~\ref{fig:RT}). 
 When there is more than one extremal surface satisfying the boundary conditions, one should choose the one with the smallest area. In the AdS context, equation~\eqref{trf} generalizes the Bekenstein-Hawking formula for the entropy of a black hole which may now be considered as a special example of~\eqref{trf}.

The prescription~\eqref{trf} entirely bypasses the computation of the R\'enyi entropies, which turn out to be much more complicated to determine. To compute $S_n$ for general $n$, one needs to find the bulk gravity geometry dual to the boundary theory on a multi-sheeted cover of the original spatial manifold, and then compute its partition function, which is much more involved than finding an extremal surface.

\section{Holographic Schwinger-Keldysh formulation} \label{sec:SK}

In this section we discuss how to compute real-time correlation functions defined on a  
Schwinger-Keldysh contour using gravity. The formulation for an equilibrium state is essential complete, but finding a prescription applicable to general non-equilibrium situations is still an open problem.

\subsection{General remarks}

\begin{figure}[!h]
\begin{center}
\includegraphics[width=8cm]{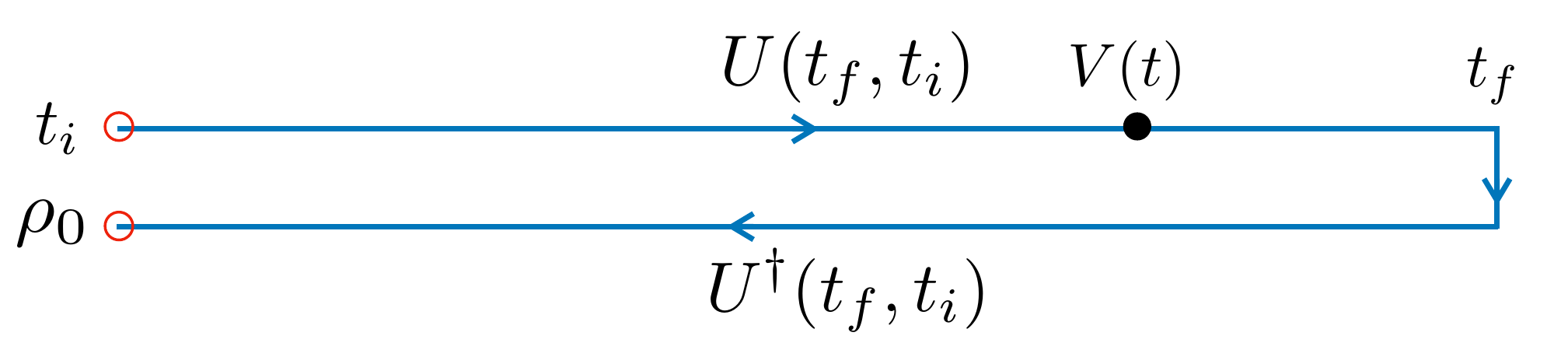}\\
\end{center}
\caption{Schwinger-Keldysh contour.}
 \label{fig:sk}
\end{figure}

In the standard formulation of quantum many-body physics, real-time response and fluctuation functions in a state 
given by a density matrix $\rho_0$  can be 
obtained from path integrals on a Schwinger-Keldysh contour (or closed time path) as indicated in 
Fig.~\ref{fig:sk}. The central object is the generating functional 
\be\label{gen0}\begin{gathered}
e^{W [\phi_{1i}, \phi_{2i}]}
= \Tr \left[\rho_0 \sP e^{ i \int d t \, (\sO_{1i} (t) \phi_{1i} (t) - \sO_{2i} (t) \phi_{2i} (t)) }\ri]\,,
\end{gathered}
\ee
where $\sO_i$ denote generic operators and $\phi_i$ their corresponding sources. Note that $\sO_{1i}$ and $\sO_{2i}$
are the same operator, with subscripts $1,2$ only indicating the segments of the contour in which they are inserted, while 
$\phi_{1i}$ and $\phi_{2i}$ are distinct fields.  $\sP$ indicates that the operators are path ordered.
The minus sign in the second term comes from the reversed time integration for the second (lower) segment. 

Given a time-dependent gravity solution, there does not yet exists a fully general procedure to compute~\eqref{gen0}. 
When the state $\rho_0$ can be prepared by a Euclidean path integral, one can write~\eqref{gen0} as a path 
integral involving some Euclidean and some Lorentzian segments. In this case one can obtain a corresponding gravity spacetime 
by patching together different pieces:  one associates with each real-time branch of the contour a 
Lorentzian spacetime, and with each imaginary-time branch a Euclidean spacetime. Different branches are joined together using patching conditions, roughly,  the bulk fields and their derivatives should be continuous.   
 With the full path integration contour represented on the gravity side, the generating 
functional~\eqref{gen0} can then be obtained using the standard procedure of integrating over the bulk fields with sources as boundary conditions. See~\cite{Skenderis:2008dh,Skenderis:2008dg} for discussions, as well as Fig. \ref{fig:SkenderisVanRees} for more detail on a specific example of such a construction. This approach is conceptually straightforward, but in practice tedious to carry out even for a thermal equilibrium computation. 
For a general non-equilibrium state, it is not clear how to set up $\rho_0$ as initial/final conditions even when  the corresponding bulk gravity solution is known.

Fortunately, for many questions of interest, there exist methods which take advantage of the analytic structure 
of the relevant gravity solutions. We first discuss the computation of~\eqref{gen0} in a thermal ensemble~\cite{Herzog:2002pc,Son:2009vu}, and then a more general proposal which applies to any spacetime with an analytic horizon (which does not have to be thermal)~\cite{hongPaoloMikeToAppear}.

\subsection{Momentum space formulation for two-point functions in a thermal state\label{sec.MomentumSpace}} 

\begin{figure*}[t] 
\begin{center}
\includegraphics[width=0.7\textwidth]{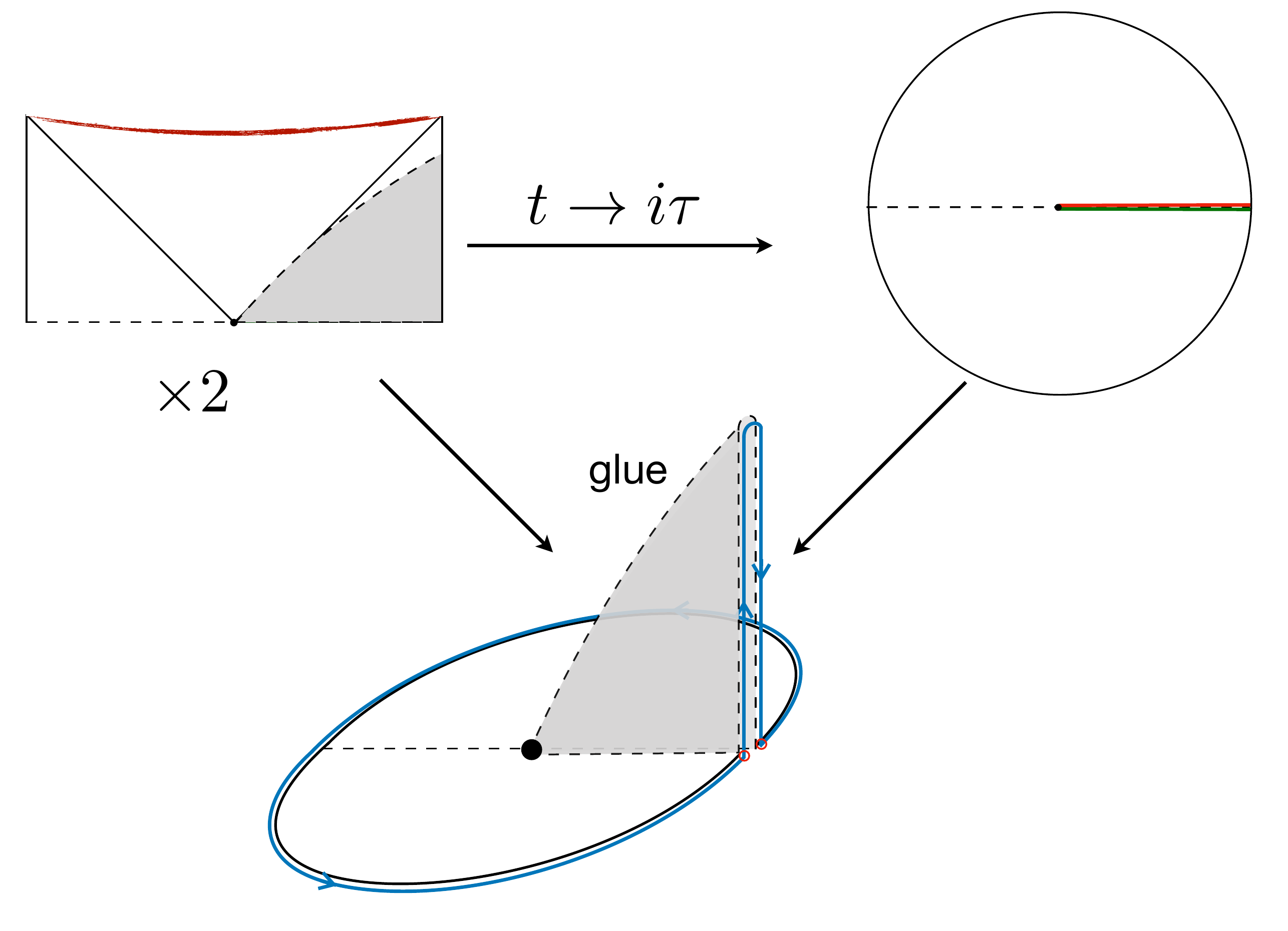}
\end{center}
\caption{Bulk geometry corresponding to the Schwinger-Keldysh contour in Fig. \ref{fig:sk1}. The top left geometry is the upper part of the Lorentzian eternal black hole (see Fig. \ref{fig:bh}) that has been cut along the moment of time symmetry indicated by the dashed line piercing the bifurcaction surface at the bottom. The part of this slice we are interested in is the region shaded in gray, located entirely in the right exterior region. The diagram on the top right depicts the Euclidean version of the eternal black-hole geometry, which also has the slice of (Euclidean) time symmetry indicated. We glue one copy each of the gray shaded Lorentzian region to the Euclidean section, gluing as indicated one along the red and and one along the green curve. For each gluing we impose that all metric fields as well any propagating matter extend to $C^1$ functions on the whole geometry \cite{Skenderis:2008dh,Skenderis:2008dg,vanRees:2009rw}. We similarly glue the two Lorentzian parts together at the dashed curve which starts out from the bifurcation surface just hugging the future horizon. The latter operation has the effect of `folding over' the Lorentzian geometry, as suggested by the folded form of the Schwinger-Keldysh contour itself. The blue boundary curve represents exactly the Schwinger-Keldysh contour of Fig.~\ref{fig:sk1}.
\label{fig:SkenderisVanRees}}
\end{figure*}   

\begin{figure}[!h]
\begin{center}
\includegraphics[width=8cm]{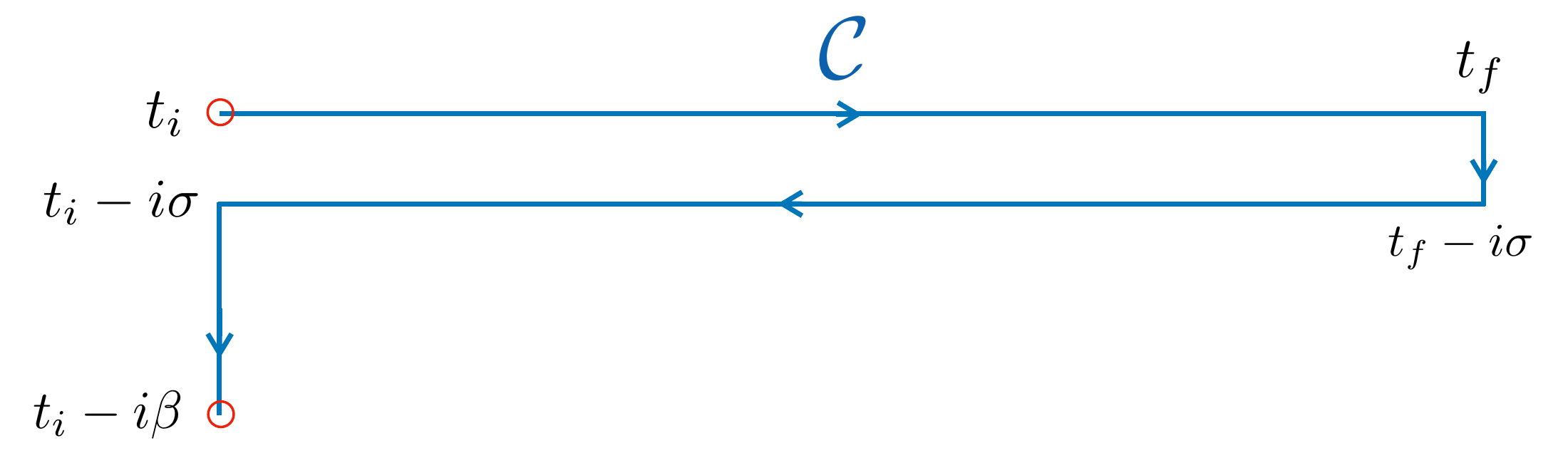}\\
\end{center}
\caption{Standard Schwinger-Keldysh contour for thermal field theory correlation functions.}
 \label{fig:sk1}
\end{figure}

Let us now restrict to a thermal ensemble with 
\be \label{them}
\rho_0 = {1 \ov Z} e^{-\beta H} , \quad Z = \Tr e^{-\beta H} \ ,
\ee
which is time-translation invariant. In addition to the contour of Fig.~\ref{fig:sk} 
one could move 
part of the Euclidean segment which represents $e^{-\beta H}$ to other times, such as $t = \infty$ as indicated 
in Fig.~\ref{fig:sk1} with a general $\sig \in [0, \beta)$.  Correlation functions obtained using different choices of $\sigma$
are different, but they can be related by simple analytic continuations, and thus encode 
the same physical information. 

On the gravity side,~\eqref{them} is described by an eternal black hole geometry, with a metric 
\be \label{bhm}
ds^2 = - f (r) dt^2 + {1 \ov f (r)} dr^2 + r^2 d \vec x^2  \ .
\ee
The detailed form of function $f$ is not important except that it has a zero at some $r=r_0$, which is the location of the event horizon, and the inverse Hawking temperature is
\be 
\beta = {4 \pi \ov f' (r_0)} \ .
\ee
The coordinates $r$ and $t$ are appropriate for $r > r_0$, and become singular at the horizon $r=r_0$. 
By using the so-called Kruskal coordinates,  one could extend the black hole geometry past $r=r_0$, with 
the Penrose diagram of maximally extended spacetime geometry\footnote{While this section gives all necessary details needed for the present purpose, for completeness we provide more details on the Kruskal construction in appendix \ref{app.Kruskal}.} shown in Fig.~\ref{fig:bh} .

\begin{figure}[!h]
\begin{center}
\includegraphics[width=0.8\columnwidth]{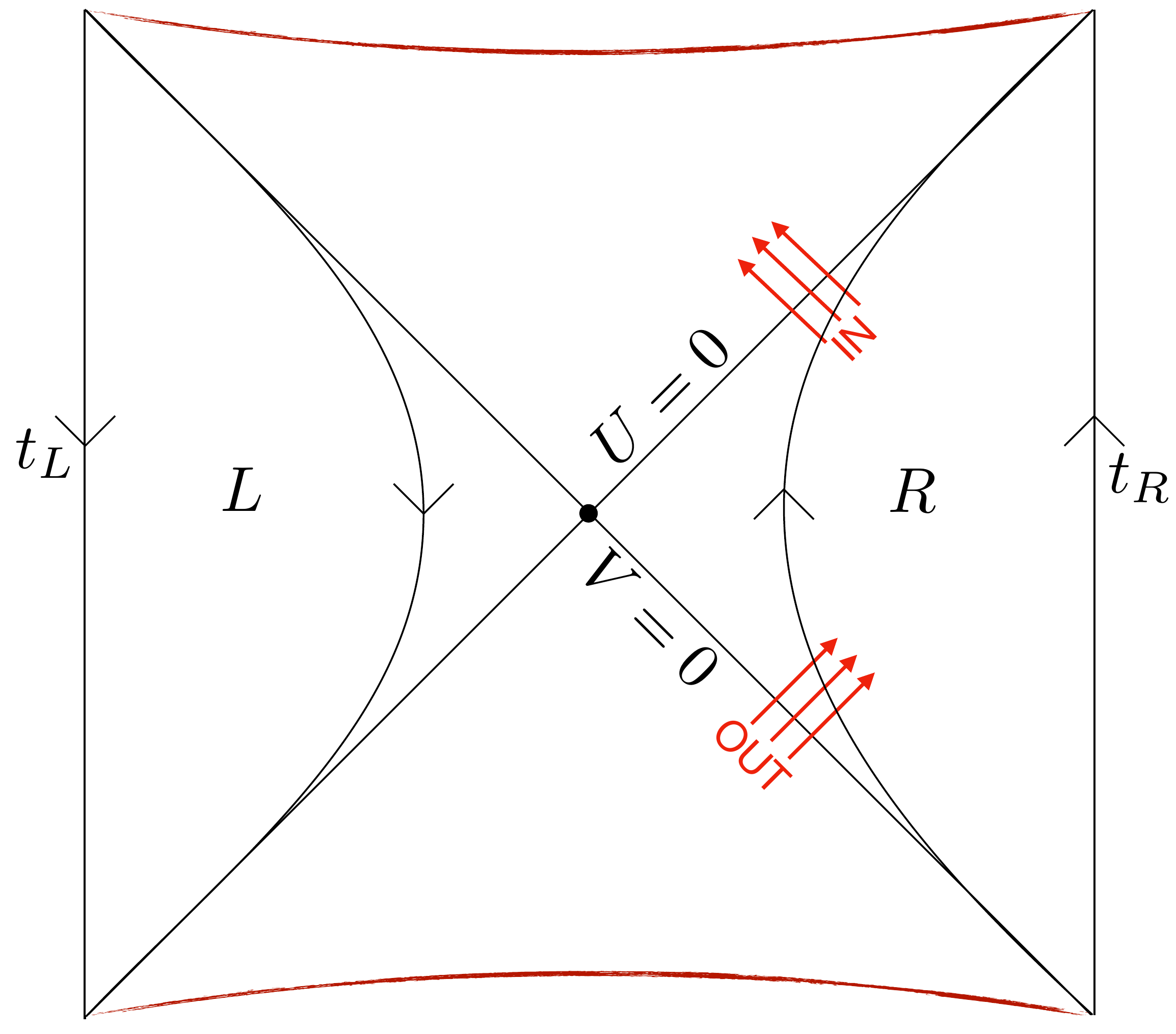}\\
\end{center}
\caption{Penrose-Carter diagram of the eternal Schwarzschild AdS black hole. The red thickened lines at the top and bottom represent the singularities, while the solid lines at $45$ degree angles are the horizons which meet at the bifurcation surface at the center of the diagram. The two asymptotic regions outside the black hole are indicated by `$L$' and `$R$'.  For convenience a review of the construction of this geometry is given in appendix \ref{app.Kruskal}. As reviewed there, the global Kruskal time runs in the same direction as $t_R$, but opposite to $t_L$, as indicated along two constant $r$ slices in each quadrant above.}
 \label{fig:bh}
\end{figure}

More explicitly, we introduce the tortoise coordinates,
\be
r_* = \int {dr \ov f} \,,
\ee
which has the near horizon behavior (as $r\to r_0$)
\be 
r_*  = {\beta \ov 4 \pi } \log (r - r_0) + \cdots \ .
\ee
Define 
\be 
v = t + r_*, \qquad u = t - r_* ,
\ee
then for the $R$ region we have 
\be\label{u1}
U = - e^{- {2 \pi \ov \beta} u}, \qquad V = e^{ {2 \pi \ov \beta} v}, 
\ee
while for the $L$ region
\be\label{u2}
 U = e^{- {2 \pi \ov \beta} u}, \qquad V = - e^{ {2 \pi \ov \beta} v} \ .
\ee
Writing $U = T -X,  V = T+X$, we may view $T$ as a ``global time'' which covers all regions of the black hole 
spacetime.  Note that under $t \to t + c, c >0$ we have 
\be 
U \to e^{-{2 \pi \ov \beta} c} U, \qquad  V \to e^{{2 \pi \ov \beta} c} V
\ee
which has opposite effects in $R$ and $L$ regions: In the $R$ region it increases $T$ while in $L$ it decreases $T$. 
Thus $t$ runs in the same direction as $T$ in the $R$ region and in the opposite direction as $T$ in the $L$ region. 

Now observe that the $L$ region expression~\eqref{u2} can be obtained from~\eqref{u1} by taking $t \to t - {i \beta \ov 2} $, 
and given that $t$ runs opposite in two sides, we can thus identify the full extended black hole 
geometry Fig.~\ref{fig:bh} as the gravity dual of the Schwinger-Keldysh contour of Fig.~\ref{fig:sk1} with $\sig = {\beta \ov 2}$. 
It then follows that for $\sig = {\beta \ov 2}$, the generating functional~\eqref{gen0} can be computed
in the fully extended black hole geometry using the usual procedure, with the bulk Feynman propagator defined 
in terms of $T$-ordering (this corresponds to the so-called Hartle-Hawking vacuum $\ket{HH}$ for a black hole spacetime), i.e. 
\be\label{inmg}
e^{W [\phi_{1}, \phi_{2}]} = \vev{HH,+\infty | HH, -\infty}_{\phi_1, \phi_2} \ .
\ee
 As an illustration, let us consider a scalar operator $\sO$ dual to a massive scalar field $\chi$ 
to quadratic level in sources $\phi_{1,2}$. For this purpose it is enough to consider the bulk quadratic action for $\chi$
\be \label{gact}
S = - \ha \int d^{d+1} x \, \left( (\p \chi)^2 + m^2 \chi^2 \ri), 
\ee
and~\eqref{gen0} is obtained by evaluating the action on the solution of $\chi (x)$ which satisfies the boundary conditions (up to some factors of $r^{\alpha_1}$, which we suppress here)
\be \label{bds}
\lim_{r \to \infty} \chi (r, x) |_{R} = \phi_1 (x), \quad  \lim_{r \to \infty} \chi (r, x) |_L = \phi_2 (x) \ .
\ee
Such a solution can be expanded as 
\be \label{bsol}
\phi (r,k) = a(k) \chi_1 (r,k) + b(k) \chi_2 (r,k)
\ee
where $\chi_1, \chi_2$ denote a basis of independent solutions to the wave equation of~\eqref{gact}.
They should be considered as defined on the fully extended black hole spacetime, and 
can be obtained by patching together the solutions in the $L$ and $R$ quadrants~\cite{Herzog:2002pc,Son:2009vu}.

It is convenient to write an independent basis of solutions to the corresponding wave equation in the $R$-region in terms of their near-horizon behavior 

\be \label{nm}
\chi_1 = e^{- i \om u},  \qquad \chi_2 = e^{- i \om v} \,,
\ee
where $\chi_1$ ($\chi_2$) describes a wave coming out of (falling into) the horizon (see Fig. \ref{fig:bh}). 
To obtain a global solution defined on a full Cauchy slice of the Kruskal geometry, we need to analytically 
continue~\eqref{nm} to the $L$ region, with the help of Kruskal coordinates $U, V$. Since in the $R$-region, 
$u = -{\beta \ov 2 \pi} \log (-U)$ and $v = {\beta \ov 2 \pi} \log V$, solutions~\eqref{nm} have branch points at the future ($U=0$) and past horizons ($V=0$) respectively. To perform analytical continuation to the $L$ quadrant with~\eqref{u2} one has to decide whether to go around 
the branch points along the upper or lower half complex planes of $U$ and $V$. 
Going around in the lower (upper) half plane means ``negative'' (``positive'') frequency solutions 
with respect to global time $T$.

To compute the right hand side of~\eqref{inmg}, the solution~\eqref{bsol} should behave as a  Feynman propagator ordered in terms of $T$. In fact it can be considered as a bulk Feynman propagator from one boundary point to another.  
Thus the solution~\eqref{bsol}  should  contain only negative frequency modes in $T$ in the future of $T$ and only positive frequency 
modes in $T$ in the past. This means that we should analytically continue $\chi_2$ 
(which lives on the future horizon) in the lower half $V$ plane, while continuing $\chi_1$ (which lives on the past horizon)  in the upper half $U$ plane (i.e. positive frequency solution).  More explicitly, we have the analytic continuation from~\eqref{u1} to~\eqref{u2} 
\bea 
\log (-U) = \log (e^{-\pi i} U ) = - \pi i + \log U \\
\log V = \log (- e^{-\pi i} V  ) =  - \pi i + \log (-V)
\eea
and thus in the $L$ region, the basis of solutions~\eqref{nm} becomes 
\be \label{inm}
\chi_1 \to e^{\ha \beta \om} \chi_1, \qquad \chi_2 \to e^{-\ha \beta \om} \chi_2 \ .
\ee
This completes the specification of the analytic continuation procedure. Imposing the boundary conditions~\eqref{bds} one can obtain the coefficients $a(k), b(k)$ in~\eqref{bsol} as 
\be 
a(k) = {\chi_2 - \chi_1 e^{-\ha \beta \om} \ov \chi_{1 \infty} \sinh{\beta \om \ov 2}}, \quad
b(k) = {\chi_1 e^{\ha \beta \om} - \chi_2 \ov \chi_{2 \infty} \sinh{\beta \om \ov 2}}
\ee
where all fields on right hand sides are in momentum space and $\chi_{1\infty} (k), \chi_{2\infty} (k)$
are defined by
\be 
\lim_{r \to \infty}  \chi_{1,2} (r, k) = r^{\De-d} \chi_{1,2\infty} (k) \ .
\ee

The generating functional for other choices of $\sig \in (0, \beta)$ can be obtained by 
generalizing the prescription~\eqref{inm} as 
\be \label{gebs}
\chi_1 \to e^{(\beta- \sig) \om } \chi_1, \qquad \chi_2 \to e^{ - \sig \om} \chi_2 \ 
\ee
where the extra factor $e^{\beta \om}$ for $\chi_1$ comes from analytically continuing $U$ in the upper half plane. 
The above prescription can also be understood in coordinate space as follows. 
From 
\be 
u = v - 2r_* = v - {\beta \ov 2 \pi} \log (r-r_0) + \cdots 
\ee
we have 
\be 
\chi_1 = e^{- i \om (v - {\beta \ov 2 \pi} \log (r-r_0) )}, \quad \chi_2 = e^{- i \om v} 
\ee
and the analytic continuation procedure~\eqref{gebs} may be phrased in coordinate language as 
taking $v \to v - i \sig$ while at the same time continuing  $r-r_0 \to e^{-2 \pi i} (r-r_0)$.

Let us conclude by mentioning another construction of relevance to the discussion of this section. In  \cite{Skenderis:2008dh,Skenderis:2008dg} the authors advocate a gluing procedure which allows one to construct bulk spacetimes dual to general contours with both Euclidean and Lorentzian sections. This prescription is illustrated on the example of the standard thermal SK contour (illustrated in Fig. \ref{fig:sk1}) in Fig. \ref{fig:SkenderisVanRees}.

\subsection{A non-equilibrium prescription} 

We will now discuss a proposal~\cite{hongPaoloMikeToAppear} which generalizes the prescription in the previous subsection 
to a general time-dependent gravity geometry which has an analytic horizon. 
Note that due to lack of time translation symmetry,  one should consider the contour of Fig.~\ref{fig:sk}. 

\begin{figure}[!h]
\begin{center}
\includegraphics[width=8cm]{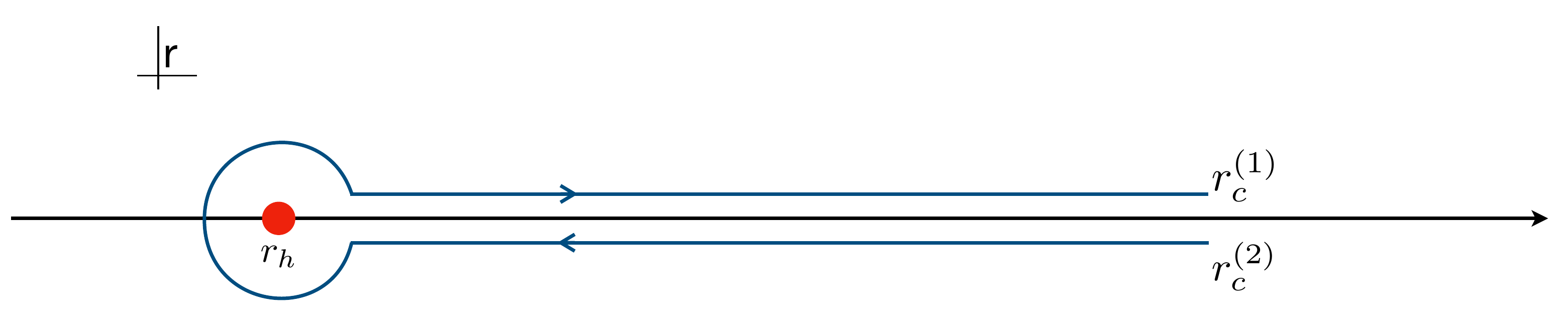} \end{center}
\caption{}
 \label{fig:contour}
\end{figure}

For $\sig =0$, the analytic continuation procedure discussed at the end of last subsection for a thermal state can be 
phrased as in Fig.~\ref{fig:contour}: one considers a complexified black hole spacetime with its two copies connected 
at the horizon $r=r_0$ via a clockwise $ 2 \pi$ rotation. In the second patch we change the orientation of $r$ direction. 
For the full complexified spacetime to have the same orientation, such a procedure also effectively reverses the orientation of the $v$ direction in the second copy. Now note that such analytic continuation can be defined for a general spacetime with a
future horizon, describing a general non-equilibrium state. Note that one does no analytic continuation in Eddington-Finkelstein time $v$ which is already regular at the future horizon.  As alluded to before, one could obtain correlation functions defined on the Schwinger-Keldysh contour for such a state by working with the complexified spacetime of Fig.~\ref{fig:contour}. 

Having introduced the idea of holography, its formalism as well as some advanced topics on the calculation of non-equilibrium correlation functions, we would now like to move on to some concrete examples illustrating the power of the holographic approach. We start with an interesting class of non-equilibrium phenomena, namely nonequilibrium steady states.

\section{Nonequilibrium Steady States}
Nonequilibrium steady states (NESS) arise for certain systems subject to external forcing, for example an applied electric field or a heat gradient, or a gradient of a chemical potential. If the system is able to adjust itself in such a way as to establish a balance between the resulting currents and the applied forcing it will settle into a stationary state. More generally, one can view such a steady state as an intermediate-time description of a system, on scales small compared to the ultimate equilibration time. If the latter can be made parametrically large then the steady-state persists, and often an effective thermodynamic description may apply in the intermediate regime. We shall see examples of this in the context of large-$N$ theories with holographic duals. In this review we focus on three main classes of steady states, namely current driven, heat driven and momentum driven NESS.
\subsubsection{Current driven NESS}
Let us assume that a certain system of interest can be divided into two sectors, for example one sector of $N_{\rm f}$ flavors of quarks\footnote{i.e. fermions in the fundamental representation of the gauge group.} and a second one consisting of $SU(N)$ gluons. We then imagine applying an external field to only one of the two components, say the quarks. It is not even necessary to assume that the two sectors are different in nature, as in the aforementioned case. For example, the paper \cite{green2005nonlinear} considers minimally coupling two scalar fields $\Psi = \Phi^1 + i \Phi^2$ in an O(N) model to a $U(1)$ gauge potential $\nabla \Psi \rightarrow (\nabla - i e A)\Psi$.

In such cases the system naturally splits up into the probe sector, the quarks in the former and the charged field $\Psi$ in the latter example, and the rest of the system, namely the gluons in the first example and the remaining $N-2$ scalar fields $\Phi^I$ in the latter. If $N$ is taken to be very large, we arrive at a situation where the `rest of the system' can absorb momentum and energy from the probe sector at a rate of ${\cal O}(1)$, while the reverse processes are suppressed by inverse powers of $N$. Then at large $N$ the `rest of the system' will act as a bath for the probe sector. As a result, even if the whole system is in a pure state, we shall recover an effective thermal description of the NESS. We shall see that the probe system, subject to the continuous drive, despite being out of equilibrium shows aspects of equilibrium thermodynamics, such as an effective temperature $T_{*}$ seen by all fluctuations, complemented by a detailed fluctuation-dissipation relation. It would be extremely nontrivial to understand this situation purely from a many-body perspective as it involves balancing the current, the production of Schwinger pairs and the scattering, leading to a relaxation into a steady state.

However, we will now explore how this extremely complicated interplay of scattering, dissipation and conductivity can be studied analytically with the help of holographic duality in a simple manner. As a prototypical example, we will consider the D3D5 brane intersection \cite{Sonner:2012if}, and will then briefly describe how to abstract these insights into a more general framework. In the probe limit, this brane intersection gives rise to a $2+1$ dimensional theory of fundamental fermions coupled to an adjoint sector \cite{DeWolfe:2001pq,Erdmenger:2002ex}, in other words we have an explicit example of the first kind we mentioned above. It is this $2+1$ dimensional theory that we will analyze.
Let us first define our system more precisely.

In the relevant limit ($N_{\rm f} \ll N$) this system is described as follows: we let the $D3$ branes backreact and take the near-horizon limit, resulting in the metric of a black hole in $AdS_5 \times S^5$,
\be\label{eq.BackgroundMetric}
ds^2=\frac{u^2}{\ell^2} \left(  -f(u) dt^2 + d\mathbf{x}^2\right)+\frac{\ell^2 du^2}{f(u) u^2} + \ell^2 d\Omega_5^2
\ee
where $ds^2=G_{ab}dX^a dX^b$, with $f(u) = 1 - \frac{u_h^4}{u^4}$, recognizing the black-hole metric introduced in \eqref{equi1}. The dynamics of the $N_{\rm f}$ D5 branes is then described by the Dirac-Born-Infeld (DBI) action, essentially a non-linear generalization of standard $U(1)$ electromagnetism in terms of a gauge potential $A$ and its field strength $F=dA$, living
in the background \eqref{eq.BackgroundMetric}.
There also is a so-called Wess-Zumino contribution to the action, which does not play a r\^ole and so we can safely ignore it in this analysis. From the point of view of the field theory the gauge field gives rise to an electric field with respect to a $U(1)$ symmetry acting on the fundamental degrees of freedom.
We proceed by applying an external electric field in the $x$ direction by choosing the gauge potential
\be\label{eq.D5gaugeAnsatz}
A = - \left(E t + A_x  \right)dx\,.
\ee
In principle we must solve the full set of equations of motion of the brane intersection, which is a complicated set of partial differential equations. 
In the present context one finds that this task can be reduced to finding two functions $\left\{\theta(u), A_x(u)   \right\}$, where $\theta(u)$ is a single angle describing the position of the $S^2$ within the $S^5$.  For the sake of simplicity, let us consider the case $\theta =0$, corresponding to massless fundamental fermions. Then the equations have a first integral, introducing a constant of integration, which turns out to be essentially the boundary current expectation value. Using the precise relation between the expectation value of the current and this integration constant, allows to deduce the conductivity relation
\be
E = \sigma_{(2+1)} \langle j \rangle 
\ee
with $\sigma_{(2+1)}$ a constant. It is important to note hat no restriction was made on the smallness of the applied field, so that this result represents the full non-linear response to an arbitrary field. The fact that the response is linear in the source is an accident of the two-dimensional setup. As we shall see below, more generally the current response to an arbitrarily strong electric field is a non-linear function.

 In fact the holographic dual geometry emerging from the above analysis turns out to hold the key to a simple intuitive description of current-driven NESS. To reveal this story, it turns out to be enlightening to study the spectrum of fluctuations around the steady state. Let us thus expand a given field $\Phi$ as
 \be
 \Phi = \Phi_{\rm steady} + \varphi
\ee
 to find that its equation of motion, formally from an action of the type
 \be\label{eq.fluctAction}
S_{\rm steady}  = \int d^6 \xi \sqrt{-{\cal S}} {\cal S}^{ab}\partial_a \Phi \partial_b \Phi\,,
 \ee
 where ${\cal S}^{ab}$ is an effective metric, and {\it not} the background metric introduced in \eqref{eq.BackgroundMetric}. Said differently, the NESS gives rise to an effective geometry, different from the background, leading directly to an effective thermodynamic description of the the non-equilibrium physics of the system.  As we have seen before, the dual geometry, and thus the metric, encode a great deal of non-trivial physical information about the field theory, and this is no different for the effective metric ${\cal S}_{ab}$. It is thus natural to proceed by elaborating the properties of this object.
  Technically speaking, it is the so-called open string metric (OSM), defined as $\gamma_{ab} = P[G]_{ab} + F_{ab}$ with $P[G]_{ab}$ being the pull back of the ambient metric onto the brane, while $\gamma^{ab} = {\cal S}^{(ab)} + {\cal A}^{[ab]}$, respectively, having separated symmetric, $ {\cal S}^{(ab)}$, and antisymmetric, ${\cal A}^{[ab]}$, parts of the inverse of $\gamma^{ab}$.

Since fluctuations see the metric ${\cal S}_{ab}$ and not the background metric, $G_{ab}$, their correlation functions are determined by the properties of this effective metric. This has important physical consequences. While this mechanism occurs rather widely, with the precise form of the metric in various cases presented in  \cite{Gursoy:2010aa,Sonner:2012if,Nakamura:2013yqa,Kundu:2013eba,Kundu:2015qda}, we will content ourselves with the specific example of the D3D5 system at hand, giving some important physical expression for the general case at the end. In this case, the effective metric takes the simple form \cite{green2005nonlinear}
\be
ds^2|_{\rm eff} = -\frac{u^4-u_*^4}{(\ell u)^2}dt^2 + \frac{(\ell u)^2}{u^4-u_*^4}du^2 + ds^2|_{\rm trans.}\,,
\ee
suppressing, as indicated, the transverse directions. This metric has a horizon at $u=u_*$ of temperature
\be
\pi T_{*} = \left[( \pi T)^4 + E^2 \ell^{-4} \right]^{1/4}\,,
\ee
different from background temperature $T$ of the metric \reef{eq.BackgroundMetric}. In particular it can be non-zero even if $T=0$.
 {\it It follows from the fluctuation action \reef{eq.fluctAction} that two-point functions of the brane excitations will be thermal at temperature $T_*$.}
 This temperature, however, has more far-reaching consequences, namely that fluctuations in the NESS satisfy an exact fluctuation-dissipation relation with respect to $T_*$:
\be\label{eq.fluctuationDissipation}
G_{\rm sym}(\omega,k) = -(1+2n_*){\rm Im}G_{R}(\omega, k)\,,
\ee
where $G_{\rm sym}$ is the symmetric (`Keldysh') Green function and $G_R$ the usual retarded Green function. The symbol $n_*$ denotes the Bose-Einstein distribution function at the effective temperature $T_*$. {\it Thus, despite the far-from equilibrium nature of the NESS, certain observables are exactly thermal with respect to an effective temperature $T_*$, determined by the applied field.} Schematically:
\bea
\text{ equilibrium geometry} \,\,&\leftrightarrow&\,\, \text{thermodynamics}\nonumber\\
\text{{\it effective} geometry} \,\,&\leftrightarrow&\,\, \text{{\it effective} thermodynamics}\,.\nonumber
\eea

 As an illustration of the power of this relation, \cite{Sonner:2012if} derived an expression of the current noise in the NESS valid for any value of the applied field, 
 \bea
 S_j &=& - \int_{-\infty}^\infty d\omega dt e^{i\omega t}\coth \left(  \omega / 2 T_*\right){\rm Im} \langle j(\omega) j(-\omega) \rangle_R\nonumber\\
 &=& 4 \sigma_{(2+1)}T_*\,.
 \eea 
It should be noted that in the integrand we have used the fluctuation-dissipation relation to express the Keldysh correlation function in terms of the imaginary part of the retarded current-current correlation function, and thus have made crucial use of the properties of the horizon of the effective metric.  
 Furthermore, \cite{Nakamura:2013yqa} determined the distribution of momentum fluctuations in certain cases, demonstrating explicitly that they were thermal at temperature $T_*$. The interpretation of the horizon entropy of the OSM appears to be more subtle. Several authors have proposed to be related to entanglement, e.g. between Schwinger pairs created from the vaccuum \cite{Sonner:2013mba}, or between fundamental and adjoint degrees of freedom \cite{Das:2010yw} or a combination of both. Recently \cite{Kundu:2015qda} has argued that there is simply no entropic interpretation for the open string horizon. The full physical interpretation of the open-string horizon constitutes and interesting open problem for the future.

The analysis here can be carried out rather generally for a $D(q+1+n)$ brane embedded in the background created by $N$ $Dp$ $(p<7)$ branes  \cite{Nakamura:2013yqa}. This has the interesting property that the temperature of the world-volume horizon is not necessarily higher than that of the background. For example a single $D2$ brane probing a stack of $D4$ branes ($n=0, q=1$) has $T_* < T$ for any $E\neq 0$. In all cases a detailed fluctuation dissipation relation of the form of Eq. \reef{eq.fluctuationDissipation} can be established for the appropriate $T_*$.

\subsubsection{Nonequilibrium heat flow}
A different class of steady states involves systems driven away from equilibrium by a temperature differential. One can imagine setting up such a situation by bringing two CFTs in contact\footnote{An intriguing alternative scenario arising from coupling to two heat baths at different temperatures is elaborated in \cite{Bakas:2015hdc}. Here the boundary metric is non-trivial, sourced by a time dependent deformation of the Hamiltionian. The system neverthless approaches a conformal, self-similar steady state in the late-time limit.}, each thermalized at its own temperature $T_R \neq T_L$. One is then interested in the heat flow, and in particular whether a steady state with nonvanishing heat current can be established in the interface region \cite{bernard2012energy,Chang:2013gba,Bhaseen:2013ypa,Doyon:2014qsa,Amado:2015uza,Herzog:2016hob}. This question has a beautifully simple answer in 2D CFT, where relativistic hydrodynamics together with conformal invariance dictate the energy flow. Under these conditions a universal form of the heat current, namely \cite{bernard2012energy}
\be
\langle  T^{tx} \rangle_s = \frac{\pi c}{12} \left( T_L^2 - T_R^2 \right)\,,
\ee
follows from the above-mentioned constraints subject to the initial conditions.  In more than one spatial dimension conformal symmetry alone is no longer powerful enough to establish a similar result. However, under certain assumptions it can be argued that CFTs with a holographic dual are asymptotic (in time) to a simple steady state configuration with similar properties. More precisely \cite{Bhaseen:2013ypa} argue that a steady state region forms, with current
\be
\langle T^{tx}_d \rangle_s \propto \frac{T_L^{d+1} - T_R^{d+1}}{u_L + u_R}\,,
\ee
where
\be
u_{L,R} = \frac{1}{d}\sqrt{\frac{\chi + d^{\pm 1}}{\chi + d^{\mp 1}}}\,,\qquad \chi = \left( T_L / T_R \right)^{(d+1)/2}\,.
\ee
These higher dimensional results follow from combining conformal relativistic hydrodynamics with the holographic insight that the only regular solutions with a constant and homogenous stress tensor are the boosted black branes  \cite{Bhaseen:2013ypa}. In field theory terms, this means that the NESS region is described by a boosted density matrix
\be\label{eq.BoostedDensityMatrix}
\rho_s = e^{-\beta \cosh\theta H + \beta \sinh\theta P^x}\,,
\ee
where temperature $T = \beta^{-1}$ and rapidity $\theta$ can be shown to be given by
\be
T = \sqrt{T_L T_R}\,,\quad \theta = \frac{\chi-1}{\sqrt{(\chi + d)(\chi + d^{-1})}}\,.
\ee
In fact, the holographically motivated assumption that the NESS is described by the boosted density matrix \eqref{eq.BoostedDensityMatrix} allows one to get control over arbitrary connected correlations of the energy current $c_n = \langle J_{\rm tot}^{n-1} T^{tx}|_{x=0}\rangle^c_{s}$ in the NESS region. This is achieved by showing that the generating function $F(z) = \sum z^n c_n/n!$ satisfies the simple equation
\be
\frac{dF(z)}{dz} = J_E(\beta_L - z, \beta_R + z)\,,\quad J_E = \langle T^{tx} \rangle_s\,,
\ee
which allows one to extract the full fluctuation spectrum of the strongly coupled NESS in arbitrary dimension. This again generalizes a two-dimennsional CFT result \cite{bernard2012energy} to higher dimensions. We emphasize again that the crucial step was to appeal to holographic duality, an in particular to properties of the bulk Einstein equations to argue for the boosted form \eqref{eq.BoostedDensityMatrix} of the density matrix in the stead-state region. The original work was performed on the assumption that the steady state region forms after the local quench in between to shock waves that emanate from the interface, but some later treatments corrected this picture slightly by replacing the shock moving towards the hotter reservoir by a rarefaction wave, as required by entropic considerations \cite{Spillane:2015daa,Lucas:2015hnv}, and as had been appreciated in previous hydrodynamic work \cite{taub1948relativistic}. It was found numerically that the analytic predictions for the NESS region do not get drastically modified, given that the temperature differential between the two reservoirs is not very large, and given that the rarefaction region does not eliminate the steady state region entirely. The interested reader can find a more detailed review of the above-mentioned developments in \cite{Bernard:2016nci}.

It is also interesting to remark that since the publication of  \cite{Bhaseen:2013ypa}, the work  \cite{Glorioso:2015vrc} has given a general classification of holographic spacetimes with constant homogenous stress tensor, extending the range of candidate NESS beyond boosted black branes. This opens the exciting possibility to explore the properties of the steady states associated with these new solutions, along the lines of  \cite{Bhaseen:2013ypa}.

\subsubsection{Flows over obstacles}\label{sec:NESS}
Many interesting NESS arise in situations where a fluid, specifically here a quantum liquid, is forced to flow in the presence of an obstacle. Familiar, albeit classical, examples include the flow of air around an airfoil or the bow of a ship or even the flow of a gas through a gravitational well, as is often relevant in astrophysical settings. In this section we review universal aspects of flows of quantum liquids for theories with holographic duals \cite{Figueras:2012rb,Sonner:2017jcf,Novak:2018pnv}.
We imagine the field theory subject to boundary conditions at spatial infinity, prescribing a simple, homogenous flow at temperature $T_{L}$ and with flow velocity $u^\mu_L$. This sets up a non-zero, momentum-carrying, flow, which is then disturbed by an obstacle, which we place at some finite position in the downstream region. The flow will then approach a second asymptotic, homogeneous flow far downstream from the obstacle, with parameters $T_R$ and $u^\mu_R$.

The main point of interest is the case of a stationary flow, which will be simple asymptotically far from the obstacle, but strongly non-linearly disturbed in its vicinity.  The spatial transition between these two behaviors is universal and elegantly described in holography.

\begin{figure}[h]
\begin{center}
\includegraphics[width=\columnwidth]{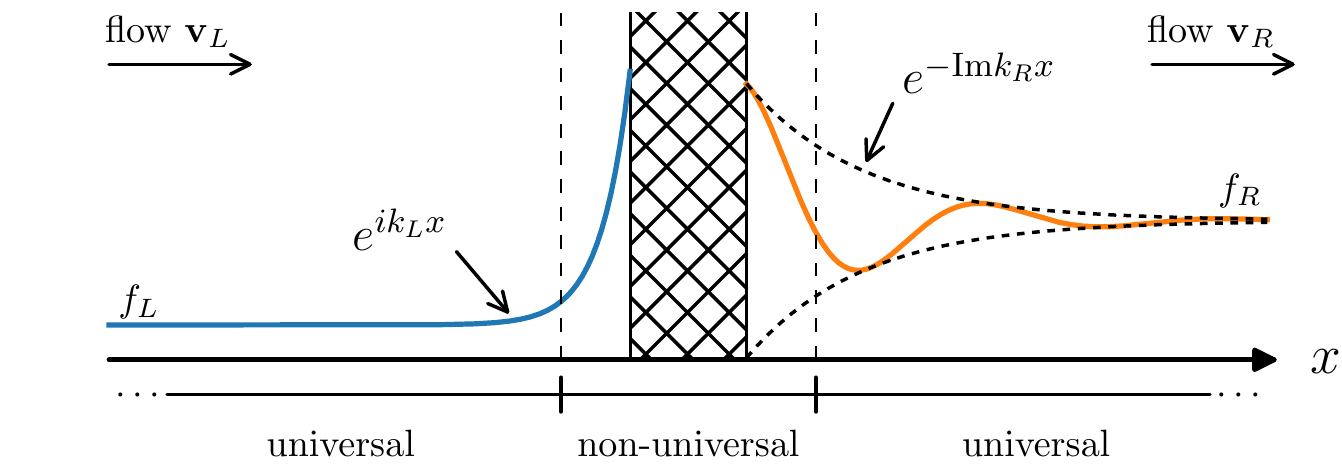}\\
\end{center}
\caption{Schematic description of flow-over-obstacle NESS. Approaching asymptotically to the left and to the right the flow returns to equilibrium, the precise shape being dictated by the spatial collective modes defined in this section. It may be helpful to think of this situation as a spatial quench, with the $x$-direction taking on the r\^ole of time. Figure taken from  \cite{Novak:2018pnv}.}
 \label{fig:schemaNESS}
\end{figure}
A particularly simple picture results for co-dimension one obstacles, when, as emphasized in Figure \ref{fig:schemaNESS}, one can view the flow as a kind of spatial quench \cite{Figueras:2012rb}. This was analyzed in detail in \cite{Sonner:2017jcf,Novak:2018pnv} and we now summarize the salient features. In analogy with temporal relaxation in holography, where as we have mentioned (and shall see explicitly in subsequent sections), universal late-time behavior can be efficiently accessed via the quasinormal modes of the system, the NESS of this section approach the asymptotic spatial equilibria at a rate governed by a spatial version of quasinormal modes\footnote{Here we are considering stationary systems which depend non-trivially on space, but it may be helpful to anticipate a close analogy between what is being developed here and the temporal relaxation of generic holographic systems via their quasinormal modes, as described in \ref{sec:QuantumQuenches} below:  essentially the former are related to the latter  by exchanging the r\^oles of $\omega$ and $k$.}, termed `spatial collective modes' in  \cite{Sonner:2017jcf,Novak:2018pnv}. In fact, similar modes have been encountered in the context of plasma absorption \cite{Amado:2007pv}, as well as holographic superconductors \cite{Maeda:2009wv,Sonner:2014tca} and hydrodynamic shocks \cite{Khlebnikov:2010yt,Khlebnikov:2011ka}. This appearance of SCM physics can be summarized as follows.
\begin{enumerate}
\item A system is set up with an asymptotic flow velocity $v_L$ far away on the left of a co-dimension one obstacle. 
\item In some region the flow is disturbed by an obstacle. 
\item Downstream from the obstacle, the disturbed flow approaches again a steady homogenous flow with generically different parameters. The spatial profile of how the flow connects the two homogeneous asymptotic flows to the strongly non-linear region in the vicinity of the obstacle is described universally by the leading spatial collective mode (SCM) $k^*_{0,\,L/R}$ on either side of the obstacle.
\end{enumerate}

 The SCM introduced here are solutions to the linearized bulk equations around backgrounds describing homogeneous and isotropic flows, in other words boosted black branes. The obstacle will excite perturbations of energy density, flow velocity field, etc. (depending on the conserved quantities present in the setup), e.g. $\varepsilon(x^\mu) = \varepsilon + \delta \varepsilon e^{ik_\mu x^\mu}$ with zero frequency, i.e. $k^\mu = (0,\bk)$. These correspond to bulk modes that are regular at the future horizon, but which also have the novel feature that they obey regularity conditions as one or the other spatial asymptotic region is approached. Such solutions typically fall into a discrete set of modes
\be
k  = k^*_n (\bv; \omega) \in \mathbb{C} \qquad (n \in \mathbb{Z})
\ee
which show up as analytic properties of correlation functions in the complex momentum plane, where poles in the UHP define modes that describe the spatial equilibrium as one asymptotic region is approached (in the conventions of \cite{Sonner:2017jcf,Novak:2018pnv} to the right), while poles in the LHP give modes that describe the decay from the obstacle toward the other spatial asymptotic region. The physics of this spatial relaxation is universal, i.e. it depends on the theory one is interested in, but not  on the shape of the obstacle, and it encodes physically interesting information. For example, the spatial decay of the shear mode obeys the dispersion relation  \cite{Sonner:2017jcf}
\be
k = -i s/\eta \bv \cdot \bk + \cdots
\ee
where $\bv$ is the asymptotic background flow velocity that is approached. What is interesting is that a spatial feature of the system, for example the decay to the right asymptotic region in  Fig. \ref{fig:schemaNESS} is given by the shear viscosity over entropy density ratio $\eta/s$. This suggests that NESS would be attractive experimental setups to quantitively determine this quantity in the lab. Typical scales for graphene at charge neutrality can be estimated, resulting in values of $\sim 1 \mu m$ at standard temperature for the parameters reported by \cite{torre2015nonlocal}.

\section{Quantum quenches\label{sec:QuantumQuenches}}

Probing the coherent dynamics of many-body quantum systems which were initially in far-from equilibrium states has become both experimentally accessible and a theoretically fruitful area of study\footnote{For reviews on this topic the reader may consult, for example, \cite{Calabrese:2009qy,RevModPhys.83.863}. }. The simplest way to set up such a state is a quantum quench, which describes a non-adiabatic process of  disturbing a quantum many-body system by an external force or changing its parameters. 
Despite its simplicity, the post-quench evolution yields rich dynamical behavior.
We will review several canonical holographic examples below, focusing on global quenches. Local quenches also contain a plethora of interesting physics, as described for example in \cite{eisler2007evolution,Calabrese:2007mtj,stephan2011local,Asplund:2011cq,Asplund:2013zba,Nozaki:2013wia,Asplund:2014coa,Nozaki:2014hna,David:2016pzn}. 

\subsection{Thermalization, AdS Dynamics, and entanglement growth}

Thermal equilibrium states in AdS are described by stationary black branes as discussed earlier in Sec.~\ref{sec.AspectsOfDuality}. The study of thermalization therefore  maps to the dynamics of black hole formation and equilibration. From general properties of the black hole formation we can delineate a number of qualitative phases of non-equilibrium dynamics: 
\begin{enumerate}
\item A given initial state is disturbed, for example by abruptly changing boundary conditions, producing a far-from-equilibrium initial states. 
\item As a result of strong gravitational dynamics, an (apparent) horizon is formed in the bulk, or a previously existing horizon is deformed. From the boundary perspective, the horizon formation may be interpreted as local equilibration, where non-conserved quantities equilibrate locally, while conserved quantities and nonlocal correlations have not yet settled into their equilibrium values.  
\item At late times the (new) horizon equilibrates via quasi normal ring down and eventually settles to the new stationary state.
From the boundary theory perspective, at this stage expectation values of boundary theory physical observables (i.e one-point functions) will have settled into their equilibrium values. 

\item Even after one-point functions have reached equilibrium values, the system could still  be far-from-equilibrium when we probe it using nonlocal quantum observables such as correlation functions involving widely separated spatial points or the entanglement entropy for a large region. For a noncompact system, such equilibration of nonlocal observables essentially persists for ever as one increases the ``size'' of a non-local observable to infinity. 

\end{enumerate}

For a global quench, where the initial non-equilibrium state is spatially homogeneous, there is no energy or momentum flow in its subsequent evolution, and thus naively nothing happens after local equilibration. But studies of nonlocal quantum observables, which were initiated by  Calabrese and Cardy in~\cite{Calabrese:2005in,Calabrese:2006rx}, have revealed striking insights into quantum dynamics of the system (item 4 above). 
By tuning a parameter of a $(1+1)$-dimensional gapped system  to 
criticality, 
Calabrese and Cardy found that~\cite{Calabrese:2005in,calabrese2009entanglement,calabrese2006time}  the entanglement entropy for a segment of size $2 R$ grows with time linearly as
\be \label{ekko}
\De S (t, R) = 2 \, t \, s_{\rm eq}, \qquad t < R
\ee
and saturates at the equilibrium value at a sharp saturation time $t_s = R$. In~\eqref{ekko},
 $\De S$ denotes difference of the entanglement entropy from that at $t=0$ and $s_{\rm eq}$ is the equilibrium thermal entropy density.

The simplicity and elegance of~\eqref{ekko} motivated many studies in holographic systems, see e.g.~\cite{AbajoArrastia:2010yt,Albash:2010mv,Balasubramanian:2010ce,Balasubramanian:2011ur,Asplund:2011cq,Keranen:2011xs,Erdmenger:2012xu,Baron:2012fv,Caceres:2012em,Galante:2012pv,Li:2013sia,Arefeva:2013wma,Caputa:2013eka,Hartman:2013qma,Liu:2013iza,Liu:2013qca,Bai:2014tla,Anous:2016kss,Casini:2015zua,Mezei:2016zxg,Arefeva:2017pho,Anous:2017tza,Yeh:2014mfa,Mezei:2018jco}, especially in higher dimensions. 

A particularly simple example of a global quench is the Vaidya solution, which describes 
the gravitational collapse of a uniform shell of null matter, i.e. a thin shell of matter collapsing at the speed of light. 
From the boundary perspective, the solution describes the thermalization process following a sudden injection of uniform energy density into the system. 
In ingoing Eddington coordinates the corresponding metric can be written as
\be\label{eq.Vaidya}
ds^2 = \frac{L^2}{z^2} \left( -f(v,z)dv^2 - 2 dv dz + d\vec{x}^2 \right)\,,
\ee
where $v$ is a null coordinate and $f(v,z)$ is a general profile function, which depends on the details of the quench protocol, described by the mass function $m(v)$\,,
\be
f(v,z) = 1-m(v) z^d\,.
\ee
In the limit that the boundary source giving rise to this metric is applied for an infinitesimally short time, this function takes the form of a step function $m(v) = \frac{M}{2} \left[1+\theta(v)\right]$. Another frequently used protocol is $m(v)= \frac{M}{2} \left[1+ \tanh(v/v_0)\right]$.  

From the prescription of computing holographic entanglement entropy discussed in Sec.~\ref{sec:RTsurface}, 
in order to calculate the entanglement entropy of a subregion $\Sig$ in the boundary theory dual to the Vaidya geomtry, one needs to find the extremal codimension two surfaces of the metric (\ref{eq.Vaidya}). 
Denoting the characteristic size of the region as $R$, one finds that for $R$ much larger than the local equilibration time $\eql$
the time evolution of entanglement entropy  is characterized by four different scaling regimes~\cite{Liu:2013iza,Liu:2013qca}:

\ben 

\item Pre-local-equilibration growth: for $t \ll \eql$, 
 \be \label{quda}
\De S_\Sig (t) = {\pi \ov d-1} \sE A_\Sig t^2 + \cdots
\ee
where  $\sE$ is the energy density and $A_{\Sig}$ is the area of $\Sig$. This result is independent of the shape of $\Sig$, the spacetime dimension $d$.

\item Post-local-equilibration linear growth:
for $R \gg t \gg \eql$, we find a universal linear growth~\cite{Hartman:2013qma,Liu:2013iza}   
\be \label{line}
\De S_\Sig (t) =\Vee s_{\rm eq}  A_\Sig t + \cdots
\ee
where $\Vee$ has dimensions of velocity, often referred to as the entangling velocity or tsunami velocity. 
$\Vee$ is {\it independent} of the shape of $\Sig$, but does depend on the nature of the final equilibrium state. 
For an equilibrium state with no chemical potential, one finds that 
\be \label{schwv}
\Vee^{(\rm S)} = {(\eta -1)^{\ha (\eta -1)} \ov \eta^{\ha \eta}}  \ , \quad \eta \equiv  {2 (d-1) \ov d}  \ .
\ee
Turning on a chemical potential tends to reduce $\Vee$. 
Note that $\Vee^{(S)} = 1$ for $d=2$ and monotonically decreases with $d$.

\item  A saturation regime in which the entanglement entropy saturates at its equilibrium value. 
The saturation can be either ``continuous'' or ``discontinuous'' depending on whether the 
time derivative of the entanglement entropy is continuous at saturation. 
In the large $R$ limit, the saturation time $t_S$ can be written as 
\be 
t_S = {R \ov v_S} 
\ee
where $v_S$ is a constant depending on the shape of $\Sig$. $v_S$ is often referred to as the saturation velocity. 
For example,  for $\Sig$ a spherical region $v_S = v_B$ where $v_B$ is the so-called butterfly velocity, while for 
a parallel strip region $v_S = \Vee$.

\item In the $R \to \infty$ limit, there exists another scaling regime between the linear growth and saturation, 
in which the evolution of the entanglement entropy becomes insensitive to the shape and size of the region. 

\een 
These results are generic for all holographic systems in the sense that they are insensitive to the specific details of the system as well as those of the quench. The  scaling regimes were obtained by  identifying various geometric regimes for the bulk extremal surface. An important observation was the existence of a family of ``critical extremal surfaces'' which lie behind the horizon and separate extremal surfaces that reach the boundary from those which fall into the black hole singularity. 
In the large size limit, one finds that the time evolution of entanglement entropy is controlled by these critical extremal surfaces \cite{Liu:2013iza,Liu:2013qca,Mezei:2016zxg,Mezei:2016wfz,Mezei:2018jco}.

Collectively, these regimes suggest that the evolution of entanglement entropy can be captured by 
the picture of an entanglement wave propagating inward from the boundary of the entangled region, which was called an ``entanglement tsunami" (see also~\cite{kimhuse}). In other words, entanglement propagates ballistically even in systems without 
quasiparticles. 

Quantum quenches have also been discussed in a variety of other contexts, see for example \cite{Buchel:2013lla,Das:2014jna,Das:2014hqa,Das:2015jka,Das:2016lla}, which initially numerically observed interesting scaling results, both in the limit of fast but smooth quenches. In fact, by developing a near-boundary expansion adapted to the rapid quench problem, these results can be analytically shown to reflect the UV conformal fixed point of the dual theory, giving an excellent match to the numerically observed behavior. This motivated the authors \cite{Das:2014jna,Das:2014hqa,Das:2015jka,Das:2016lla} to establish analogous results purely from a field theory perspective,  in the context of CFTs and free theories.

More generally whether given initial states will eventually thermalize and if so, how fast and in what manner depends both on the nature of the system of interest (for example, many-body localized vs. ergodic or integrable vs. chaotic) and on properties of the initial state itself. In holographic contexts,  there are general theorems saying that sufficiently massive and compact objects will  collapse to form a black hole, which imply sufficiently excited non-equilibrium states generically thermalize. This is consistent with the standard lore regarding thermalization for a non-integrable quantum system. 
If instead we focus on initial states which are represented by small bulk initial data, the process of eventual thermalization becomes more intricate. This involves crucially the physics and geometry of AdS and in particular the presence of the timelike boundary which reflects outgoing modes {\it back into the bulk}. In this way small initial disturbances can be non-linearly amplified by successive reflections and eventually lead to the formation of a horizon \cite{HolzegelDafermos,Bizon:2011gg}. This reflects a more intricate path to thermalization of the field theory, and it is interesting to speculate what physics of the dual field theory corresponds to this behavior. The presence of stable regions within the space of initial data suggests a possible relation to a quantum many-body version of the classical KAM theorem, in the sense that not all initial data immediately become fully ergodic.

\subsection{Dynamics of Superconductors and Superfluids}
An additional ingredient in many condensed matter systems is an order parameter for a  certain symmetry breaking pattern. So far, the majority of studies have focused on breaking a global continuous symmetry of the dual field theory \cite{Hartnoll:2008kx}, i.e. the case of a holographic superfluid, but many other candidates for phases showing interesting order have been explored, for example spatially modulated phases \cite{Domokos:2007kt,Nakamura:2009tf,Donos:2011bh}. It has long been of interested to investigate the fate of broken symmetry states dynamically, and in particular for superfluids and superconductors. We now review work in this direction using holographic duality.
\subsubsection{Quenches of holographic superfluids}
The order parameter translates, via the holographic dictionary, into a charged matter field propagating in the bulk space-time, $\psi$, while the $U(1)$ current is represented by a Maxwell field, $A$, as outlined in the introductory sections of this review. The model action, also known as the bottom-up holographic superfluid \cite{Hartnoll:2008kx}, is thus given as
\be\label{eq.superfluid}
S = S_{\rm grav} +S_{\rm Max} +  \int d^{d+1}x\sqrt{-g} \left(- |D\psi|^2 - V(|\psi|) \right)\,,
\ee
where $S_{\rm grav}$ and $S_{\rm Max}$ were given in \reef{eq.EHaction} and \eqref{Max}. Hence, in order to study the dynamics of a holographic superconductor, one is led to study the time development of Einstein's equations with negative cosmological constant, sourced by an energy momentum tensor made up from the field strength $F$ as well as the complex scalar $\psi$.

 \subsubsection{Rapid Quenches and Dynamics of Symmetry Breaking}
Interesting phenomena arise when systems whose ground state involves a broken symmetry are displaced far from equilibrium by a sudden quench. In this case an immediate question of interest concerns the subsequent behavior of the order parameter in the long-time limit. 
By studying the dynamics following a coupling quench of the integrable BCS Hamiltonian (the `Richardson model') \cite{barankov2004collective,yuzbashyan2006relaxation,yuzbashyan2005solution,PhysRevLett.93.130403} identified a regime of persistent oscillations of the order parameter, followed by a regime of damped oscillations for stronger quenches. The analysis is valid in the collisionless regime on time scales shorter than the energy relaxation scale. In \cite{Bhaseen:2012gg} rapid quenches of superfluids were studied holographically, capturing the time evolution in its entirety. The starting point is the action (\ref{eq.superfluid}) for $d=3$, and for the simple potential $
V(\psi) = m^2 |\psi|^2$ with $m^2 \ell^2 = -2$. In order to utilize a characteristic solution scheme (see Section \ref{sec.Characteristic} for more details), the metric is chosen as
 \be
 ds^2 = \frac{\ell^2}{z^2} \left[  -F dt^2 - 2 dt dz + S^2 (dx_1^2 + dx_2^2)\right]\,,
 \ee
where $F(t,z)$ and $S(t,z)$ are nontrivial metric functions depending on time and the holographic bulk direction $z$, while $\psi = \psi(t,x)$, $A_t = A_t(t,z)$. 
 Near the $AdS$ boundary the matter fields have an expansion
\bea
\psi &=& z \psi_1(t) + z^2 \psi_2(t)\,\nonumber\\
A_t &= &\mu(t) - z \rho(t) + \cdots\,.
\eea
Here $\psi_1(t)$ denotes the time-dependent source of the order parameter, while $\mu$ is the chemical potential. Then the expectation value of the charge density can be found by holographically renormalizing the asymptotic behavior of the bulk fields resulting in the expressions
 \bea
 \langle J_t(t)  \rangle &=&\frac{\rho(t) - \dot \mu(t)}{2\kappa^2}\,\nonumber\\
 \langle {\cal O}(t) \rangle &=& \frac{\psi_2(t) + \dot\psi_1(t) + 2i\mu(t) \psi_1(t)}{2\kappa^2}\,.
 \eea
We note that the second expression given here corrects a typo in \cite{Bhaseen:2012gg}.

The order parameter field starts in an equilibrium state and is then quenched at time zero by applying a Gaussian profile to the source
\be
\psi_1(t) = \bar\delta e^{-(t/\bar \tau)^2}
\ee
where $\bar \delta$ characterizes the strength of the quench, and $\bar \tau$ the time scale. In order to explore the nonequilibrium phase diagram, $\bar\delta$ is varied, while $\tau$ is kept fixed at a scale.
\begin{figure}[h!]
\begin{center}
\includegraphics[width=\columnwidth]{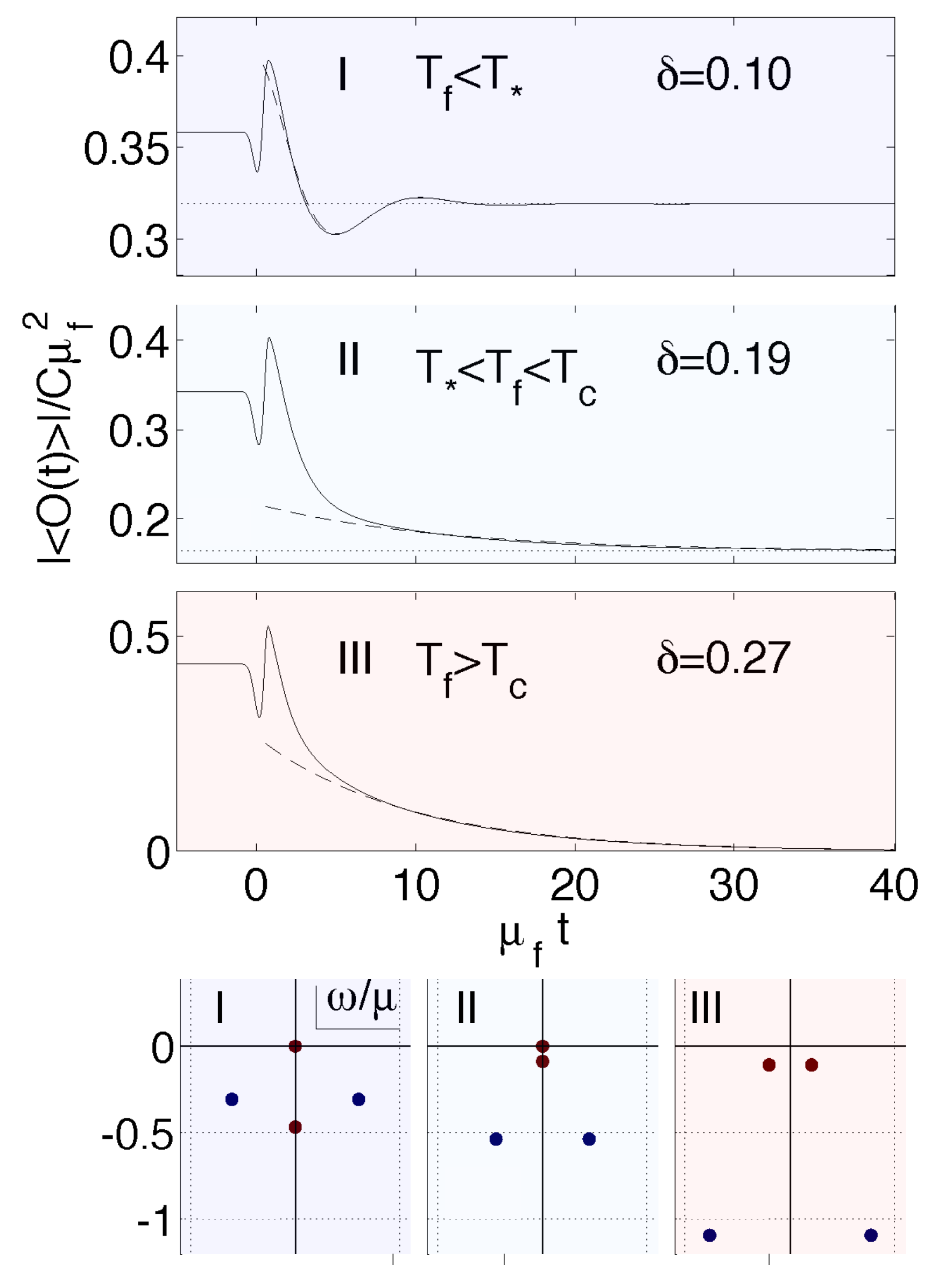}
\vskip1em
\caption{Three regimes, $I, II, III$, of the nonequilibrium phase diagram of a holographic superfluid. The location of the poles of the two-point function in each regime is shown in the bottom row. On the gravity side these correspond to the quasinormal modes of the complex order parameter field. Figure taken from \cite{Bhaseen:2012gg}.
\label{fig.QuenchQNM}}
\end{center}
\end{figure}
 Since the system fully thermalizes at late times, one can characterize each quench by the final temperature $T_f$ it attains asymptotically. The resulting dynamics falls into one of three regimes. Two of them (I \& II) lie on either side of a non-equilibrium phase transition, characterized by an emergent temperature $T_*$. This behavior is strikingly similar to the one observed in the Richardson model \cite{barankov2004collective,yuzbashyan2006relaxation,yuzbashyan2005solution,PhysRevLett.93.130403}. In more detail, we have
\begin{enumerate}[I]
\item Weak quench ($T_f < T_* < T_c$): the order parameter relaxes to a nonzero value with damped oscillation.
\item Intermediate quench ($T_* < T_f < T_c$):  the order parameter relaxes to a nonzero value via pure exponential decay.
\item Strong quench ($T_f > T_c$): the order parameter relaxes to zero via pure exponential decay.
\end{enumerate}
But in this case the full power of holography allows to physically characterize this system from a complementary perspective and identifying the physical mechanism behind the transition. Having identified the emergent final temperature, $T_f$, one may decompose the dynamics in terms of damped collective oscillations of the many-body system. These manifest themselves as poles in thermal correlations functions, and are encoded holographically in terms of the quasi-normal modes of the final-state black hole. It was demonstrated in \cite{Bhaseen:2012gg} that the transition at $T_*$ can be seen in the collective excitation spectrum, as the exchange of dominance of the leading poles in the two-point correlation function of the order parameter $G(\omega, \mathbf{k}=0) = \langle {\cal O}^\dagger(\omega) {\cal O}(-\omega) \rangle$. Related results have been obtained using the $\varepsilon$-expansion in \cite{Podolsky:2012pv,Katan:2015qfa}.

This, together with the exchange of dominance of poles (see Fig. \ref{fig.QuenchQNM}), makes the transition from one phase to the other clear. In fact the on-axis pole giving rise to this phenomenon is the so-called amplitude or Higgs mode of the superfluid, recently measured at the SI transition, as reported in \cite{endres2012higgs}. It is interesting to note that a study of the many-body dynamics of the relaxation of antiferromagnetic order in the XXZ model yields results in striking resemblance to those discussed here \cite{PhysRevLett.102.130603}, with the order parameter undergoing exponentially damped decay or exponentially damped decaying oscillations, towards its final equilibrium state.

\subsection{Scaling Laws for finite-rate quenches and holographic turbulence}
\begin{figure}[!h]
\begin{center}
\includegraphics[width=\columnwidth]{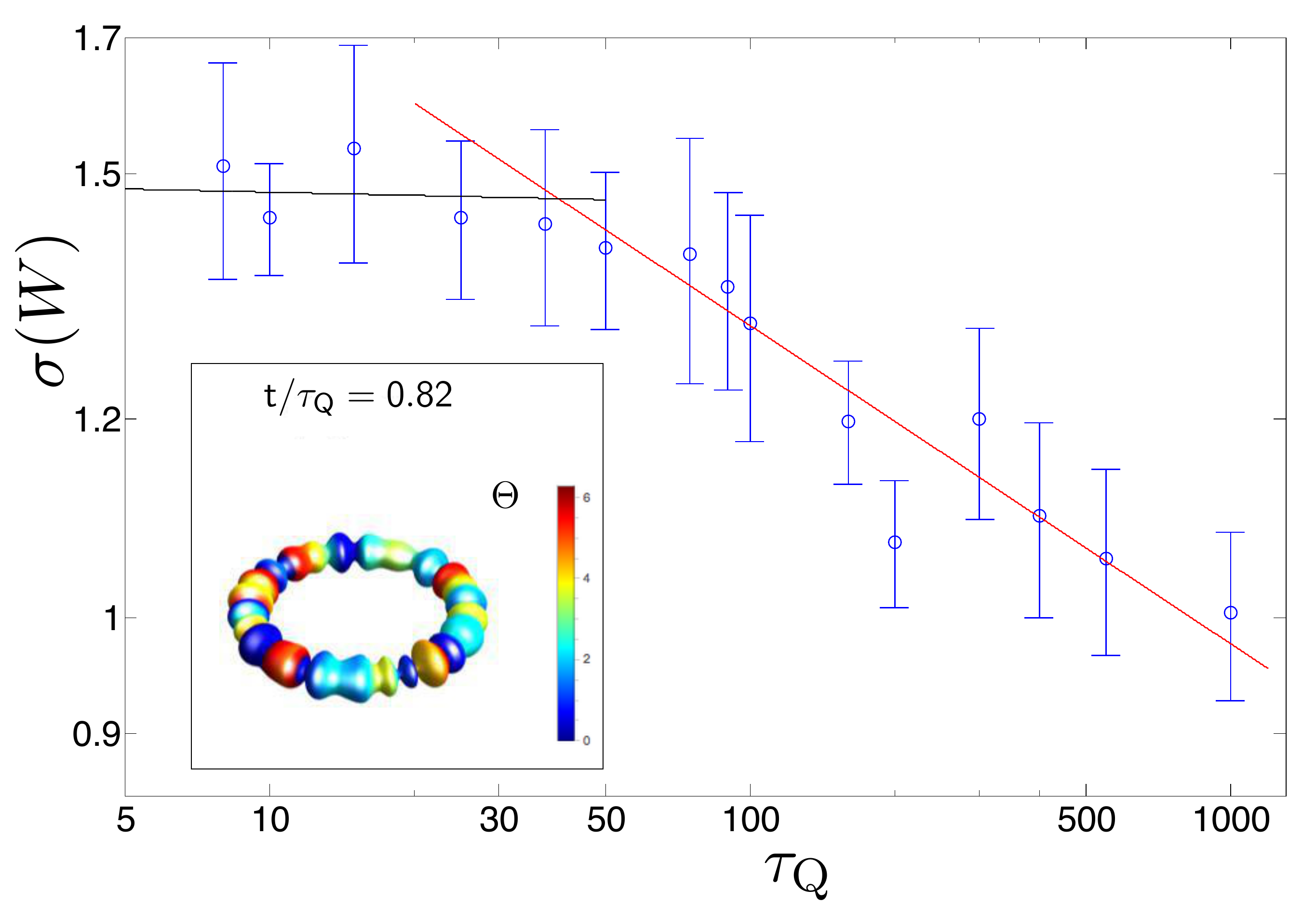}\\
\end{center}
\caption{KZ scaling law of the winding number of the condensate of a $1+1$ dimensional superfluid on a ring as a function of the quench rate $\tau_Q$. A KZ scaling regime can be observed for large values of $\tau_Q$ preceded by a plateau region. The inset shows a snapshot of the phase gradient ($\Theta$) and magnitude (thickness as a function of angle around the ring) of the condensate for an example quench. Figures taken from  \cite{Sonner:2014tca}. A similar scaling preceded by a plateau region was observed in \cite{Chesler:2014gya} for vortex statistics in a $2+1$ superfluid. Figures taken from \cite{Sonner:2014tca}. }
 \label{fig:KZRing}
\end{figure}

Above we noted that despite the generally intricate nature of nonequilibrium dynamics, sometimes scaling results can be obtained, for example in the limit of very fast quenches \cite{Das:2014jna,Das:2014hqa,Das:2015jka,Buchel:2013gba}. In fact, obtaining scaling laws as a function of quench parameters is a subject with a venerable history, and we will now explore this issue for symmetry breaking quenches. Indeed a paradigmatic nonequilibrium phenomenon manifests itself if symmetry-breaking critical points are crossed at a finite rate $\tau_Q$ \cite{Zurek:1985qw,Kibble:1976sj}. The critical point can be either a thermal phase transition, or a quantum-critical point \cite{polkovnikov2005universal,PhysRevLett.95.105701}.  
In such situations the symmetry breaking order parameter will take uncorrelated expectation values in regions separated more than a certain distance, and their eventual resolution results in the creation of topological defects -- under the condition that the vacuum manifold allows them. The number and distribution of topological defects has been proposed to follow a scaling relation, the so-called Kibble-Zurek (KZ) scaling  \cite{Zurek:1985qw,Kibble:1976sj}, whose form is determined by equilibrium critical exponents. When a second-order critical point (or a quantum-critical point) is approached at the finite rate $\tau_Q$, the instantaneous correlation length $\xi(t)$ and relaxation time $\tau(t)$ evolve as
\be
\xi(t) = \frac{\xi_0}{|\epsilon(t)|^\nu}\,,\qquad \tau(t) = \frac{\tau_0}{|\epsilon(t)|^{z\nu}}
\ee
where $\epsilon(t) = t/\tau_Q$ parametrizes the distance to the relevant critical point as a function of time. One then posits that the system will loose its ability to adiabatically follow the externally imposed change at the instant $\hat t$ where the remaining time to cross the critical point equals the equilibration time scale $\tau_Q$, i.e. $\tau[\epsilon(\hat t)] = \tau_Q$. The system will then be effectively frozen during the interval $(-\hat t, \hat t)$, where
\be
\hat t \sim \left(\tau_0 \tau_Q^{z\nu}  \right)^{\frac{1}{1+z\nu}}\,,\qquad\textrm{with}\qquad \hat \xi \sim \xi_0 \left( \frac{\tau_Q}{\tau_0}  \right)^{\frac{\nu}{1+z\nu}}\,\,.
\ee
Since different parts of the system of size $\sim\hat\xi$ are no longer able to communicate, one expects that the order parameter will take on uncorrelated values on patches of size $\sim\hat\xi$ and thus that the density $d$ of topological defects after the quench through the critical point will scale approximately as
\be\label{eq.KZscaling}
d \sim \hat \xi^{n-D}\,,
\ee
where $D$ is the spatial dimension of the system and $n$ is the spatial dimension of the defect.
Following on from the general scaling theory \cite{Zurek:1985qw,Kibble:1976sj}, the Kibble-Zurek mechanism has been studied in a variety of model systems \cite{delCampo:2013nla}. The study of dynamical defect formation necessitates a solution of many-body dynamics far from equilibrium, often hopelessly out of reach, but, as we have seen many times above, a task for which holography is well suited  \cite{Basu:2011ft,Basu:2012gg,Gao:2012aw,Garcia-Garcia:2013rha,Basu:2013soa,Sonner:2014tca,Chesler:2014gya}.  The study   \cite{Sonner:2014tca}  focuses on winding-number statistics of a superfluid ring (see Figure \ref{fig:KZRing}), while \cite{Chesler:2014gya} investigate vortex formation in a 2D superfluid. These works were able to confirm the validity of the predicted scaling laws, establishing the applicability of the KZ scaling law for strongly coupled systems without quasiparticles. However, \cite{Sonner:2014tca,Chesler:2014gya} were also able to extract accurate values for the pre-factor, which under certain conditions can deviate significantly from the KZ prediction  \cite{Chesler:2014gya}. 

We have argued that holographic duality allows us to extract a simple intuitive picture of the complicated many-body dynamics, by thinking about the quasi-normal and spatial-conformal modes of the system. Whenever the deviations of the order parameter from its equilibrium value (at the instantaneous value of $\epsilon(t)$) are small, we can investigate the system using bulk linear equations. The response is then governed by the {\it leading poles} in the complex frequency plane. When the frozen system enters the parameter regime where the broken symmetry is favored, one finds an exponential growth regime governed by the time-scale  \cite{Chesler:2014gya} $[{\rm Im}(\omega_0)]^{-1}>0$ of the leading unstable quasinormal mode $\omega_0(\epsilon, k)$, computed about the supercooled un-condensed state. This leads to the exponentially growing contribution,
\be
C(t,q) =\zeta \int^t_{\hat t}  dt'|H(q)|^2 e^{2\int^t_{t''}{\rm Im}\mathfrak{w}_0(\epsilon(t''),q)dt''} \,,
\ee
to the Fourier transform of the correlation function $C(t,\mathbf{r}):= \langle \psi^*(t,\bx + \mathbf{r}), \psi^*(t,\bx)\rangle$, where $H(q)$ is a slowly varying function of momentum whose details we will not need, and $\zeta$ parametrizes the typical amplitude of the noise correlation in the system. Translated into real space, the leading quasinormal mode analysis predicts
\be
C(t,r) = \tilde\epsilon(t) e^{a_2 \bar t^{1+z\nu}}e^{-\frac{r^2}{\ell_{\rm co}(t)^2}}
\ee
in terms of the reduced time $\bar t := t/\hat t$, and where $\ell_{\rm co} \sim \bar t^{\frac{1+(z-2)\nu}{2}}$ is the time-dependent coarsening length.  The ${\cal O}(1)$ parameter $a_2$ is not universal, but the results below do not depend on its precise form. This result has interesting consequences, namely it predicts that the system may undergo a parametrically large amount of coarsening already before a well-defined condensate forms. Let us denote this latter time-scale as $t_{\rm eq}$. A large amount of early coarsening happens whenever the the timescales $\hat t$ and $t_{\rm eq}$ are parametrically different, which concretely means that the parameter
\be
R\gg 1 \qquad\textrm{with}\qquad R \sim \zeta^{-1}\tau_Q^{\frac{(d-z)\nu-2\beta}{1+\nu z}}\,,
\ee
where $\beta$ is the condensate critical exponent, $|\psi |^2\sim \epsilon^{2\beta}$ near the phase transition.
Holographic systems have $\zeta \sim \frac{1}{N^2}$ and thus fall into the class of theories that are expected to undergo a parametrically large amount of coarsening before $t_{\rm eq}$. This was numerically confirmed in \cite{Chesler:2014gya}. A general lesson emerging from these explicit holographic results on finite-rate quenches is the good agreement with the scaling form \eqref{eq.KZscaling} predicted by KZ, even in the strongly coupled regime, whereas the numerical prefactor following from general KZ arguments (see \cite{delCampo:2013nla} for a discussion) clearly has to be taken with a grain of salt, as illustrated by the detailed comparison in  \cite{Sonner:2014tca} and \cite{Chesler:2014gya}.

Yet another nonequilibrium paradigm crucially involving inhomogeneous configurations of the field theory is the topic of {\it turbulence}. Holography furnishes us with a formalism whose hydrodynamic limit is well understood, while at the same time comprising a fully UV complete description of physics on all lenghth-scales. This allows one to capture turbulent effects beyond hydrodynamics within a well-defined framework amenable to numerical and analytical analysis. Such a study was undertaken in \cite{Adams:2013vsa} for ordinary fluids and \cite{Adams:2012pj} for superfluids. Having commented on ordinary turbulence previously, we concern ourselves here with the superfluid case. \vskip0.5em

For field theories with broken symmetries it is more natural for vorticity to be carried by quantum vortices of the superfluid, which extend as vortex lines into the holographic bulk. Such a system is described by the theory \eqref{eq.superfluid}, although the concrete studies of \cite{Adams:2012pj} and \cite{Ewerz:2014tua} were performed in the probe limit. Interestingly the superfluid dual to a bulk black hole with scalar condensate shows turbulent $k^{-5/3}$ Kolmogorov scaling with a direct energy cascade, which has a beautiful holographic interpretation: energy is transferred from long wavelengths to short wavelengths until they reach the size of a typical vortex core. Since this field-theory vortex extends as a vortex line into the bulk, where it punches a whole through the condensate shielding the horizon, this mode can now efficiently dissipate into the horizon through the vortex tube. This explanation of the direct cascade via dissipation at the vortex scale into the horizon was verified in \cite{Adams:2012pj}, by measuring locally the energy flux through the horizon.

 \section{Numerical Techniques for AdS/CMT away from equilibrium}\label{sec:num}

In the semi-classical large-$N$ limit, the task of studying the exact time dependent physics of a quantum field theory is translated into the task of solving a set of Einstein-matter equations (\ref{eq.superfluid}) from a given initial configuration, subject to suitable boundary conditions. GR, as the name suggests is a generally covariant theory, and it is non-trivial to understand how it gives rise to a system of equations that propagate given initial configurations forward in time. This is, however, necessary, not least from the point of view of numerically solving the Einstein equations. Given the rich and complicated gauge structure of the theory, it is no surprise that there are different schemes for doing so, several of which have found successful application in the past years.
 
  Here we review two different classes of schemes which have been successfully employed in the context of AdS/CMT, namely
  \begin{enumerate}
  \item the characteristic method, which propagates data given on a lightlike slice.
  \item the (generalized) harmonic scheme, which propagates initial data given on a spacelike slice.
  \end{enumerate}
 While the latter has seen widespread application in asymptotically flat gravity, notably in the first stable evolution of the inspiral problem \cite{Pretorius:2005gq}, the former has seen much success in the context of non equilibrium holography, i.e. largely in AdS space. While giving rise to very efficient solution methods, choosing to evolve along characteristics comes with a price: in the presence of focussing these characteristics can converge, eventually forming what are known as caustics. In such cases the characteristic `time' variable is no longer single valued, leading to a breakdown of the method. Despite this limitation the approach has proved very fruitful in nonequilbirium AdS simulations, since in the cases of physical interest no such caustics have formed outside of apparent horizons, and therefore can be excised from the computational domain. This is essentially the case because physically interesting situations, from the AdS/CMT point of view, almost always involve so-called `large' black holes, with an in-fall time that is short on the typical time-scales of the evolution. Caustics are thus almost guaranteed to form only behind any horizons.
 
 Another major advantage of the characteristic scheme over the Cauchy scheme is the ease with which the singularity can be excised. Excision can be achieved by letting the numerical grid end somewhere just inside the horizon, essentially by stopping the radial integration. This is equivalent to specifying free boundary conditions at a regular point behind the horizon, which at the same time ensures regularity {\it at} the horizon, and does not lead to any artificial boundary effects on the exterior region due to causality.

The main technical difficulty in achieving a well-posed evolution of the Einstein equations (plus suitable matter) stems from the diffeomorphism invariance of the theory. A pedagogical introduction is given in the books \cite{hawking1973large,wald2010general}. The general ideas behind defining a well-posed evolution for a theory with a local gauge (coordinate) redundancy can be illustrated more easily using the example of Maxwell's equations that enjoy a much more simple gauge invariance as compared to GR. Furthermore this simple example illustrates the main issues present in the more complicated gravitational case. We therefore describe in the following the two major choices of evolution schemes, on the example of Maxwell's equations. This is also a useful excercise, as it is an essential ingredient in the study of a holographic superfluid in the probe limit, and has been used in the recent studies \cite{Adams:2012pj,Sonner:2014tca,Chesler:2014gya}. After giving an account of the conceptual issues using the example of evolving Maxwell equations (with sources) on asymptotically AdS spacetimes, we will outline the corresponding evolution problems for Einstein gravity.    

\subsection{The Cauchy Scheme}\label{sec.CauchyScheme}
 The most adaptable approach is given by Cauchy-like evolution, particularly when one deals with the gravitational case. The reason is simple: the global choice of coordinates employed in the characteristic scheme\footnote{One is of course free, in principle, to allow for the freedom to dynamically adapt the gauge also within the class of choices suitable for characteristic evolution, and this may indeed be interesting to pursue. However, as of writing of this review, this has not been explored in numerical holography.} may become degenerate as evolution proceeds. The evolution may lead to caustics, i.e. loci where several null rays intersect, and at such points characteristic evolution is ill-defined. After these general remarks, let us now start by explaining Cauchy evolution schemes starting with the example of Maxwell theory.
The evolution equations are second order partial differential equations. One therefore expects that initial data should correspond to a set of functions at the initial time, as well as their first derivatives. We shall see that this expectation is correct, up to an important detail, namely that the initial data themselves are not completely free, but rather must satisfy certain constraints.

Let us write the metric of the asymptotically AdS$_{d+1}$ spacetime in the (`Schwarzschild-like') form
\be\label{eq.SchwarzschildAdS}
ds^2 = \frac{\ell^2}{z^2}\left( \frac{dz^2}{f(z)} - f(z)dt^2 + \sum_idx^i dx^i \right)\,,
\ee
where $f(z) =1 - (z/z_h)^d$. 
Initial data is specified on a constant time surface $\Sigma_0$ at $t-t_0=0$ with timelike unit normal $n_a = \sqrt{-g_{tt}} \delta_{at}$.
Maxwell's equations with a source then take the form
\be\label{eq.Maxwell}
{\cal E}^b\:=\nabla_a F^{ab} -j^b=0\,,
\ee
although for simplicity we will for now on use the source-free equations $j^b = 0$. Adding back the sources is straightforward.
At this point it is convenient to introduce the notation $\left\{ x^A \right\} = \left\{z,x^i  \right\}$ for the coordinates on the spacelike $\Sigma_0$. 
If the equations in the present decomposition are to propagate the degrees of freedom contained in $A^a$, we immediately run into a problem: the component equation along the unit normal $n_a {\cal E}^a$ does not contain second time derivatives, while all other orthogonal components do have second time derivatives. This means that we only have $d$ dynamical equations for $d+1$ evolution variables. As is well known this is not a fundamental problem, but merely a complication in the formulation of the evolution equations due to gauge invariance.

In fact, as we shall see now, the time-like component of Maxwell's equation $n_a{\cal E}^a$ gives precisely the constraint equation on $\Sigma_0$, which must be satisfied by admissible initial data. This constraint equation is nothing but the differential form of the Gauss Law on $\Sigma_0$:
\be
\nabla_i \left( \nabla^i A^t - \nabla^t A_i \right) = 0 \qquad \Leftrightarrow \qquad D \cdot \mathbf{E} = 0\,,
\ee
where $D$ is the covariant derivative on $\Sigma_0$ and $\mathbf{E}$ is the electric field, whose components are defined by the round brackets in the equation above. We remark in passing that the magnetic constraint $D\cdot\mathbf{B}=0$ is satisfied identically.
Since we have already stated that the problem of the timelike component is related to gauge invariance, it is not surprising that one way to proceed from here is to pick a specific gauge. In the case at hand a standard choice is the covariant Lorenz gauge
\be
\nabla_bA^b = 0\,,
\ee
but more general gauge conditions where the right hand side is an arbitrary source function,
\be \label{eq.GeneralizedLorenz}
C:=\nabla_a A^a - \Phi(A^b)=0\,,
\ee
 are possible. The analogous choice in the case of the Einstein equations is at the heart of the (generalized) harmonic scheme. An example of the scheme \eqref{eq.GeneralizedLorenz} has been implemented in the works of \cite{Withers:2014sja,Donos:2015eew}, who numerically solved the Einstein-Maxwell system using the so-called DeTurck approach.
Let us first choose the standard Lorenz gauge, $\Phi=0$. With this choice we can formulate a well-posed initial value problem as follows: start with the Maxwell equations, \eqref{eq.Maxwell},
written as
\be\label{eq.MaxwellLorenzGauge}
\nabla^2 A_a = R_a{}^bA_b +  \nabla^b \nabla_a A^a\,,
\ee
where the last term vanishes in Lorentz gauge.
The curvature term on the right hand side is present in our chosen aAdS background.
It is equations of this form which can be shown to have well-posed initial value formulations on globally hyperbolic background spaces (see e.g. \cite{wald2010general}). Of course, AdS is not globally hyperbolic, so one needs to specify in addition suitable boundary conditions. We shall return to this issue below.

 For the case at hand one specifies initial data $\left( A_b, \partial_t A_b\right)$ on $\Sigma_0$, subject to the initial value constraint $D\cdot \mathbf{E}=0$. By a gauge transformation we may bring this initial data into the Lorenz gauge. Alternatively one can specify initial data only for the `physical components', $\left( A_B, \partial_t A_B \right)$, and then determine the remaining components from the others, via the gauge condition. The evolution equations in hyperbolic form can be used to time-evolve the initial data -- again subject to suitable boundary conditions in the case of AdS. One can show that the solution stays in the Lorenz gauge, if and only if the initial data satisfies the gauge condition on $\Sigma_0$, and that $\partial_t \left(  \nabla_bA^b\right)\bigr|_{\Sigma_0} =0$. The latter condition is equivalent to the initial value constraint $0=\nabla^a F_{a0}$, as can be seen from (\ref{eq.MaxwellLorenzGauge}). 
 
  It is instructive to count how many degrees of freedom (per spatial point) are actually propagated in this way. Naively we have $d+1$ Klein-Gordon-type equations, but the initial data constraint immediately eliminates one degree of freedom reducing the total to $d$. The gauge invariance introduces another free function, the gauge parameter, removing one further degree of freedom, so that in effect the Maxwell equations propagate $d-1$ degrees of freedom.

Let us now allow for a general source $\Phi$, assumed, for the time being\footnote{Later in section \ref{sec.CauchyAdS} we shall allow such sources to obey their own dynamical equations.}, to be a specified function. A convenient trick in order to proceed is to add a term $\nabla^a C$ to Maxwell's equations to obtain
\be\label{eq.MaxwellGenHarmonic}
\nabla_b F^{ba} + \nabla^aC=0\,.
\ee
This evidently reduces to Maxwell's equations when the gauge condition is satisfied, $C=0$. By similar manipulations as above, one sees that the principal part of (\ref{eq.MaxwellGenHarmonic}) is $\nabla^2A_a$, i.e. all components of $A_a$ satisfy hyperbolic equations, as desired. One can now show that $C=0$ everywhere, if $C$ vanishes on $\Sigma_0$ and $\partial_t C\bigr|_{\Sigma_0}=0$. The latter condition is, again, equivalent to the initial data constraint. In other words, the initial value problem is well posed, given that the initial data satisfy the constraint, and that the gauge function $C$ vanishes on $\Sigma_0$.

 In AdS, however, we still need to consider the issue of suitable conditions at the timelike boundary. In general the precise form of the asymptotic boundary condition depends on the dimension of the spacetime, and on the requirements of the physical problem under study\footnote{For example, one may choose to have all sources turned off, or one may want to specify a given profile for a certain source, as described in Section \ref{sec.AspectsOfDuality}.}, so here we will be schematic. Suppose a function, for example one of the components of $A_a$ or a component of some other matter field, has asymptotic behavior (c.f. \eqref{asymp})
 \be
 \phi(z,\,\ldots) = \phi_0 + \phi_1 z + \phi_2 z^2 + \cdots\,,
 \ee
 which for simplicity we assume to proceed in integer powers. These asymptotics encode source and expectation value behavior as described in Section \ref{sec.AspectsOfDuality}, so that the field-theory source corresponds to a term $a(x^\mu)z^{\alpha_1} = \phi_0(x^\mu)$ in the expansion above.
 Then our goal  typically is to set the first few terms (those which depend on the data $a(x^\mu)$ alone)  in this series to zero, so that the leading-order boundary behavior is given by the `vev' term $\phi_{\alpha_2}(x^\mu)z^{\alpha_2}$. This can be achieved by defining a rescaled function $\bar \phi = \phi z^{\alpha_1-1}$ and imposing a homogeneous Dirichlet boundary condition on the rescaled field $\bar \phi(z=0)=0$. In AdS/CFT terms, such a boundary condition is equivalent to demanding that the source of the dual operator vanish, while its expectation value will be determined by the dynamics. A similar approach, with rescaling by appropriate powers, would impose inhomogeneous Dirichlet boundary conditions to specify a non-trivial source function, $\phi_{\alpha_1}(x^\mu) \equiv a(x^\mu)$, if so desired. It is essential to ensure that the gauge function $\Phi (A^b)$ is chosen in such a way as not to interfere with the prescribed boundary behavior. While this is solved on a case by case basis in the exisiting literature, to the best of our knowledge, no systematic study of this issue has been undertaken. It would be useful to investigate this important issue further, in particular for the case of gravitational dynamics to be addressed below.
 
 Boundary conditions in the interior are usually determined by regularity conditions on fields at the various degenerate points of the background, such as horizons, or axes of symmetry. The position of a horizon can straightforwardly be inferred from the form of the background metric in the present case. This issue is more subtle in the full gravitational problem, and described in detail in the literature, for example in \cite{Bantilan:2012vu}.
 
\subsubsection{Summary}
Thus the Cauchy method proceeds as follows:
\begin{enumerate}
\item At the initial surface $\Sigma_0$, i.e. at $t = t_0$ one sets up initial data consisting of the fields $A_a$ and their derivatives $\partial_t A_a$, subject to the initial data constraint $D\cdot \mathbf{E}=0$, $D\cdot \mathbf{B}=0$.
\item By a gauge transformation on $\Sigma_0$ one brings the initial data into the desired form, \eqref{eq.GeneralizedLorenz}. As explained above, the form of the evolution equations now guarantees that the solution remains in the chosen gauge for all time, given that the initial data satisfy the constraint equation.
\item The Maxwell Equations in this gauge form a set of hyperbolic differential equations, which can be stepped forward in time using any finite difference approximation, such as fixed-order Runge-Kutta, for example, making sure that appropriate boundary and regularity conditions are imposed at each step (see for example \cite{Chesler:2013lia,Zhang:2016coy}).
\end{enumerate}
We thus have constructed a second-order evolution scheme along a timelike direction $t$, starting from constrained initial data. We now explain how an analogous scheme can be formulated for the Einstein Equations in AdS, again starting from constrained initial data.
\subsubsection{The Cauchy Method for $AdS$ Gravity}\label{sec.CauchyAdS}
We are now interested in solving Einstein's Equations
\be\label{eq.EEq}
0={\cal E}_{ab}:=G_{ab} + \Lambda g_{ab} - T_{ab} 
\ee
in asymptotically AdS spacetimes.
A Cauchy scheme for AdS gravity was numerically implemented in \cite{Bantilan:2012vu}, based on the seminal work of \cite{Pretorius:2005gq} and we largely follow their treatment here. The authors of \cite{Heller:2012je} also present a Cauchy-like scheme for dynamics in AdS, and use it to study Bjorken flow in the strongly coupled field theory. Strictly speaking, the name Cauchy scheme is a misnomer, since AdS has no Cauchy surface due to its time-like boundary. One has instead an initial-boundary value problem. The method we describe here most closely resembles Cauchy schemes in flat space, and so we follow the naming convention of \cite{Bantilan:2012vu}. We recommend the treatment in \cite{wald2010general} (Chapter 10) as a pedagogical introduction to the Cauchy problem in General Relativity.

The idea is to choose an initial time surface $\Sigma_0$ -- roughly speaking a generalization of the notion of the $t=t_0$ hypersurface above -- and specify initial data for the Einstein Equations, in the same sense as was done above for Maxwell. Let us denote the timelike normal to this surface $n_a$. The natural object to consider as initial data is then the functional form of the metric at the initial time. That is to say, one specifies a Riemannian metric $h_{AB}$ and its derivative away from $\Sigma_0$, $\partial_t h_{AB}$, which in fact is nothing but the extrinsic curvature $K_{AB}$ of $\Sigma_0$. In the full, evolved, spacetime with metric $g_{ab}$, the Riemannian metric on $\Sigma_0$ will be thought of as the induced metric $h_{AB} = g_{AB} + n_A n_B$. Evidently this data leaves the remaining $d+1$ components of the metric undetermined. Luckily these are balanced by the $d+1$ free functions to specify coordinates (diffeomorphisms), suggesting again that the apparent problem lies in the gauge freedom of the equations. Moreover, the components $n_a {\cal E}^{ab}$ of Einstein's Equations do not contain any second time derivatives and thus do not serve to propagate any physical degrees of freedom. Instead they give rise to constraints on initial data. Since, compared to the Maxwell case above, there is a further free index in the projection $n_a{\cal E}^{ab}$, there are now two kinds of constraints: one along $n_a$ and $d$ perpendicular to it. The former is called the Hamiltonian constraint, while the latter are often called the momentum constraints. With the help of the Gauss-Codacci relations these can be expressed as
\be
{}^{(d)}R + K^2 - K_{ab}K^{ab} - 2 \Lambda = \rho_E\,,
\ee
for the Hamiltonian constraint and
\be
D_b K^{ba} - h^{ab}D_b K = j^a\,,
\ee
for the momentum constraints. The right-hand sides, $\rho_E = T_{ab}n^a n^b$ and $j^a = -T_{bc}n^b h^{ac}$, vanish for pure gravity, and otherwise take on the corresponding values appropriate for sources of energy-momentum that make up $T_{ab}$, projected on $\Sigma_0$.

Since, again, the source of the complications is diffemorphism invariance, one should construct evolution equations in a suitable gauge, the analog of the Lorenz gauge procedure described above. This is exactly what is achieved in the (generalized) harmonic scheme.

This scheme renders the Einstein Equations hyperbolic in the following way. One chooses coordinates $x^a$, satisfying the wave equation with some specific source,
\be
C^a :=H^a - \nabla^2 x^a=0\,,
\ee
where $H^a$ can either be a known function, in which case the original harmonic scheme \cite{choquetbruhat1962cauchy} is included as the special case $H^a=0$, or we specify separate evolution equations for the sources. Schematically
\be
{\cal L}_a \left[ H^a \right]=0\qquad [\text{no summation}]\,.
\ee
To see how this renders the Einstein Equations hyperbolic, we use the same trick as in Eq. (\ref{eq.MaxwellGenHarmonic}) above. That is we start with the fundamental equations (for the sake of convenience, the trace-removed Einstein Equations, $R_{ab} = \bar T_{ab}$), and subtract a constraint term
\be
R_{ab} - \nabla_{(a}C_{b)} - \bar T_{ab} =0\,.
\ee
Evidently, when the coordinates satisfy the gauge condition $C_a=0$, this is equivalent to the Einstein Equations. The presence of the $\nabla_{(a} C_{b)}$ term serves to subtract an unwanted $\nabla_{(a} \nabla^2 x_{b)}$ term from the Ricci tensor, so that the principle part of the equations becomes
\be
{\rm P}\left[ R_{ab} - \nabla_{(a}C_{b)}  \right] = -\frac{1}{2}g^{ab}g_{cd,ab} \,,
\ee
showing that we have a hyperbolic system, of the type that admits a well-posed initial value problem, amenable to numerical solution techniques.

 The art of such schemes consists in choosing an appropriate set of source functions, or more generally evolution equation for $H^a$ to achieve stable numerical evolution \cite{Pretorius:2004jg,Pretorius:2005gq}. Typically this is done in such as way as to choose evolution equations for the constraints which ensure hat any potential growing modes which would violate the constraints are instead damped \cite{Gundlach:2005eh}. However, this could potentially be subtle in empty AdS space, where small constraint violating modes can be amplified by successive reflections off the timelike boundary.

Appropriate boundary conditions are specified, as we saw above, by rescaling the evolution variables to eliminate unwanted asymptotic components via the imposition of Dirichlet conditions at the boundary. Internal boundary conditions follow from regularity. A detailed description of boundary conditions for evolution of $AdS_5$ gravity are given in \cite{Bantilan:2012vu} and may be used as a guide for other dimensional setups as well. The mathematically rigorous state of the art concerning well posedness of AdS evolution can be found, for example in \cite{Holzegel:2011qk,Holzegel:2013vwa,Holzegel:2015swa,Holzegel:2015swa}.

\subsection{The Characteristic Scheme}\label{sec.Characteristic}
\begin{figure}[h]
\begin{center}
\includegraphics[width=0.9\columnwidth]{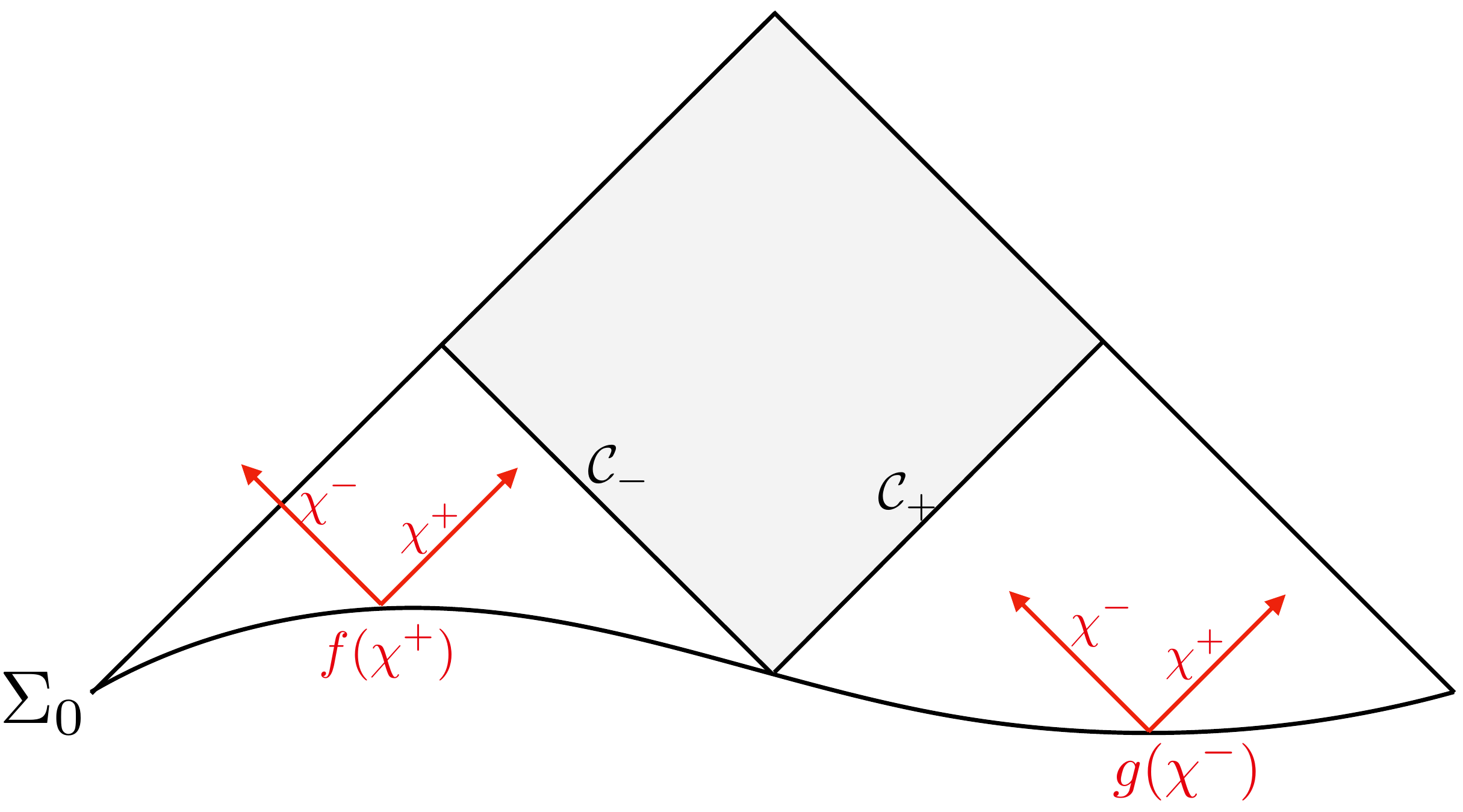}\\
\end{center}
\caption{Characteristic (double-null) vs. Cauchy formulation of the wave equation in Minkowski space. The gray area is determined by data given on the null cone ${\cal C} = {\cal C}_- \cup {\cal C}_+$. Data on the timeslice $\Sigma_0$ determine the solution on the entire triangular region whose base is $\Sigma_0$. However, one can see that the gray area only depends on purely ingoing data $g(\chi^-)$ on one half of the Cauchy slice $\Sigma_0$ and purely outgoing data $f(\chi^+)$ on the other half, in other words on exactly half the amount of data specified on $\Sigma_0$. This cutting in half of the necessary data is already a sign of the improved efficiency of characteristic methods (so long as one is in fact only interested in the region inside the development of ${\cal C} = {\cal C}_- \cup {\cal C}_+$).}
 \label{fig:CharacteristicMinkowski}
\end{figure}
\subsubsection{Intuition from the wave equation}
In this approach we use an ingoing null direction as the evolution variable. Furthermore, and in distinction to the Cauchy scheme of section \ref{sec.CauchyScheme}, we fix the gauge explicitly. In gravity (see below), this means that we chose a global set of coordinates for the entire evolution. Consequently this characteristic scheme is less adaptable than the generalized harmonic scheme, and in particular breaks down for spacetimes which contain caustics, an issue which we will return to below. For the time being we mention that these disadvantages are very often compensated by improvements in performance and stability \cite{Chesler:2013lia}.

We first describe the ideas somewhat schematically before going on to fill in the details and explicit equations. Let us start with a simple analogy, the wave equation in $1+1$ flat space,
\be
\partial_t^2 \phi = \partial_x^2\phi\, .
\ee
This equation has the general solution
\be
\phi(t,x) = f(t-x) + g(t+x)\,,
\ee
where $f$ and $g$ are arbitrary functions. The former describes an arbitrary right-moving or `in-going' wave, while the latter describes a general left-moving or 'out-going' wave. The curves $\chi_\mp(t,x) = t\mp x $ are known as in/out-going characteristics. In fact the wave operator factorizes along these characteristics into
\be
\partial_x^2 - \partial_t^2 = (\partial_t + \partial_x)(\partial_t - \partial_x)\,
\ee
and each of $f,g$ satisfies a first-order equation. One can convince oneself that the full solution of the wave equation on a causal diamond in Minkowski spacetime can be constructed by combining a left-moving solution $f(\chi^+)$ and a right-moving solution $g(\chi^-)$ in the way shown in Fig. \ref{fig:CharacteristicMinkowski}, so that effectively one only ever solves first-order equations in $\chi^\pm$, respectively. It is thus possible to construct an efficient scheme taking advantage of the factorization of the wave operator along characteristics, which is also at the core of the characteristic approach on AdS. Since AdS has a timelike boundary it actually turns out to be more convenient to work in terms of a single null coordinate, $u$, instead of the double-null formulation in terms of $\chi^\pm$ we just outlined.
\subsubsection{Characteristic method for Maxwell in AdS}
Let us now proceed to a more detailed description of the characteristic scheme in AdS using ingoing Eddington coordinates. Characteristics of Eq. (\ref{eq.SchwarzschildAdS}) are given by
\be
u = \int f^{-1}dz - t\,,\qquad v = \int f^{-1}dz + t\,.
\ee
Since we are interested in characteristic evolution, we rewrite the metric in terms of $u$, the so-called ingoing Eddington-Finkelstein form
\be\label{eq.backgroundEF}
ds^2 = \frac{\ell^2}{z^2} \left[ -f(z)du^2 - 2 du dz + \sum_{i,j=1}^{d-1} q_{ij}dx^i dx^j  \right]\,.
\ee
Poincar\'e AdS is the special case $z_h\rightarrow \infty$ and $q_{ij}$ can parametrize the metric on a $d-1$ sphere or $d-1$ flat space and $q_{ik}q^{kj} = \delta_i{}^j$. For a finite value of $z_h$ this background describes a black brane. This spacetime has a conformal boundary ${\cal B}$ at $z\rightarrow 0$ with unit normal $n_{\alpha} = \frac{\delta_{z\alpha}}{z \sqrt{f}}$. For simplicity we will from now on set $\ell=1$ by a choice of units. We define an initial null surface ${\cal N}|_{u=u_0}:={\cal N}_0$ via the coordinate condition $u = u_0$. We have the normal $k_\alpha = K \delta_{\alpha u}$ with arbitrary prefactor.  This surface is spanned by the null rays $x^i$ and the coordinate $z$ varying along the null rays. By convention we give the set of coordinates $\left\{ u,x^i \right\}$ the label $\left\{ x^\mu \right\}$, while the total coordinate, including the radial $z$ direction is labeled $X^a$.  We are interested in the Maxwell equations \eqref{eq.Maxwell}, where $j$ includes the sources in the theory, such as the charged scalar in Eq. (\ref{eq.superfluid}). For the purpose of the present analysis we need not concern ourselves with their own dynamics, as all the subtleties and techniques we want to illustrate reside in the gauge sector.

In order to have a well-posed problem it is necessary to choose a gauge. We will pick the axial gauge
\be\label{eq.nullGauge}
A_z = 0\,,
\ee
which, as we shall see, is both convenient from a holographic perspective and well suited for the characteristic evolution scheme. There is still some residual gauge freedom, as a gauge transformation whose parameter does not depend on $z$ preserves Eq. (\ref{eq.nullGauge}). That is, we may still make a transformation
\be\label{eq.MaxResidual}
A_\mu \rightarrow A_\mu + \partial_\mu \lambda(x)\,.
\ee
We will fix this residual gauge freedom by giving a condition on a single $z$ slice.

The equations then decompose as follows. Firstly we have the scalar auxiliary equation
\be
{\cal E}^z = 0\,,\label{eq.aux}
\ee
which will impose a condition on a given $z$ slice. One can show that the auxiliary equation is satisfied iff it is satisfied at a single $z$ slice, and the evolution equations for the other components are satisfied. It is convenient to choose the boundary slice to impose Eq. (\ref{eq.aux}), where it will be identically satisfied if the fields have regular behavior at $z=0$.
\begin{figure}[h]
\begin{center}
\includegraphics[width=0.6\columnwidth]{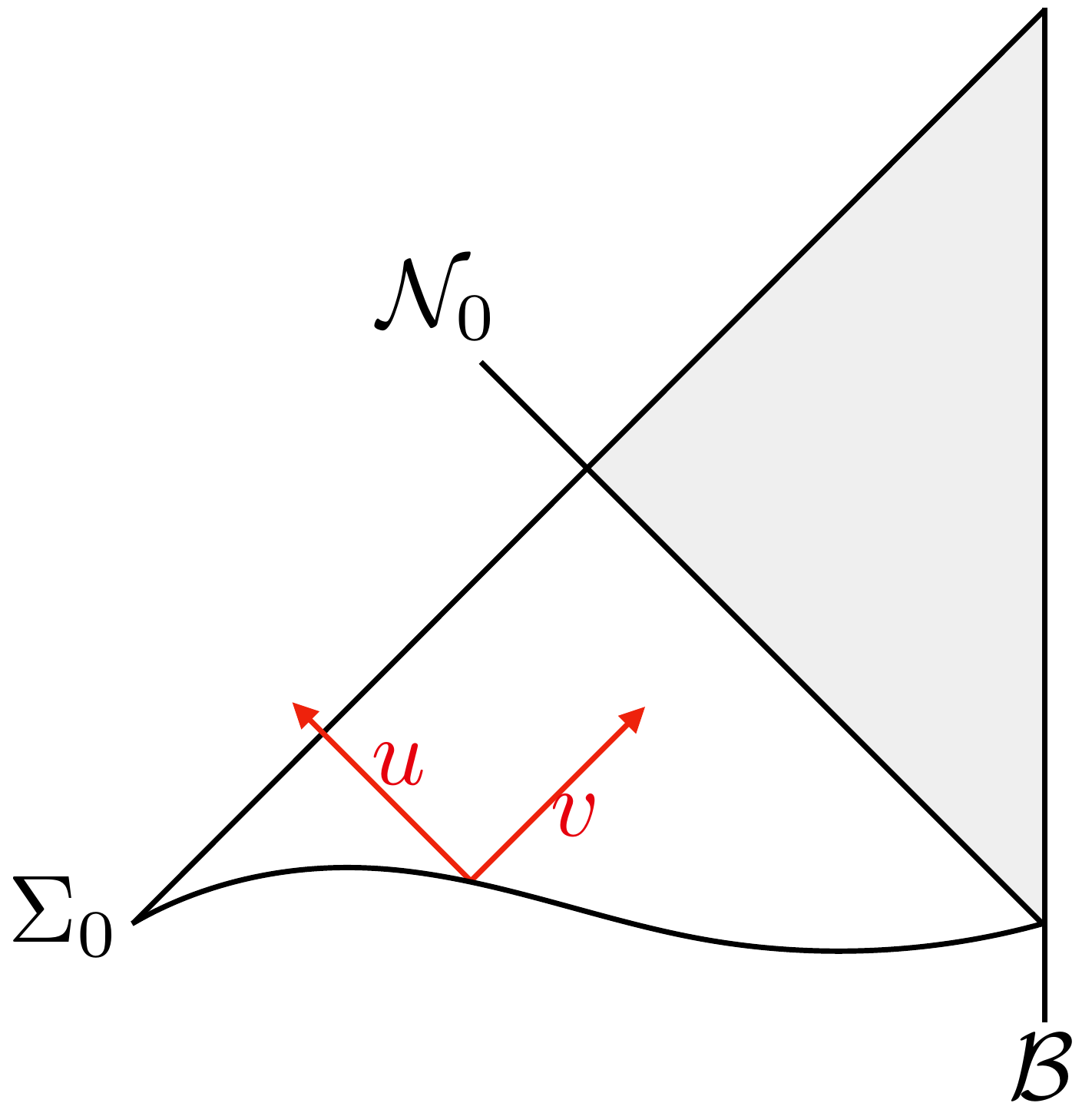}\\
\end{center}
\caption{Characteristic vs. Cauchy formulation of the wave equation in Anti de Sitter space. The gray area is fully determined by data given on the null slice ${\cal N}_0$, as well as the boundary conditions at ${\cal B}$. Data on the timeslice $\Sigma_0$, together with boundary conditions at ${\cal B}$ determine the entire causal future of $\Sigma_0$, but it is easy to see that that the gray area only depends on the purely outgoing data on the Cauchy slice $\Sigma_0$.}
 \label{fig:CharacteristicAdS}
\end{figure}
Secondly we have the main equations
\be
{\cal E}^\mu = 0\,,
\ee
which in turn decompose into a scalar hypersurface equation on ${\cal N}_{0}$, namely ${\cal E}^u=0$, 
\bea
 {\cal D}_{zz}\left(A_{u}\right) = H_u\left[A_i\,,D^{(n)}_i A_j\right]\,,\nonumber\\ \label{eq.hypersurface}
\eea
where ${\cal D}_{zz}$ is a second order differential operator whose details depend on the dimension $d$ as does the precise structure of $H_u$.
We also have the remaining vector evolution equations, ${\cal E}^i=0$,
\be\label{eq.evolution}
{\cal D}_z \partial_u A_i = H_i \left[ A_u, A_{u,z},A_i\,,D^{(n)}_i A_j   \right]\,,
\ee
where ${\cal D}_z$ is a first-order differential operator whose details depend on the dimension $d$, as does $H_i$. In $d=3$, i.e. for $AdS_4$, it is simply $\partial_z$. 
From Eqs. (\ref{eq.evolution}) we can determine the first $u-$derivatives of $A_i$ away from ${\cal N}_0$ and therefore the evolution of the system. An extremely useful aspect of the characteristic scheme is the nested structure of the equations, which we now explain on the Maxwell example.

We note that $H_u$  in equation \eqref{eq.hypersurface} only depends on the components $A_i$ and their derivatives in the hypersurface direction $D^{(n)}_jA_i$. Similarly Eqs (\ref{eq.evolution}) depend on the aforementioned quantities, as well as quantities like $A_{u,z}$ which are determined by Eq. (\ref{eq.hypersurface}). That is, after solving Eq. (\ref{eq.hypersurface}),  Eqs (\ref{eq.evolution}) can be evaluated from the knowledge of the functions $A_i$ solely on ${\cal N}_0$ by integration along $z$. Once $A_u$ has been determined, it can be substituted into $H_i$ in Eq. (\ref{eq.evolution}) which determine the  time derivatives of $A_i$ by integrating $H_i$ along $z$. Given this knowledge we can use a finite difference approximation to progress to the next null surface ${\cal N}|_{u_0+\delta u}$ and so on. This convenient nested evolution structure is a general feature of characteristic schemes, and appears again in the characteristic formulation of Einstein's equations \cite{winicour2001characteristic,Chesler:2013lia}. The reader may wonder why this scheme seems to make do with less initial data than the generalized harmonic scheme, where we need to specific a function and its derivative for each degree of freedom. This is related to the fact, illustrated in Figs. \ref{fig:CharacteristicMinkowski} and \ref{fig:CharacteristicAdS}, that half the initial data specified on a Cauchy surface $\Sigma_0$ does not influence the region, which is fully determined in the characteristic scheme and shaded in gray in both figures.

This structure, together with the convenient global choice of gauge makes such schemes are very efficient, as compared to the generalized harmonic evolution. On the other hand, the latter is more adaptable as it allows to adjust the gauge choice dynamically during evolution, which for certain kinds of problems may even become a necessity.

Finally, returning to the issue of boundary conditions, we note that each integration of Eq. (\ref{eq.evolution}) along a $z$-slice leaves the freedom to add an arbitrary function of $\left\{u, x^i  \right\}$. This allows us to set boundary conditions for the fields. 
Let us now investigate the structure of the equations in more detail to explicitly see how this works. We start by focusing on the auxiliary equation. Considering the covariant derivative $0=\nabla_a {\cal E}^a$, current conservation, and imposing ${\cal E}^\mu=0$, implies the relation
\be
z\partial_z{\cal E}^z = (d+1){\cal E}^z\,.
\ee
Thus, if ${\cal E}^z  = 0$ on some $z$-slice, it is zero throughout as claimed above. Writing out ${\cal E}^z$ explicitly in the background (\ref{eq.backgroundEF}), one obtains
\be\label{eq.auxiliaryAdS4}
{\cal E}^z = \frac{z^4}{\ell^4}\left(\partial_u F_{zu} + \partial_i  q^{ij} \left( fF_{jz} - F_{ju} \right) \right) - j^z\,.
\ee
We see that if fields are regular, meaning the term in parentheses grows at most as $z^{-2}$ as $z\rightarrow 0$, and if the current vanishes as ${\cal O}(z^2)$ in the same limit, the condition $\partial_z{\cal E}^z=0$ is met identically at $z=0$. The former condition is manifestly obeyed, while the implied decay condition on the sources making up $j^z$ can be phrased in field-theory terms by saying that the operator dual to the matter sources must be marginal or relevant in the field theory (recall our discussion in section \ref{sec.AspectsOfDuality}).

We conclude this section with a summary of the characteristic evolution scheme for electromagnetism, before turning to the gravity case.
\subsubsection{Summary}
 As described above, the characteristic method proceeds as follows:
\begin{enumerate}
\item At the initial surface $u = u_0$ one sets up arbitrary initial data for the dynamical fields $A_i$. By a choice of gauge we arrange for $A_z=0$.
\item Using the hypersurface equation one determines $A_u$ on the initial data surface by radial integration. The function $h(x)$ is used to fix the remaining gauge freedom, while $g(x)$ is used to prescribe boundary data for $A_u$.
\item Using the null evolution equations the time derivatives $A_i{}_{,u}$ are determined from the data of $A_i$ and $A_u$ on the initial surface via radial integration. The functions $g_i(x)$ are used to prescribe boundary data for $A_i$. 
\item The fields $A_i$ are propagated to the next surface $u = u_0 + \delta u$ using a suitable time evolution scheme, such as Runge-Kutta finite difference integration.
\item The procedure is repeated  at the $u_0 + \delta u$ surface and so on.
\end{enumerate}
We thus have constructed an evolution scheme along null characteristics with a boundary value constraint, but no constraints on initial data $A_i\bigr|_{u = u_0}$.
\subsubsection{The Characteristic Method for $AdS$ Gravity}
We now apply a characteristic evolution scheme to Einstein's Equations \eqref{eq.EEq}.  We emphasize conceptual points and in particular build on the analogy with the Maxwell case we treated above. A more detailed technical treatment can be found in \cite{Chesler:2013lia}, which we follow closely.
As mentioned above, the characteristic method for gravity evolution takes the approach of making a global choice of the coordinate system (gauge fixing). The metric takes the form
\be
ds^2 = \frac{\ell^2}{z^2} \left[g_{\mu\nu}(x,z) dx^\mu dx^\nu + 2 w_\mu(x) dx^\mu dz \right]\,.
\ee
Just as the choice of radial gauge in the Maxwell example, this choice of metric leaves some residual gauge freedom. The first is the direct analog of (\ref{eq.MaxResidual}) above, that is we may still transform by diffeomorphisms which depend on $x$ only:
\be
x^\mu \rightarrow f^\mu (x)\,.
\ee
Again this condition can be fixed on a single $z$ slice, and one often uses it to set $w_\mu = -\delta^{0}_\mu$, although other choices are possible. We still  have a further freedom, not present in the Maxwell case, namely we may send
\be\label{eq.RadialShift}
z \rightarrow z -\tfrac{z^2}{\ell^2} \delta \lambda(x)\,.
\ee
In cases where an apparent horizon exists, it is computationally convenient to use this remaining gauge freedom to set its coordinate radius to a fixed position $z_h$ \cite{Chesler:2013lia}. 

We can then decompose the equations in a manner similar to the case above, according to their index symmetries, i.e. into scalar ($2 A := - \frac{\ell^2}{z^2}g_{00}$), vector ($F_i := -\frac{\ell^2}{z^2}g_{0i}$) and tensor $(G_{ij}:=\frac{\ell^2}{z^2}g_{ij})$ components. These tensor components can be further decomposed into the trace part, which is itself a scalar, and a traceless tensor part by writing
\be
G_{ij} = \Sigma^2 \hat g_{ij}
\ee
and imposing $\hat g_{ij}$ to have unit determinant. The Einstein Equations then take the form of a set of linear radial ODE, which we show here schematically. The precise form of the equations, including the source terms, can be found in \cite{Chesler:2013lia}, but we would like to particularly emphasize their nested form
\bea
{\cal D}_{zz} \Sigma &=&H_{\Sigma}[\hat g] \label{eq.SigmaODE}\\
\left({\cal D}_{zz}\right)^j{}_i F_j &=& H^F_i[\hat g,\Sigma] \label{eq.FODE}\\
{\cal D}^A_{zz} A &=& H^A[\hat g,\Sigma, F, d_+\Sigma, d_+ \hat g_{ij}] \label{eq.AODE}\,.
\eea
In addition there are linear first order radial ODE for the null time derivatives of $\Sigma, \hat g_{ij}, F_j$:
\bea
{\cal D}_z d_+ \Sigma &=& H_{d_+\Sigma}[\hat g, \Sigma, F] \label{eq.SigmaTime}\\
\left({\cal D}_z\right)^{kl}_{ij}d_+ \hat g_{kl} &=& H_{d_+ \hat g}[\hat g,\Sigma,F,d_+\Sigma]_{ij} \label{eq.gTime} \\
\left( {\cal D}_z \right)^{j}_i d_+ F_j &=& H_{d_+F}[\hat g, \Sigma,F,d_+\Sigma, d_+\hat g, A]_i\label{eq.FTime}\,.
\eea
As above, we have adopted the shift-invariant derivative of \cite{Chesler:2013lia}, such that $d_+ = \partial_u -\tfrac{z^2}{\ell^2} A \partial_z$. Note the nested structure of Equations (\ref{eq.SigmaODE}-\ref{eq.FTime}), whereby we can successively solve for the unknown function on a given $z$-slice, using the function known up to a given point in order to compute the right-hand side terms of the next equation in the nested sequence of integrations.

Lastly, there is a second order (in null derivatives) ODE for $\Sigma$:
\be\label{eq.SigmaddODE}
d_+ \left( d_+ \Sigma \right) = H_{d^2_+ \Sigma}[\hat g, \Sigma, F, d_+\Sigma, d_+ \hat g,A]\,.
\ee
It is convenient to regard $A$ as an auxiliary field and $\Sigma, F_i$ and $\hat g$ as the dynamical propagating degrees of freedom. With this choice \footnote{Other choices are possible, for example one may regard all of $A, F_i, \Sigma$ as auxiliary.}, the equation (\ref{eq.AODE}) determines, via radial integration, the field $A$ from the knowledge of $\hat g_{ij}, \Sigma, F$ and their null derivatives at a given (null) hypersurface.

If Eqs. (\ref{eq.SigmaODE}),  (\ref{eq.FODE}) are satisfied on the initial (null) time slice, then they are satisfied everywhere, given that the dynamical equations (\ref{eq.SigmaTime} - \ref{eq.FTime}) are satisfied. They are thus initial data constraints.

If Eq. (\ref{eq.SigmaddODE}) is imposed on a single $z$ slice, it holds throughout the bulk, again assuming the remaining equations hold. This can be shown from the gravitational Bianchi identities, together with the conservation of the gravitational energy-momentum tensor, in direct analogy with he $U(1)$ case where conservation of $J^a$ took its place. Hence this equation gives a boundary value constraint.

Then Eqs. (\ref{eq.SigmaTime} - \ref{eq.FTime}) determine the first order (null) time derivatives of the propagating fields. Notice that the nested structure of these equations allows to solving them one by one on a given null hypersurface. We can then use any convenient time stepping algorithm to determine the dynamical fields on the next timeslice, where the procedure can be repeated as above.

To some extent we are free to choose which of the functions are considered auxiliary and which are dynamical. Correspondingly the interpretation of some constraint equations changes. A slight variation to the scheme above is to view only $\hat g_{ij}$ as propagating, and treat $\Sigma, F_i$ and $A$ as auxiliary.  We then have no constraint on initial data, but instead three boundary value constraints, namely Eqs. (\ref{eq.SigmaTime}), (\ref{eq.FTime}), and (\ref{eq.SigmaddODE}). It is this choice which resembles most closely the characteristic scheme in the EM case, described above.
\subsection*{Acknowledgements}
We would like to thank Benjamin Withers for his comments on a preliminary version of this draft. The work of JS has been supported by the Fonds National Suisse de la Recherche Scientifique (Schweizerischer Nationalfonds zur F\"orderung der wissenschaftlichen Forschung) through Project Grant 200021 162796 as well as the NCCR 51NF40-141869 ``The Mathematics of Physics" (SwissMAP). HL is partially supported by the Office of High Energy Physics of U. S. Department of Energy under grant Contract Number  DE-SC0012567.  H. L. would also like to thank  Galileo Galilei Institute for Theoretical Physics for the hospitality during the workshop ``Entanglement in Quantum Systems'' and the Simons Foundation for partial support during the completion of this work.

\appendix
\section{Kruskal extension of eternal AdS black hole\label{app.Kruskal}}
For completeness we review in this appendix the notion of the maximal extension of the eternal Schwarzschild black hole in AdS, which was used in Section \ref{sec.MomentumSpace} above. In particular we wish to give a more geometric perspective than in the bulk, explaining in more details how the full geometry depicted in Fig. \ref{fig:bh} is constructed. For convenience to the reader this section has some redundant elements which were already covered in the main part of this review.

 We remind the reader that the manifold depicted in Fig. \ref{fig:bh}  serves as the geometric dual of the thermal state at inverse temperature $\beta$ of the field theory. The fact, as we review here, that this manifold has a maximal extension with {\it two} asymptotic boundaries (see Fig. \ref{fig:bh}) is the simplest manifestation of a Schwinger-Keldysh like doubled contour within AdS/CFT. The starting point is the metric \eqref{bhm}, which we repeat here for convenience
\be
ds^2 = - f (r) dt^2 + {1 \ov f (r)} dr^2 + r^2 d \vec x^2  \ .
\ee
This metric covers the exterior region of a black hole in AdS, but has a coordinate singularity at $r=r_0$, where $f(r_0)=0$. To explore the geometry beyond this locus one defines a new set of coordinates. To this end, we start by defining the tortoise coordinate
\be
r_* = \int\frac{dr}{f(r)}
\ee
in terms of which we set $u = t - r_*$ and $v =  t + r_*$.
In terms of $u$ and $v$ the metric is still degenerate at the point $r=r_0$, but we are only one step from defining a coordinate system in which this apparently singular behavior is removed. Let
\be
U = - e^{-\frac{2\pi}{\beta} u}\,,\qquad V = e^{\frac{2\pi}{\beta} v}\,,
\ee
so that
\be\label{eq.KruskalMetricAdSBH}
ds^2 = -\left( \frac{\beta}{2\pi}  \right)^2 f e^{-\frac{4\pi}{\beta}r_*} dU dV + r^2 d \vec x^2 \,,
\ee
which is now completely regular at the horizon.
By taking the usual range $r\in [0,\infty)$ and $t \in \mathbb{R}$, we have $r_* \in\mathbb{R}$, and thus
\be
U \in (-\infty,0 ]\,, \qquad V \in [0,\infty)\qquad \textrm{`Right'}
\ee
We have labeled this range of coordinates as the `right' region of the spacetime, as it covers precisely the triangular region labeled `R' in the diagram of Fig. \eqref{fig:bh}.
However, there is absolutely nothing preventing us from considering (i.e. extending) the metric \eqref{eq.KruskalMetricAdSBH} for all real values $U,V \in \mathbb{R}$. This is precisely what is referred to as the Kruskal extension of the Schwarzschild solution and was used explicitly in section \ref{sec.MomentumSpace}. In addition to the `right' exterior region we started with, this extended spacetime also contains a left exterior region for which
\be
U \in [0,\infty)\,,\qquad V \in (-\infty,0]\qquad \textrm{`Left'}
\ee
This is the triangular region labeled `L' in Fig. \eqref{fig:bh}. In addition to these two regions there also are the two interior regions
\bea
U &\in& [0,\infty)\,,V \in [0,\infty) \qquad \textrm{Future Interior} \nonumber\\
U &\in& (-\infty,0]\,,V \in (-\infty,0] \qquad \textrm{Past Interior}\nonumber\\
\eea
Finally, since $U$ and $V$ are null directions, one often uses timelike and spacelike combinations
\be
T = \tfrac{1}{2}\left( V + U \right)\,,\qquad X = \tfrac{1}{2}\left( V - U \right)\,.
\ee
Tracing back the various definitions that led us here, we can deduce that the direction of Kruskal time $T$ coincides with the direction of $t$, the `Schwarzschild time' in the right exterior, while it is opposed to it in the left exterior region, essentially due to the sign difference in $U$ and $V$ between those regions.

\bibliographystyle{utphys}
\bibliography{adscmtbib}{}

\providecommand{\href}[2]{#2}\begingroup\raggedright\begin{thebibliography}{100}

\bibitem{Maldacena:1997re}
J.~M. Maldacena, ``{The Large N limit of superconformal field theories and
  supergravity},'' {\em Adv.Theor.Math.Phys.} {\bfseries 2} (1998) 231--252,
\href{http://arxiv.org/abs/hep-th/9711200}{{\ttfamily arXiv:hep-th/9711200
  [hep-th]}}.

\bibitem{Gubser:1998bc}
S.~Gubser, I.~R. Klebanov, and A.~M. Polyakov, ``{Gauge theory correlators from
  noncritical string theory},''
  \href{http://dx.doi.org/10.1016/S0370-2693(98)00377-3}{{\em Phys.Lett.}
  {\bfseries B428} (1998) 105--114},
\href{http://arxiv.org/abs/hep-th/9802109}{{\ttfamily arXiv:hep-th/9802109
  [hep-th]}}.

\bibitem{Witten:1998qj}
E.~Witten, ``{Anti-de Sitter space and holography},'' {\em
  Adv.Theor.Math.Phys.} {\bfseries 2} (1998) 253--291,
\href{http://arxiv.org/abs/hep-th/9802150}{{\ttfamily arXiv:hep-th/9802150
  [hep-th]}}.

\bibitem{casalderrey2014gauge}
J.~Casalderrey-Solana, {\em Gauge/string duality, hot QCD and heavy ion
  collisions}.
\newblock Cambridge University Press, 2014.

\bibitem{zaanen2015holographic}
J.~Zaanen, Y.~Liu, Y.-W. Sun, and K.~Schalm, {\em Holographic duality in
  condensed matter physics}.
\newblock Cambridge University Press, 2015.

\bibitem{Son:2007vk}
D.~T. Son and A.~O. Starinets, ``{Viscosity, Black Holes, and Quantum Field
  Theory},''
  \href{http://dx.doi.org/10.1146/annurev.nucl.57.090506.123120}{{\em Ann. Rev.
  Nucl. Part. Sci.} {\bfseries 57} (2007) 95--118},
\href{http://arxiv.org/abs/0704.0240}{{\ttfamily arXiv:0704.0240 [hep-th]}}.

\bibitem{Hartnoll:2009sz}
S.~A. Hartnoll, ``{Lectures on holographic methods for condensed matter
  physics},'' \href{http://dx.doi.org/10.1088/0264-9381/26/22/224002}{{\em
  Class.Quant.Grav.} {\bfseries 26} (2009) 224002},
\href{http://arxiv.org/abs/0903.3246}{{\ttfamily arXiv:0903.3246 [hep-th]}}.

\bibitem{Herzog:2009xv}
C.~P. Herzog, ``{Lectures on Holographic Superfluidity and
  Superconductivity},''
  \href{http://dx.doi.org/10.1088/1751-8113/42/34/343001}{{\em J.Phys.}
  {\bfseries A42} (2009) 343001},
\href{http://arxiv.org/abs/0904.1975}{{\ttfamily arXiv:0904.1975 [hep-th]}}.

\bibitem{McGreevy:2009xe}
J.~McGreevy, ``{Holographic duality with a view toward many-body physics},''
  \href{http://dx.doi.org/10.1155/2010/723105}{{\em Adv.High Energy Phys.}
  {\bfseries 2010} (2010) 723105},
\href{http://arxiv.org/abs/0909.0518}{{\ttfamily arXiv:0909.0518 [hep-th]}}.

\bibitem{Adams:2012th}
A.~Adams, L.~D. Carr, T.~SchÃ¤fer, P.~Steinberg, and J.~E. Thomas,
  ``{Strongly Correlated Quantum Fluids: Ultracold Quantum Gases, Quantum
  Chromodynamic Plasmas, and Holographic Duality},''
  \href{http://dx.doi.org/10.1088/1367-2630/14/11/115009}{{\em New J. Phys.}
  {\bfseries 14} (2012) 115009},
\href{http://arxiv.org/abs/1205.5180}{{\ttfamily arXiv:1205.5180 [hep-th]}}.

\bibitem{DeWolfe:2013cua}
O.~DeWolfe, S.~S. Gubser, C.~Rosen, and D.~Teaney, ``{Heavy ions and string
  theory},'' \href{http://dx.doi.org/10.1016/j.ppnp.2013.11.001}{{\em Prog.
  Part. Nucl. Phys.} {\bfseries 75} (2014) 86--132},
\href{http://arxiv.org/abs/1304.7794}{{\ttfamily arXiv:1304.7794 [hep-th]}}.

\bibitem{Hartnoll:2016apf}
S.~A. Hartnoll, A.~Lucas, and S.~Sachdev, ``{Holographic quantum matter},''
\href{http://arxiv.org/abs/1612.07324}{{\ttfamily arXiv:1612.07324 [hep-th]}}.

\bibitem{Hubeny:2010ry}
V.~E. Hubeny and M.~Rangamani, ``{A Holographic view on physics out of
  equilibrium},'' \href{http://dx.doi.org/10.1155/2010/297916}{{\em Adv. High
  Energy Phys.} {\bfseries 2010} (2010) 297916},
\href{http://arxiv.org/abs/1006.3675}{{\ttfamily arXiv:1006.3675 [hep-th]}}.

\bibitem{Adams:2013vsa}
A.~Adams, P.~M. Chesler, and H.~Liu, ``{Holographic turbulence},''
  \href{http://dx.doi.org/10.1103/PhysRevLett.112.151602}{{\em Phys. Rev.
  Lett.} {\bfseries 112} no.~15, (2014) 151602},
\href{http://arxiv.org/abs/1307.7267}{{\ttfamily arXiv:1307.7267 [hep-th]}}.

\bibitem{Green:2013zba}
S.~R. Green, F.~Carrasco, and L.~Lehner, ``{Holographic Path to the Turbulent
  Side of Gravity},'' \href{http://dx.doi.org/10.1103/PhysRevX.4.011001}{{\em
  Phys. Rev.} {\bfseries X4} no.~1, (2014) 011001},
\href{http://arxiv.org/abs/1309.7940}{{\ttfamily arXiv:1309.7940 [hep-th]}}.

\bibitem{Donos:2011bh}
A.~Donos and J.~P. Gauntlett, ``{Holographic striped phases},''
  \href{http://dx.doi.org/10.1007/JHEP08(2011)140}{{\em JHEP} {\bfseries 08}
  (2011) 140},
\href{http://arxiv.org/abs/1106.2004}{{\ttfamily arXiv:1106.2004 [hep-th]}}.

\bibitem{Donos:2012wi}
A.~Donos and J.~P. Gauntlett, ``{Black holes dual to helical current phases},''
  \href{http://dx.doi.org/10.1103/PhysRevD.86.064010}{{\em Phys. Rev.}
  {\bfseries D86} (2012) 064010},
\href{http://arxiv.org/abs/1204.1734}{{\ttfamily arXiv:1204.1734 [hep-th]}}.

\bibitem{Withers:2014sja}
B.~Withers, ``{Holographic Checkerboards},''
  \href{http://dx.doi.org/10.1007/JHEP09(2014)102}{{\em JHEP} {\bfseries 09}
  (2014) 102},
\href{http://arxiv.org/abs/1407.1085}{{\ttfamily arXiv:1407.1085 [hep-th]}}.

\bibitem{Horowitz:2012ky}
G.~T. Horowitz, J.~E. Santos, and D.~Tong, ``{Optical Conductivity with
  Holographic Lattices},''
  \href{http://dx.doi.org/10.1007/JHEP07(2012)168}{{\em JHEP} {\bfseries 07}
  (2012) 168},
\href{http://arxiv.org/abs/1204.0519}{{\ttfamily arXiv:1204.0519 [hep-th]}}.

\bibitem{Donos:2013eha}
A.~Donos and J.~P. Gauntlett, ``{Holographic Q-lattices},''
  \href{http://dx.doi.org/10.1007/JHEP04(2014)040}{{\em JHEP} {\bfseries 04}
  (2014) 040},
\href{http://arxiv.org/abs/1311.3292}{{\ttfamily arXiv:1311.3292 [hep-th]}}.

\bibitem{Davison:2014lua}
R.~A. Davison and B.~Gout\'eraux, ``{Momentum dissipation and effective
  theories of coherent and incoherent transport},''
  \href{http://dx.doi.org/10.1007/JHEP01(2015)039}{{\em JHEP} {\bfseries 01}
  (2015) 039},
\href{http://arxiv.org/abs/1411.1062}{{\ttfamily arXiv:1411.1062 [hep-th]}}.

\bibitem{Lucas:2015vna}
A.~Lucas, ``{Conductivity of a strange metal: from holography to memory
  functions},'' \href{http://dx.doi.org/10.1007/JHEP03(2015)071}{{\em JHEP}
  {\bfseries 03} (2015) 071},
\href{http://arxiv.org/abs/1501.05656}{{\ttfamily arXiv:1501.05656 [hep-th]}}.

\bibitem{Donos:2015gia}
A.~Donos and J.~P. Gauntlett, ``{Navier-Stokes Equations on Black Hole Horizons
  and DC Thermoelectric Conductivity},''
  \href{http://dx.doi.org/10.1103/PhysRevD.92.121901}{{\em Phys. Rev.}
  {\bfseries D92} no.~12, (2015) 121901},
\href{http://arxiv.org/abs/1506.01360}{{\ttfamily arXiv:1506.01360 [hep-th]}}.

\bibitem{Lucas:2015pxa}
A.~Lucas and S.~Sachdev, ``{Memory matrix theory of magnetotransport in strange
  metals},'' \href{http://dx.doi.org/10.1103/PhysRevB.91.195122}{{\em Phys.
  Rev.} {\bfseries B91} no.~19, (2015) 195122},
\href{http://arxiv.org/abs/1502.04704}{{\ttfamily arXiv:1502.04704
  [cond-mat.str-el]}}.

\bibitem{Lucas:2017idv}
A.~Lucas and K.~C. Fong, ``{Hydrodynamics of electrons in graphene},''
  \href{http://dx.doi.org/10.1088/1361-648X/aaa274}{{\em J. Phys. Condens.
  Matter} {\bfseries 30} (2018) 053001},
\href{http://arxiv.org/abs/1710.08425}{{\ttfamily arXiv:1710.08425
  [cond-mat.str-el]}}.

\bibitem{Vegh:2013sk}
D.~Vegh, ``{Holography without translational symmetry},''
\href{http://arxiv.org/abs/1301.0537}{{\ttfamily arXiv:1301.0537 [hep-th]}}.

\bibitem{Blake:2013bqa}
M.~Blake and D.~Tong, ``{Universal Resistivity from Holographic Massive
  Gravity},'' \href{http://dx.doi.org/10.1103/PhysRevD.88.106004}{{\em Phys.
  Rev.} {\bfseries D88} no.~10, (2013) 106004},
\href{http://arxiv.org/abs/1308.4970}{{\ttfamily arXiv:1308.4970 [hep-th]}}.

\bibitem{Blake:2013owa}
M.~Blake, D.~Tong, and D.~Vegh, ``{Holographic Lattices Give the Graviton an
  Effective Mass},''
  \href{http://dx.doi.org/10.1103/PhysRevLett.112.071602}{{\em Phys. Rev.
  Lett.} {\bfseries 112} no.~7, (2014) 071602},
\href{http://arxiv.org/abs/1310.3832}{{\ttfamily arXiv:1310.3832 [hep-th]}}.

\bibitem{Davison:2013jba}
R.~A. Davison, ``{Momentum relaxation in holographic massive gravity},''
  \href{http://dx.doi.org/10.1103/PhysRevD.88.086003}{{\em Phys. Rev.}
  {\bfseries D88} (2013) 086003},
\href{http://arxiv.org/abs/1306.5792}{{\ttfamily arXiv:1306.5792 [hep-th]}}.

\bibitem{Aharony:1999ti}
O.~Aharony, S.~S. Gubser, J.~M. Maldacena, H.~Ooguri, and Y.~Oz, ``{Large N
  field theories, string theory and gravity},''
  \href{http://dx.doi.org/10.1016/S0370-1573(99)00083-6}{{\em Phys. Rept.}
  {\bfseries 323} (2000) 183--386},
\href{http://arxiv.org/abs/hep-th/9905111}{{\ttfamily arXiv:hep-th/9905111
  [hep-th]}}.

\bibitem{DHoker:2002nbb}
E.~D'Hoker and D.~Z. Freedman, ``{Supersymmetric gauge theories and the AdS /
  CFT correspondence},'' in {\em {Strings, Branes and Extra Dimensions: TASI
  2001: Proceedings}}, pp.~3--158.
\newblock 2002.
\newblock
\href{http://arxiv.org/abs/hep-th/0201253}{{\ttfamily arXiv:hep-th/0201253
  [hep-th]}}.
\newblock

\bibitem{nuastase2015introduction}
H.~N{\u{a}}stase, {\em Introduction to the ADS/CFT Correspondence}.
\newblock Cambridge University Press, 2015.

\bibitem{ammon2015gauge}
M.~Ammon and J.~Erdmenger, {\em Gauge/gravity duality: Foundations and
  applications}.
\newblock Cambridge University Press, 2015.

\bibitem{Klebanov:1999tb}
I.~R. Klebanov and E.~Witten, ``{AdS / CFT correspondence and symmetry
  breaking},'' \href{http://dx.doi.org/10.1016/S0550-3213(99)00387-9}{{\em
  Nucl. Phys.} {\bfseries B556} (1999) 89--114},
\href{http://arxiv.org/abs/hep-th/9905104}{{\ttfamily arXiv:hep-th/9905104
  [hep-th]}}.

\bibitem{deHaro:2000vlm}
S.~de~Haro, S.~N. Solodukhin, and K.~Skenderis, ``{Holographic reconstruction
  of space-time and renormalization in the AdS / CFT correspondence},''
  \href{http://dx.doi.org/10.1007/s002200100381}{{\em Commun. Math. Phys.}
  {\bfseries 217} (2001) 595--622},
\href{http://arxiv.org/abs/hep-th/0002230}{{\ttfamily arXiv:hep-th/0002230
  [hep-th]}}.

\bibitem{Son:2002sd}
D.~T. Son and A.~O. Starinets, ``{Minkowski space correlators in AdS / CFT
  correspondence: Recipe and applications},''
  \href{http://dx.doi.org/10.1088/1126-6708/2002/09/042}{{\em JHEP} {\bfseries
  09} (2002) 042},
\href{http://arxiv.org/abs/hep-th/0205051}{{\ttfamily arXiv:hep-th/0205051
  [hep-th]}}.

\bibitem{Policastro:2002se}
G.~Policastro, D.~T. Son, and A.~O. Starinets, ``{From AdS / CFT correspondence
  to hydrodynamics},''
  \href{http://dx.doi.org/10.1088/1126-6708/2002/09/043}{{\em JHEP} {\bfseries
  09} (2002) 043},
\href{http://arxiv.org/abs/hep-th/0205052}{{\ttfamily arXiv:hep-th/0205052
  [hep-th]}}.

\bibitem{Herzog:2007ij}
C.~P. Herzog, P.~Kovtun, S.~Sachdev, and D.~T. Son, ``{Quantum critical
  transport, duality, and M-theory},''
  \href{http://dx.doi.org/10.1103/PhysRevD.75.085020}{{\em Phys. Rev.}
  {\bfseries D75} (2007) 085020},
\href{http://arxiv.org/abs/hep-th/0701036}{{\ttfamily arXiv:hep-th/0701036
  [hep-th]}}.

\bibitem{Holzhey:1994we}
C.~Holzhey, F.~Larsen, and F.~Wilczek, ``{Geometric and renormalized entropy in
  conformal field theory},''
  \href{http://dx.doi.org/10.1016/0550-3213(94)90402-2}{{\em Nucl. Phys.}
  {\bfseries B424} (1994) 443--467},
\href{http://arxiv.org/abs/hep-th/9403108}{{\ttfamily arXiv:hep-th/9403108
  [hep-th]}}.

\bibitem{Calabrese:2004eu}
P.~Calabrese and J.~L. Cardy, ``{Entanglement entropy and quantum field
  theory},'' \href{http://dx.doi.org/10.1088/1742-5468/2004/06/P06002}{{\em J.
  Stat. Mech.} {\bfseries 0406} (2004) P06002},
\href{http://arxiv.org/abs/hep-th/0405152}{{\ttfamily arXiv:hep-th/0405152
  [hep-th]}}.

\bibitem{Ryu:2006bv}
S.~Ryu and T.~Takayanagi, ``{Holographic derivation of entanglement entropy
  from AdS/CFT},'' \href{http://dx.doi.org/10.1103/PhysRevLett.96.181602}{{\em
  Phys. Rev. Lett.} {\bfseries 96} (2006) 181602},
\href{http://arxiv.org/abs/hep-th/0603001}{{\ttfamily arXiv:hep-th/0603001
  [hep-th]}}.

\bibitem{Hubeny:2007xt}
V.~E. Hubeny, M.~Rangamani, and T.~Takayanagi, ``{A Covariant holographic
  entanglement entropy proposal},''
  \href{http://dx.doi.org/10.1088/1126-6708/2007/07/062}{{\em JHEP} {\bfseries
  07} (2007) 062},
\href{http://arxiv.org/abs/0705.0016}{{\ttfamily arXiv:0705.0016 [hep-th]}}.

\bibitem{Skenderis:2008dh}
K.~Skenderis and B.~C. van Rees, ``{Real-time gauge/gravity duality},''
  \href{http://dx.doi.org/10.1103/PhysRevLett.101.081601}{{\em Phys. Rev.
  Lett.} {\bfseries 101} (2008) 081601},
\href{http://arxiv.org/abs/0805.0150}{{\ttfamily arXiv:0805.0150 [hep-th]}}.

\bibitem{Skenderis:2008dg}
K.~Skenderis and B.~C. van Rees, ``{Real-time gauge/gravity duality:
  Prescription, Renormalization and Examples},''
  \href{http://dx.doi.org/10.1088/1126-6708/2009/05/085}{{\em JHEP} {\bfseries
  05} (2009) 085},
\href{http://arxiv.org/abs/0812.2909}{{\ttfamily arXiv:0812.2909 [hep-th]}}.

\bibitem{Herzog:2002pc}
C.~P. Herzog and D.~T. Son, ``{Schwinger-Keldysh propagators from AdS/CFT
  correspondence},''
  \href{http://dx.doi.org/10.1088/1126-6708/2003/03/046}{{\em JHEP} {\bfseries
  03} (2003) 046},
\href{http://arxiv.org/abs/hep-th/0212072}{{\ttfamily arXiv:hep-th/0212072
  [hep-th]}}.

\bibitem{Son:2009vu}
D.~T. Son and D.~Teaney, ``{Thermal Noise and Stochastic Strings in AdS/CFT},''
  \href{http://dx.doi.org/10.1088/1126-6708/2009/07/021}{{\em JHEP} {\bfseries
  07} (2009) 021},
\href{http://arxiv.org/abs/0901.2338}{{\ttfamily arXiv:0901.2338 [hep-th]}}.

\bibitem{hongPaoloMikeToAppear}
P.~Glorioso, M.~Crossley, and H.~Liu, ``Non-equilibrium effective field theory
  from holography,'' {\em to appear} .

\bibitem{vanRees:2009rw}
B.~C. van Rees, ``{Real-time gauge/gravity duality and ingoing boundary
  conditions},''
  \href{http://dx.doi.org/10.1016/j.nuclphysbps.2009.07.078}{{\em Nucl. Phys.
  Proc. Suppl.} {\bfseries 192-193} (2009) 193--196},
\href{http://arxiv.org/abs/0902.4010}{{\ttfamily arXiv:0902.4010 [hep-th]}}.

\bibitem{green2005nonlinear}
A.~Green and S.~Sondhi, ``Nonlinear quantum critical transport and the
  schwinger mechanism for a superfluid-mott-insulator transition of bosons,''
  {\em Physical review letters} {\bfseries 95} no.~26, (2005) 267001.

\bibitem{Sonner:2012if}
J.~Sonner and A.~G. Green, ``{Hawking Radiation and Non-equilibrium Quantum
  Critical Current Noise},''
  \href{http://dx.doi.org/10.1103/PhysRevLett.109.091601}{{\em Phys. Rev.
  Lett.} {\bfseries 109} (2012) 091601},
\href{http://arxiv.org/abs/1203.4908}{{\ttfamily arXiv:1203.4908
  [cond-mat.str-el]}}.

\bibitem{DeWolfe:2001pq}
O.~DeWolfe, D.~Z. Freedman, and H.~Ooguri, ``{Holography and defect conformal
  field theories},'' \href{http://dx.doi.org/10.1103/PhysRevD.66.025009}{{\em
  Phys. Rev.} {\bfseries D66} (2002) 025009},
\href{http://arxiv.org/abs/hep-th/0111135}{{\ttfamily arXiv:hep-th/0111135
  [hep-th]}}.

\bibitem{Erdmenger:2002ex}
J.~Erdmenger, Z.~Guralnik, and I.~Kirsch, ``{Four-dimensional superconformal
  theories with interacting boundaries or defects},''
  \href{http://dx.doi.org/10.1103/PhysRevD.66.025020}{{\em Phys. Rev.}
  {\bfseries D66} (2002) 025020},
\href{http://arxiv.org/abs/hep-th/0203020}{{\ttfamily arXiv:hep-th/0203020
  [hep-th]}}.

\bibitem{Gursoy:2010aa}
U.~Gursoy, E.~Kiritsis, L.~Mazzanti, and F.~Nitti, ``{Langevin diffusion of
  heavy quarks in non-conformal holographic backgrounds},''
  \href{http://dx.doi.org/10.1007/JHEP12(2010)088}{{\em JHEP} {\bfseries 12}
  (2010) 088},
\href{http://arxiv.org/abs/1006.3261}{{\ttfamily arXiv:1006.3261 [hep-th]}}.

\bibitem{Nakamura:2013yqa}
S.~Nakamura and H.~Ooguri, ``{Out of Equilibrium Temperature from
  Holography},'' \href{http://dx.doi.org/10.1103/PhysRevD.88.126003}{{\em Phys.
  Rev.} {\bfseries D88} no.~12, (2013) 126003},
\href{http://arxiv.org/abs/1309.4089}{{\ttfamily arXiv:1309.4089 [hep-th]}}.

\bibitem{Kundu:2013eba}
A.~Kundu and S.~Kundu, ``{Steady-state Physics, Effective Temperature Dynamics
  in Holography},'' \href{http://dx.doi.org/10.1103/PhysRevD.91.046004}{{\em
  Phys. Rev.} {\bfseries D91} no.~4, (2015) 046004},
\href{http://arxiv.org/abs/1307.6607}{{\ttfamily arXiv:1307.6607 [hep-th]}}.

\bibitem{Kundu:2015qda}
A.~Kundu, ``{Effective Temperature in Steady-state Dynamics from Holography},''
  \href{http://dx.doi.org/10.1007/JHEP09(2015)042}{{\em JHEP} {\bfseries 09}
  (2015) 042},
\href{http://arxiv.org/abs/1507.00818}{{\ttfamily arXiv:1507.00818 [hep-th]}}.

\bibitem{Sonner:2013mba}
J.~Sonner, ``{Holographic Schwinger Effect and the Geometry of Entanglement},''
  \href{http://dx.doi.org/10.1103/PhysRevLett.111.211603}{{\em Phys. Rev.
  Lett.} {\bfseries 111} no.~21, (2013) 211603},
\href{http://arxiv.org/abs/1307.6850}{{\ttfamily arXiv:1307.6850 [hep-th]}}.

\bibitem{Das:2010yw}
S.~R. Das, T.~Nishioka, and T.~Takayanagi, ``{Probe Branes, Time-dependent
  Couplings and Thermalization in AdS/CFT},''
  \href{http://dx.doi.org/10.1007/JHEP07(2010)071}{{\em JHEP} {\bfseries 07}
  (2010) 071},
\href{http://arxiv.org/abs/1005.3348}{{\ttfamily arXiv:1005.3348 [hep-th]}}.

\bibitem{Bakas:2015hdc}
I.~Bakas, K.~Skenderis, and B.~Withers, ``{Self-similar equilibration of
  strongly interacting systems from holography},''
  \href{http://dx.doi.org/10.1103/PhysRevD.93.101902}{{\em Phys. Rev.}
  {\bfseries D93} no.~10, (2016) 101902},
\href{http://arxiv.org/abs/1512.09151}{{\ttfamily arXiv:1512.09151 [hep-th]}}.

\bibitem{bernard2012energy}
D.~Bernard and B.~Doyon, ``Energy flow in non-equilibrium conformal field
  theory,'' {\em Journal of Physics A: Mathematical and Theoretical} {\bfseries
  45} no.~36, (2012) 362001.

\bibitem{Chang:2013gba}
H.-C. Chang, A.~Karch, and A.~Yarom, ``{An ansatz for one dimensional steady
  state configurations},''
  \href{http://dx.doi.org/10.1088/1742-5468/2014/06/P06018}{{\em J. Stat.
  Mech.} {\bfseries 1406} no.~6, (2014) P06018},
\href{http://arxiv.org/abs/1311.2590}{{\ttfamily arXiv:1311.2590 [hep-th]}}.

\bibitem{Bhaseen:2013ypa}
M.~Bhaseen, B.~Doyon, A.~Lucas, and K.~Schalm, ``Energy flow in quantum
  critical systems far from equilibrium,'' {\em Nature Physics} {\bfseries 11}
  no.~6, (2015) 509.

\bibitem{Doyon:2014qsa}
B.~Doyon, A.~Lucas, K.~Schalm, and M.~J. Bhaseen, ``{Non-equilibrium steady
  states in the Klein-Gordon theory},''
  \href{http://dx.doi.org/10.1088/1751-8113/48/9/095002}{{\em J. Phys.}
  {\bfseries A48} no.~9, (2015) 095002},
\href{http://arxiv.org/abs/1409.6660}{{\ttfamily arXiv:1409.6660
  [cond-mat.stat-mech]}}.

\bibitem{Amado:2015uza}
I.~Amado and A.~Yarom, ``{Black brane steady states},''
  \href{http://dx.doi.org/10.1007/JHEP10(2015)015}{{\em JHEP} {\bfseries 10}
  (2015) 015},
\href{http://arxiv.org/abs/1501.01627}{{\ttfamily arXiv:1501.01627 [hep-th]}}.

\bibitem{Herzog:2016hob}
C.~P. Herzog, M.~Spillane, and A.~Yarom, ``{The holographic dual of a Riemann
  problem in a large number of dimensions},''
  \href{http://dx.doi.org/10.1007/JHEP08(2016)120}{{\em JHEP} {\bfseries 08}
  (2016) 120},
\href{http://arxiv.org/abs/1605.01404}{{\ttfamily arXiv:1605.01404 [hep-th]}}.

\bibitem{Spillane:2015daa}
M.~Spillane and C.~P. Herzog, ``{Relativistic Hydrodynamics and Non-Equilibrium
  Steady States},''
  \href{http://dx.doi.org/10.1088/1742-5468/2016/10/103208}{{\em J. Stat.
  Mech.} {\bfseries 1610} no.~10, (2016) 103208},
\href{http://arxiv.org/abs/1512.09071}{{\ttfamily arXiv:1512.09071 [hep-th]}}.

\bibitem{Lucas:2015hnv}
A.~Lucas, K.~Schalm, B.~Doyon, and M.~J. Bhaseen, ``{Shock waves, rarefaction
  waves, and nonequilibrium steady states in quantum critical systems},''
  \href{http://dx.doi.org/10.1103/PhysRevD.94.025004}{{\em Phys. Rev.}
  {\bfseries D94} no.~2, (2016) 025004},
\href{http://arxiv.org/abs/1512.09037}{{\ttfamily arXiv:1512.09037 [hep-th]}}.

\bibitem{taub1948relativistic}
A.~Taub, ``Relativistic rankine-hugoniot equations,'' {\em Physical Review}
  {\bfseries 74} no.~3, (1948) 328.

\bibitem{Bernard:2016nci}
D.~Bernard and B.~Doyon, ``{Conformal field theory out of equilibrium: a
  review},'' \href{http://dx.doi.org/10.1088/1742-5468/2016/06/064005}{{\em J.
  Stat. Mech.} {\bfseries 1606} no.~6, (2016) 064005},
\href{http://arxiv.org/abs/1603.07765}{{\ttfamily arXiv:1603.07765
  [cond-mat.stat-mech]}}.

\bibitem{Glorioso:2015vrc}
P.~Glorioso, ``{Classification of certain asymptotically AdS space-times with
  Ricci-flat boundary},'' \href{http://dx.doi.org/10.1007/JHEP12(2016)126}{{\em
  JHEP} {\bfseries 12} (2016) 126},
\href{http://arxiv.org/abs/1511.05107}{{\ttfamily arXiv:1511.05107 [gr-qc]}}.

\bibitem{Figueras:2012rb}
P.~Figueras and T.~Wiseman, ``{Stationary holographic plasma quenches and
  numerical methods for non-Killing horizons},''
  \href{http://dx.doi.org/10.1103/PhysRevLett.110.171602}{{\em Phys. Rev.
  Lett.} {\bfseries 110} (2013) 171602},
\href{http://arxiv.org/abs/1212.4498}{{\ttfamily arXiv:1212.4498 [hep-th]}}.

\bibitem{Sonner:2017jcf}
J.~Sonner and B.~Withers, ``{Universal spatial structure of nonequilibrium
  steady states},''
  \href{http://dx.doi.org/10.1103/PhysRevLett.119.161603}{{\em Phys. Rev.
  Lett.} {\bfseries 119} no.~16, (2017) 161603},
\href{http://arxiv.org/abs/1705.01950}{{\ttfamily arXiv:1705.01950 [hep-th]}}.

\bibitem{Novak:2018pnv}
I.~Novak, J.~Sonner, and B.~Withers, ``{Overcoming obstacles in nonequilibrium
  holography},''
\href{http://arxiv.org/abs/1806.08655}{{\ttfamily arXiv:1806.08655 [hep-th]}}.

\bibitem{Amado:2007pv}
I.~Amado, C.~Hoyos-Badajoz, K.~Landsteiner, and S.~Montero, ``{Absorption
  lengths in the holographic plasma},''
  \href{http://dx.doi.org/10.1088/1126-6708/2007/09/057}{{\em JHEP} {\bfseries
  09} (2007) 057},
\href{http://arxiv.org/abs/0706.2750}{{\ttfamily arXiv:0706.2750 [hep-th]}}.

\bibitem{Maeda:2009wv}
K.~Maeda, M.~Natsuume, and T.~Okamura, ``{Universality class of holographic
  superconductors},'' \href{http://dx.doi.org/10.1103/PhysRevD.79.126004}{{\em
  Phys. Rev.} {\bfseries D79} (2009) 126004},
\href{http://arxiv.org/abs/0904.1914}{{\ttfamily arXiv:0904.1914 [hep-th]}}.

\bibitem{Sonner:2014tca}
J.~Sonner, A.~del Campo, and W.~H. Zurek, ``{Universal far-from-equilibrium
  Dynamics of a Holographic Superconductor},''
\href{http://arxiv.org/abs/1406.2329}{{\ttfamily arXiv:1406.2329 [hep-th]}}.

\bibitem{Khlebnikov:2010yt}
S.~Khlebnikov, M.~Kruczenski, and G.~Michalogiorgakis, ``{Shock waves in
  strongly coupled plasmas},''
  \href{http://dx.doi.org/10.1103/PhysRevD.82.125003}{{\em Phys. Rev.}
  {\bfseries D82} (2010) 125003},
\href{http://arxiv.org/abs/1004.3803}{{\ttfamily arXiv:1004.3803 [hep-th]}}.

\bibitem{Khlebnikov:2011ka}
S.~Khlebnikov, M.~Kruczenski, and G.~Michalogiorgakis, ``{Shock waves in
  strongly coupled plasmas II},''
  \href{http://dx.doi.org/10.1007/JHEP07(2011)097}{{\em JHEP} {\bfseries 07}
  (2011) 097},
\href{http://arxiv.org/abs/1105.1355}{{\ttfamily arXiv:1105.1355 [hep-th]}}.

\bibitem{torre2015nonlocal}
I.~Torre, A.~Tomadin, A.~K. Geim, and M.~Polini, ``Nonlocal transport and the
  hydrodynamic shear viscosity in graphene,'' {\em Physical Review B}
  {\bfseries 92} no.~16, (2015) 165433.

\bibitem{Calabrese:2009qy}
P.~Calabrese and J.~Cardy, ``{Entanglement entropy and conformal field
  theory},'' \href{http://dx.doi.org/10.1088/1751-8113/42/50/504005}{{\em J.
  Phys.} {\bfseries A42} (2009) 504005},
\href{http://arxiv.org/abs/0905.4013}{{\ttfamily arXiv:0905.4013
  [cond-mat.stat-mech]}}.

\bibitem{RevModPhys.83.863}
A.~Polkovnikov, K.~Sengupta, A.~Silva, and M.~Vengalattore, ``Colloquium:
  Nonequilibrium dynamics of closed interacting quantum systems,''
  \href{http://dx.doi.org/10.1103/RevModPhys.83.863}{{\em Rev. Mod. Phys.}
  {\bfseries 83} (Aug, 2011) 863--883}.
  \url{https://link.aps.org/doi/10.1103/RevModPhys.83.863}.

\bibitem{eisler2007evolution}
V.~Eisler and I.~Peschel, ``Evolution of entanglement after a local quench,''
  {\em Journal of Statistical Mechanics: Theory and Experiment} {\bfseries
  2007} no.~06, (2007) P06005.

\bibitem{Calabrese:2007mtj}
P.~Calabrese and J.~Cardy, ``{Entanglement and correlation functions following
  a local quench: a conformal field theory approach},''
  \href{http://dx.doi.org/10.1088/1742-5468/2007/10/P10004}{{\em J. Stat.
  Mech.} {\bfseries 0710} no.~10, (2007) P10004},
\href{http://arxiv.org/abs/0708.3750}{{\ttfamily arXiv:0708.3750 [quant-ph]}}.

\bibitem{stephan2011local}
J.-M. St{\'e}phan and J.~Dubail, ``Local quantum quenches in critical
  one-dimensional systems: entanglement, the loschmidt echo, and light-cone
  effects,'' {\em Journal of Statistical Mechanics: Theory and Experiment}
  {\bfseries 2011} no.~08, (2011) P08019.

\bibitem{Asplund:2011cq}
C.~T. Asplund and S.~G. Avery, ``{Evolution of Entanglement Entropy in the
  D1-D5 Brane System},''
  \href{http://dx.doi.org/10.1103/PhysRevD.84.124053}{{\em Phys.Rev.}
  {\bfseries D84} (2011) 124053},
\href{http://arxiv.org/abs/1108.2510}{{\ttfamily arXiv:1108.2510 [hep-th]}}.

\bibitem{Asplund:2013zba}
C.~T. Asplund and A.~Bernamonti, ``{Mutual information after a local quench in
  conformal field theory},''
  \href{http://dx.doi.org/10.1103/PhysRevD.89.066015}{{\em Phys. Rev.}
  {\bfseries D89} no.~6, (2014) 066015},
\href{http://arxiv.org/abs/1311.4173}{{\ttfamily arXiv:1311.4173 [hep-th]}}.

\bibitem{Nozaki:2013wia}
M.~Nozaki, T.~Numasawa, and T.~Takayanagi, ``{Holographic Local Quenches and
  Entanglement Density},''
  \href{http://dx.doi.org/10.1007/JHEP05(2013)080}{{\em JHEP} {\bfseries 05}
  (2013) 080},
\href{http://arxiv.org/abs/1302.5703}{{\ttfamily arXiv:1302.5703 [hep-th]}}.

\bibitem{Asplund:2014coa}
C.~T. Asplund, A.~Bernamonti, F.~Galli, and T.~Hartman, ``{Holographic
  Entanglement Entropy from 2d CFT: Heavy States and Local Quenches},''
  \href{http://dx.doi.org/10.1007/JHEP02(2015)171}{{\em JHEP} {\bfseries 02}
  (2015) 171},
\href{http://arxiv.org/abs/1410.1392}{{\ttfamily arXiv:1410.1392 [hep-th]}}.

\bibitem{Nozaki:2014hna}
M.~Nozaki, T.~Numasawa, and T.~Takayanagi, ``{Quantum Entanglement of Local
  Operators in Conformal Field Theories},''
  \href{http://dx.doi.org/10.1103/PhysRevLett.112.111602}{{\em Phys. Rev.
  Lett.} {\bfseries 112} (2014) 111602},
\href{http://arxiv.org/abs/1401.0539}{{\ttfamily arXiv:1401.0539 [hep-th]}}.

\bibitem{David:2016pzn}
J.~R. David, S.~Khetrapal, and S.~P. Kumar, ``{Universal corrections to
  entanglement entropy of local quantum quenches},''
  \href{http://dx.doi.org/10.1007/JHEP08(2016)127}{{\em JHEP} {\bfseries 08}
  (2016) 127},
\href{http://arxiv.org/abs/1605.05987}{{\ttfamily arXiv:1605.05987 [hep-th]}}.

\bibitem{Calabrese:2005in}
P.~Calabrese and J.~L. Cardy, ``{Evolution of entanglement entropy in
  one-dimensional systems},''
  \href{http://dx.doi.org/10.1088/1742-5468/2005/04/P04010}{{\em J. Stat.
  Mech.} {\bfseries 0504} (2005) P04010},
\href{http://arxiv.org/abs/cond-mat/0503393}{{\ttfamily arXiv:cond-mat/0503393
  [cond-mat]}}.

\bibitem{Calabrese:2006rx}
P.~Calabrese and J.~L. Cardy, ``{Time-dependence of correlation functions
  following a quantum quench},''
  \href{http://dx.doi.org/10.1103/PhysRevLett.96.136801}{{\em Phys. Rev. Lett.}
  {\bfseries 96} (2006) 136801},
\href{http://arxiv.org/abs/cond-mat/0601225}{{\ttfamily arXiv:cond-mat/0601225
  [cond-mat]}}.

\bibitem{calabrese2009entanglement}
P.~Calabrese and J.~Cardy, ``Entanglement entropy and conformal field theory,''
  {\em Journal of Physics A: Mathematical and Theoretical} {\bfseries 42}
  no.~50, (2009) 504005.

\bibitem{calabrese2006time}
P.~Calabrese and J.~Cardy, ``Time dependence of correlation functions following
  a quantum quench,'' {\em Physical review letters} {\bfseries 96} no.~13,
  (2006) 136801.

\bibitem{AbajoArrastia:2010yt}
J.~Abajo-Arrastia, J.~Aparicio, and E.~Lopez, ``{Holographic Evolution of
  Entanglement Entropy},''
  \href{http://dx.doi.org/10.1007/JHEP11(2010)149}{{\em JHEP} {\bfseries 1011}
  (2010) 149},
\href{http://arxiv.org/abs/1006.4090}{{\ttfamily arXiv:1006.4090 [hep-th]}}.

\bibitem{Albash:2010mv}
T.~Albash and C.~V. Johnson, ``{Evolution of Holographic Entanglement Entropy
  after Thermal and Electromagnetic Quenches},''
  \href{http://dx.doi.org/10.1088/1367-2630/13/4/045017}{{\em New J.Phys.}
  {\bfseries 13} (2011) 045017},
\href{http://arxiv.org/abs/1008.3027}{{\ttfamily arXiv:1008.3027 [hep-th]}}.

\bibitem{Balasubramanian:2010ce}
V.~Balasubramanian, A.~Bernamonti, J.~de~Boer, N.~Copland, B.~Craps, {\em
  et~al.}, ``{Thermalization of Strongly Coupled Field Theories},''
  \href{http://dx.doi.org/10.1103/PhysRevLett.106.191601}{{\em Phys.Rev.Lett.}
  {\bfseries 106} (2011) 191601},
\href{http://arxiv.org/abs/1012.4753}{{\ttfamily arXiv:1012.4753 [hep-th]}}.

\bibitem{Balasubramanian:2011ur}
V.~Balasubramanian, A.~Bernamonti, J.~de~Boer, N.~Copland, B.~Craps, {\em
  et~al.}, ``{Holographic Thermalization},''
  \href{http://dx.doi.org/10.1103/PhysRevD.84.026010}{{\em Phys.Rev.}
  {\bfseries D84} (2011) 026010},
\href{http://arxiv.org/abs/1103.2683}{{\ttfamily arXiv:1103.2683 [hep-th]}}.

\bibitem{Keranen:2011xs}
V.~Keranen, E.~Keski-Vakkuri, and L.~Thorlacius, ``{Thermalization and
  entanglement following a non-relativistic holographic quench},''
  \href{http://dx.doi.org/10.1103/PhysRevD.85.026005}{{\em Phys.Rev.}
  {\bfseries D85} (2012) 026005},
\href{http://arxiv.org/abs/1110.5035}{{\ttfamily arXiv:1110.5035 [hep-th]}}.

\bibitem{Erdmenger:2012xu}
J.~Erdmenger and S.~Lin, ``{Thermalization from gauge/gravity duality:
  Evolution of singularities in unequal time correlators},''
  \href{http://dx.doi.org/10.1007/JHEP10(2012)028}{{\em JHEP} {\bfseries 1210}
  (2012) 028},
\href{http://arxiv.org/abs/1205.6873}{{\ttfamily arXiv:1205.6873 [hep-th]}}.

\bibitem{Baron:2012fv}
W.~Baron, D.~Galante, and M.~Schvellinger, ``{Dynamics of holographic
  thermalization},'' \href{http://dx.doi.org/10.1007/JHEP03(2013)070}{{\em
  JHEP} {\bfseries 1303} (2013) 070},
\href{http://arxiv.org/abs/1212.5234}{{\ttfamily arXiv:1212.5234 [hep-th]}}.

\bibitem{Caceres:2012em}
E.~Caceres and A.~Kundu, ``{Holographic Thermalization with Chemical
  Potential},'' \href{http://dx.doi.org/10.1007/JHEP09(2012)055}{{\em JHEP}
  {\bfseries 1209} (2012) 055},
\href{http://arxiv.org/abs/1205.2354}{{\ttfamily arXiv:1205.2354 [hep-th]}}.

\bibitem{Galante:2012pv}
D.~Galante and M.~Schvellinger, ``{Thermalization with a chemical potential
  from AdS spaces},'' \href{http://dx.doi.org/10.1007/JHEP07(2012)096}{{\em
  JHEP} {\bfseries 1207} (2012) 096},
\href{http://arxiv.org/abs/1205.1548}{{\ttfamily arXiv:1205.1548 [hep-th]}}.

\bibitem{Li:2013sia}
Y.-Z. Li, S.-F. Wu, Y.-Q. Wang, and G.-H. Yang, ``{Linear growth of
  entanglement entropy in holographic thermalization captured by horizon
  interiors and mutual information},''
  \href{http://dx.doi.org/10.1007/JHEP09(2013)057}{{\em JHEP} {\bfseries 1309}
  (2013) 057},
\href{http://arxiv.org/abs/1306.0210}{{\ttfamily arXiv:1306.0210 [hep-th]}}.

\bibitem{Arefeva:2013wma}
I.~Aref'eva, A.~Bagrov, and A.~S. Koshelev, ``{Holographic Thermalization from
  Kerr-AdS},''
\href{http://arxiv.org/abs/1305.3267}{{\ttfamily arXiv:1305.3267 [hep-th]}}.

\bibitem{Caputa:2013eka}
P.~Caputa, G.~Mandal, and R.~Sinha, ``{Dynamical entanglement entropy with
  angular momentum and U(1) charge},''
\href{http://arxiv.org/abs/1306.4974}{{\ttfamily arXiv:1306.4974 [hep-th]}}.

\bibitem{Hartman:2013qma}
T.~Hartman and J.~Maldacena, ``{Time Evolution of Entanglement Entropy from
  Black Hole Interiors},''
  \href{http://dx.doi.org/10.1007/JHEP05(2013)014}{{\em JHEP} {\bfseries 1305}
  (2013) 014},
\href{http://arxiv.org/abs/1303.1080}{{\ttfamily arXiv:1303.1080 [hep-th]}}.

\bibitem{Liu:2013iza}
H.~Liu and S.~J. Suh, ``{Entanglement Tsunami: Universal Scaling in Holographic
  Thermalization},''
  \href{http://dx.doi.org/10.1103/PhysRevLett.112.011601}{{\em Phys. Rev.
  Lett.} {\bfseries 112} (2014) 011601},
\href{http://arxiv.org/abs/1305.7244}{{\ttfamily arXiv:1305.7244 [hep-th]}}.

\bibitem{Liu:2013qca}
H.~Liu and S.~J. Suh, ``{Entanglement growth during thermalization in
  holographic systems},''
  \href{http://dx.doi.org/10.1103/PhysRevD.89.066012}{{\em Phys. Rev.}
  {\bfseries D89} no.~6, (2014) 066012},
\href{http://arxiv.org/abs/1311.1200}{{\ttfamily arXiv:1311.1200 [hep-th]}}.

\bibitem{Bai:2014tla}
X.~Bai, B.-H. Lee, L.~Li, J.-R. Sun, and H.-Q. Zhang, ``{Time Evolution of
  Entanglement Entropy in Quenched Holographic Superconductors},''
  \href{http://dx.doi.org/10.1007/JHEP04(2015)066}{{\em JHEP} {\bfseries 04}
  (2015) 066},
\href{http://arxiv.org/abs/1412.5500}{{\ttfamily arXiv:1412.5500 [hep-th]}}.

\bibitem{Anous:2016kss}
T.~Anous, T.~Hartman, A.~Rovai, and J.~Sonner, ``{Black Hole Collapse in the
  1/c Expansion},'' \href{http://dx.doi.org/10.1007/JHEP07(2016)123}{{\em JHEP}
  {\bfseries 07} (2016) 123},
\href{http://arxiv.org/abs/1603.04856}{{\ttfamily arXiv:1603.04856 [hep-th]}}.

\bibitem{Casini:2015zua}
H.~Casini, H.~Liu, and M.~Mezei, ``{Spread of entanglement and causality},''
  \href{http://dx.doi.org/10.1007/JHEP07(2016)077}{{\em JHEP} {\bfseries 07}
  (2016) 077},
\href{http://arxiv.org/abs/1509.05044}{{\ttfamily arXiv:1509.05044 [hep-th]}}.

\bibitem{Mezei:2016zxg}
M.~Mezei, ``{On entanglement spreading from holography},''
  \href{http://dx.doi.org/10.1007/JHEP05(2017)064}{{\em JHEP} {\bfseries 05}
  (2017) 064},
\href{http://arxiv.org/abs/1612.00082}{{\ttfamily arXiv:1612.00082 [hep-th]}}.

\bibitem{Arefeva:2017pho}
I.~{\relax Ya}. Aref'eva, M.~A. Khramtsov, and M.~D. Tikhanovskaya,
  ``{Thermalization after holographic bilocal quench},''
  \href{http://dx.doi.org/10.1007/JHEP09(2017)115}{{\em JHEP} {\bfseries 09}
  (2017) 115},
\href{http://arxiv.org/abs/1706.07390}{{\ttfamily arXiv:1706.07390 [hep-th]}}.

\bibitem{Anous:2017tza}
T.~Anous, T.~Hartman, A.~Rovai, and J.~Sonner, ``{From Conformal Blocks to Path
  Integrals in the Vaidya Geometry},''
  \href{http://dx.doi.org/10.1007/JHEP09(2017)009}{{\em JHEP} {\bfseries 09}
  (2017) 009},
\href{http://arxiv.org/abs/1706.02668}{{\ttfamily arXiv:1706.02668 [hep-th]}}.

\bibitem{Yeh:2014mfa}
C.-P. Yeh, J.-T. Hsiang, and D.-S. Lee, ``{Holographic influence functional and
  its application to decoherence induced by quantum critical theories},''
  \href{http://dx.doi.org/10.1103/PhysRevD.91.046009}{{\em Phys. Rev.}
  {\bfseries D91} no.~4, (2015) 046009},
\href{http://arxiv.org/abs/1410.7111}{{\ttfamily arXiv:1410.7111 [hep-th]}}.

\bibitem{Mezei:2018jco}
M.~Mezei, ``{Membrane theory of entanglement dynamics from holography},''
\href{http://arxiv.org/abs/1803.10244}{{\ttfamily arXiv:1803.10244 [hep-th]}}.

\bibitem{Mezei:2016wfz}
M.~Mezei and D.~Stanford, ``{On entanglement spreading in chaotic systems},''
  \href{http://dx.doi.org/10.1007/JHEP05(2017)065}{{\em JHEP} {\bfseries 05}
  (2017) 065},
\href{http://arxiv.org/abs/1608.05101}{{\ttfamily arXiv:1608.05101 [hep-th]}}.

\bibitem{kimhuse}
H.~Kim and D.~A. Huse, ``Ballistic spreading of entanglement in a diffusive
  nonintegrable system,'' {\em Physical review letters} {\bfseries 111} no.~12,
  (2013) 127205.

\bibitem{Buchel:2013lla}
A.~Buchel, L.~Lehner, R.~C. Myers, and A.~van Niekerk, ``{Quantum quenches of
  holographic plasmas},'' \href{http://dx.doi.org/10.1007/JHEP05(2013)067}{{\em
  JHEP} {\bfseries 05} (2013) 067},
\href{http://arxiv.org/abs/1302.2924}{{\ttfamily arXiv:1302.2924 [hep-th]}}.

\bibitem{Das:2014jna}
S.~R. Das, D.~A. Galante, and R.~C. Myers, ``{Universal scaling in fast quantum
  quenches in conformal field theories},''
  \href{http://dx.doi.org/10.1103/PhysRevLett.112.171601}{{\em Phys. Rev.
  Lett.} {\bfseries 112} (2014) 171601},
\href{http://arxiv.org/abs/1401.0560}{{\ttfamily arXiv:1401.0560 [hep-th]}}.

\bibitem{Das:2014hqa}
S.~R. Das, D.~A. Galante, and R.~C. Myers, ``{Universality in fast quantum
  quenches},'' \href{http://dx.doi.org/10.1007/JHEP02(2015)167}{{\em JHEP}
  {\bfseries 02} (2015) 167},
\href{http://arxiv.org/abs/1411.7710}{{\ttfamily arXiv:1411.7710 [hep-th]}}.

\bibitem{Das:2015jka}
S.~R. Das, D.~A. Galante, and R.~C. Myers, ``{Smooth and fast versus
  instantaneous quenches in quantum field theory},''
  \href{http://dx.doi.org/10.1007/JHEP08(2015)073}{{\em JHEP} {\bfseries 08}
  (2015) 073},
\href{http://arxiv.org/abs/1505.05224}{{\ttfamily arXiv:1505.05224 [hep-th]}}.

\bibitem{Das:2016lla}
S.~R. Das, D.~A. Galante, and R.~C. Myers, ``{Quantum Quenches in Free Field
  Theory: Universal Scaling at Any Rate},''
  \href{http://dx.doi.org/10.1007/JHEP05(2016)164}{{\em JHEP} {\bfseries 05}
  (2016) 164},
\href{http://arxiv.org/abs/1602.08547}{{\ttfamily arXiv:1602.08547 [hep-th]}}.

\bibitem{HolzegelDafermos}
M.~Dafermos and G.~Holzegel, ``{Conjecture of non-linear instability of AdS},''
  {\em `The problem of stability for black hole spacetimes' (talk by M.D. at
  Isaac Newton Institute, Cambridge UK)} (2006) .

\bibitem{Bizon:2011gg}
P.~Bizon and A.~Rostworowski, ``{On weakly turbulent instability of anti-de
  Sitter space},'' \href{http://dx.doi.org/10.1103/PhysRevLett.107.031102}{{\em
  Phys. Rev. Lett.} {\bfseries 107} (2011) 031102},
\href{http://arxiv.org/abs/1104.3702}{{\ttfamily arXiv:1104.3702 [gr-qc]}}.

\bibitem{Hartnoll:2008kx}
S.~A. Hartnoll, C.~P. Herzog, and G.~T. Horowitz, ``{Holographic
  Superconductors},''
  \href{http://dx.doi.org/10.1088/1126-6708/2008/12/015}{{\em JHEP} {\bfseries
  0812} (2008) 015},
\href{http://arxiv.org/abs/0810.1563}{{\ttfamily arXiv:0810.1563 [hep-th]}}.

\bibitem{Domokos:2007kt}
S.~K. Domokos and J.~A. Harvey, ``{Baryon number-induced Chern-Simons couplings
  of vector and axial-vector mesons in holographic QCD},''
  \href{http://dx.doi.org/10.1103/PhysRevLett.99.141602}{{\em Phys. Rev. Lett.}
  {\bfseries 99} (2007) 141602},
\href{http://arxiv.org/abs/0704.1604}{{\ttfamily arXiv:0704.1604 [hep-ph]}}.

\bibitem{Nakamura:2009tf}
S.~Nakamura, H.~Ooguri, and C.-S. Park, ``{Gravity Dual of Spatially Modulated
  Phase},'' \href{http://dx.doi.org/10.1103/PhysRevD.81.044018}{{\em Phys.
  Rev.} {\bfseries D81} (2010) 044018},
\href{http://arxiv.org/abs/0911.0679}{{\ttfamily arXiv:0911.0679 [hep-th]}}.

\bibitem{barankov2004collective}
R.~Barankov, L.~Levitov, and B.~Spivak, ``Collective rabi oscillations and
  solitons in a time-dependent bcs pairing problem,'' {\em Physical review
  letters} {\bfseries 93} no.~16, (2004) 160401.

\bibitem{yuzbashyan2006relaxation}
E.~A. Yuzbashyan, O.~Tsyplyatyev, and B.~L. Altshuler, ``Relaxation and
  persistent oscillations of the order parameter in fermionic condensates,''
  {\em Physical review letters} {\bfseries 96} no.~9, (2006) 097005.

\bibitem{yuzbashyan2005solution}
E.~A. Yuzbashyan, B.~L. Altshuler, V.~B. Kuznetsov, and V.~Z. Enolskii,
  ``Solution for the dynamics of the bcs and central spin problems,'' {\em
  Journal of Physics A: Mathematical and General} {\bfseries 38} no.~36, (2005)
  7831.

\bibitem{PhysRevLett.93.130403}
R.~A. Barankov and L.~S. Levitov, ``Atom-molecule coexistence and collective
  dynamics near a feshbach resonance of cold fermions,''
  \href{http://dx.doi.org/10.1103/PhysRevLett.93.130403}{{\em Phys. Rev. Lett.}
  {\bfseries 93} (Sep, 2004) 130403}.
  \url{http://link.aps.org/doi/10.1103/PhysRevLett.93.130403}.

\bibitem{Bhaseen:2012gg}
M.~Bhaseen, J.~P. Gauntlett, B.~Simons, J.~Sonner, and T.~Wiseman,
  ``{Holographic Superfluids and the Dynamics of Symmetry Breaking},''
  \href{http://dx.doi.org/10.1103/PhysRevLett.110.015301}{{\em Phys.Rev.Lett.}
  {\bfseries 110} no.~1, (2013) 015301},
\href{http://arxiv.org/abs/1207.4194}{{\ttfamily arXiv:1207.4194 [hep-th]}}.

\bibitem{Podolsky:2012pv}
D.~Podolsky and S.~Sachdev, ``{Spectral functions of the Higgs mode near
  two-dimensional quantum critical points},''
  \href{http://dx.doi.org/10.1103/PhysRevB.86.054508}{{\em Phys. Rev.}
  {\bfseries B86} (2012) 054508},
\href{http://arxiv.org/abs/1205.2700}{{\ttfamily arXiv:1205.2700
  [cond-mat.quant-gas]}}.

\bibitem{Katan:2015qfa}
Y.~T. Katan and D.~Podolsky, ``{Spectral function of the Higgs mode in 4-ε
  dimensions},'' \href{http://dx.doi.org/10.1103/PhysRevB.91.075132}{{\em Phys.
  Rev.} {\bfseries B91} no.~7, (2015) 075132},
\href{http://arxiv.org/abs/1412.4546}{{\ttfamily arXiv:1412.4546
  [cond-mat.str-el]}}.

\bibitem{endres2012higgs}
M.~Endres, T.~Fukuhara, D.~Pekker, M.~Cheneau, P.~Schau$\beta$, C.~Gross,
  E.~Demler, S.~Kuhr, and I.~Bloch, ``The higgs'amplitude mode at the
  two-dimensional superfluid/mott insulator transition,'' {\em Nature}
  {\bfseries 487} no.~7408, (2012) 454--458.

\bibitem{PhysRevLett.102.130603}
P.~Barmettler, M.~Punk, V.~Gritsev, E.~Demler, and E.~Altman, ``Relaxation of
  antiferromagnetic order in spin-$1/2$ chains following a quantum quench,''
  \href{http://dx.doi.org/10.1103/PhysRevLett.102.130603}{{\em Phys. Rev.
  Lett.} {\bfseries 102} (Apr, 2009) 130603}.
  \url{https://link.aps.org/doi/10.1103/PhysRevLett.102.130603}.

\bibitem{Buchel:2013gba}
A.~Buchel, R.~C. Myers, and A.~van Niekerk, ``{Universality of Abrupt
  Holographic Quenches},''
  \href{http://dx.doi.org/10.1103/PhysRevLett.111.201602}{{\em Phys. Rev.
  Lett.} {\bfseries 111} (2013) 201602},
\href{http://arxiv.org/abs/1307.4740}{{\ttfamily arXiv:1307.4740 [hep-th]}}.

\bibitem{Zurek:1985qw}
W.~H. Zurek, ``{Cosmological Experiments in Superfluid Helium?},''
\href{http://dx.doi.org/10.1038/317505a0}{{\em Nature} {\bfseries 317} (1985)
  505--508}.

\bibitem{Kibble:1976sj}
T.~W.~B. Kibble, ``{Topology of Cosmic Domains and Strings},''
\href{http://dx.doi.org/10.1088/0305-4470/9/8/029}{{\em J. Phys.} {\bfseries
  A9} (1976) 1387--1398}.

\bibitem{polkovnikov2005universal}
A.~Polkovnikov, ``Universal adiabatic dynamics in the vicinity of a quantum
  critical point,'' {\em Physical Review B} {\bfseries 72} no.~16, (2005)
  161201.

\bibitem{PhysRevLett.95.105701}
W.~H. Zurek, U.~Dorner, and P.~Zoller, ``Dynamics of a quantum phase
  transition,'' \href{http://dx.doi.org/10.1103/PhysRevLett.95.105701}{{\em
  Phys. Rev. Lett.} {\bfseries 95} (Sep, 2005) 105701}.
  \url{https://link.aps.org/doi/10.1103/PhysRevLett.95.105701}.

\bibitem{delCampo:2013nla}
A.~del Campo and W.~H. Zurek, ``{Universality of phase transition dynamics:
  Topological Defects from Symmetry Breaking},''
  \href{http://dx.doi.org/10.1142/S0217751X1430018X,
  10.1142/9789814583060_0002}{{\em Int. J. Mod. Phys.} {\bfseries A29} no.~8,
  (2014) 1430018}, \href{http://arxiv.org/abs/1310.1600}{{\ttfamily
  arXiv:1310.1600 [cond-mat.stat-mech]}}.
[,31(2013)].

\bibitem{Basu:2011ft}
P.~Basu and S.~R. Das, ``{Quantum Quench across a Holographic Critical
  Point},'' \href{http://dx.doi.org/10.1007/JHEP01(2012)103}{{\em JHEP}
  {\bfseries 01} (2012) 103},
\href{http://arxiv.org/abs/1109.3909}{{\ttfamily arXiv:1109.3909 [hep-th]}}.

\bibitem{Basu:2012gg}
P.~Basu, D.~Das, S.~R. Das, and T.~Nishioka, ``{Quantum Quench Across a Zero
  Temperature Holographic Superfluid Transition},''
  \href{http://dx.doi.org/10.1007/JHEP03(2013)146}{{\em JHEP} {\bfseries 03}
  (2013) 146},
\href{http://arxiv.org/abs/1211.7076}{{\ttfamily arXiv:1211.7076 [hep-th]}}.

\bibitem{Gao:2012aw}
X.~Gao, A.~M. Garcia-Garcia, H.~B. Zeng, and H.-Q. Zhang, ``{Normal modes and
  time evolution of a holographic superconductor after a quantum quench},''
  \href{http://dx.doi.org/10.1007/JHEP06(2014)019}{{\em JHEP} {\bfseries 06}
  (2014) 019},
\href{http://arxiv.org/abs/1212.1049}{{\ttfamily arXiv:1212.1049 [hep-th]}}.

\bibitem{Garcia-Garcia:2013rha}
A.~M. García-García, H.~B. Zeng, and H.~Q. Zhang, ``{A thermal quench induces
  spatial inhomogeneities in a holographic superconductor},''
  \href{http://dx.doi.org/10.1007/JHEP07(2014)096}{{\em JHEP} {\bfseries 07}
  (2014) 096},
\href{http://arxiv.org/abs/1308.5398}{{\ttfamily arXiv:1308.5398 [hep-th]}}.

\bibitem{Basu:2013soa}
P.~Basu, D.~Das, S.~R. Das, and K.~Sengupta, ``{Quantum Quench and Double Trace
  Couplings},'' \href{http://dx.doi.org/10.1007/JHEP12(2013)070}{{\em JHEP}
  {\bfseries 12} (2013) 070},
\href{http://arxiv.org/abs/1308.4061}{{\ttfamily arXiv:1308.4061 [hep-th]}}.

\bibitem{Chesler:2014gya}
P.~M. Chesler, A.~M. Garcia-Garcia, and H.~Liu, ``{Defect Formation beyond
  Kibble-Zurek Mechanism and Holography},''
  \href{http://dx.doi.org/10.1103/PhysRevX.5.021015}{{\em Phys.Rev.} {\bfseries
  X5} no.~2, (2015) 021015},
\href{http://arxiv.org/abs/1407.1862}{{\ttfamily arXiv:1407.1862 [hep-th]}}.

\bibitem{Adams:2012pj}
A.~Adams, P.~M. Chesler, and H.~Liu, ``{Holographic Vortex Liquids and
  Superfluid Turbulence},''
  \href{http://dx.doi.org/10.1126/science.1233529}{{\em Science} {\bfseries
  341} (2013) 368--372},
\href{http://arxiv.org/abs/1212.0281}{{\ttfamily arXiv:1212.0281 [hep-th]}}.

\bibitem{Ewerz:2014tua}
C.~Ewerz, T.~Gasenzer, M.~Karl, and A.~Samberg, ``{Non-Thermal Fixed Point in a
  Holographic Superfluid},''
  \href{http://dx.doi.org/10.1007/JHEP05(2015)070}{{\em JHEP} {\bfseries 05}
  (2015) 070},
\href{http://arxiv.org/abs/1410.3472}{{\ttfamily arXiv:1410.3472 [hep-th]}}.

\bibitem{Pretorius:2005gq}
F.~Pretorius, ``{Evolution of binary black hole spacetimes},''
  \href{http://dx.doi.org/10.1103/PhysRevLett.95.121101}{{\em Phys. Rev. Lett.}
  {\bfseries 95} (2005) 121101},
\href{http://arxiv.org/abs/gr-qc/0507014}{{\ttfamily arXiv:gr-qc/0507014
  [gr-qc]}}.

\bibitem{hawking1973large}
S.~W. Hawking and G.~F.~R. Ellis, {\em The large scale structure of
  space-time}, vol.~1.
\newblock Cambridge university press, 1973.

\bibitem{wald2010general}
R.~M. Wald, {\em General relativity}.
\newblock University of Chicago press, 2010.

\bibitem{Donos:2015eew}
A.~Donos and J.~P. Gauntlett, ``{Minimally packed phases in holography},''
  \href{http://dx.doi.org/10.1007/JHEP03(2016)148}{{\em JHEP} {\bfseries 03}
  (2016) 148},
\href{http://arxiv.org/abs/1512.06861}{{\ttfamily arXiv:1512.06861 [hep-th]}}.

\bibitem{Bantilan:2012vu}
H.~Bantilan, F.~Pretorius, and S.~S. Gubser, ``{Simulation of Asymptotically
  AdS5 Spacetimes with a Generalized Harmonic Evolution Scheme},''
  \href{http://dx.doi.org/10.1103/PhysRevD.85.084038}{{\em Phys. Rev.}
  {\bfseries D85} (2012) 084038},
\href{http://arxiv.org/abs/1201.2132}{{\ttfamily arXiv:1201.2132 [hep-th]}}.

\bibitem{Chesler:2013lia}
P.~M. Chesler and L.~G. Yaffe, ``{Numerical solution of gravitational dynamics
  in asymptotically anti-de Sitter spacetimes},''
  \href{http://dx.doi.org/10.1007/JHEP07(2014)086}{{\em JHEP} {\bfseries 1407}
  (2014) 086},
\href{http://arxiv.org/abs/1309.1439}{{\ttfamily arXiv:1309.1439 [hep-th]}}.

\bibitem{Zhang:2016coy}
M.~Guo, C.~Niu, Y.~Tian, and H.~Zhang, ``{Applied AdS/CFT with Numerics},''
  \href{http://dx.doi.org/10.22323/1.271.0003}{{\em PoS} {\bfseries Modave2015}
  (2016) 003},
\href{http://arxiv.org/abs/1601.00257}{{\ttfamily arXiv:1601.00257 [gr-qc]}}.

\bibitem{Heller:2012je}
M.~P. Heller, R.~A. Janik, and P.~Witaszczyk, ``{A numerical relativity
  approach to the initial value problem in asymptotically Anti-de Sitter
  spacetime for plasma thermalization - an ADM formulation},''
  \href{http://dx.doi.org/10.1103/PhysRevD.85.126002}{{\em Phys. Rev.}
  {\bfseries D85} (2012) 126002},
\href{http://arxiv.org/abs/1203.0755}{{\ttfamily arXiv:1203.0755 [hep-th]}}.

\bibitem{choquetbruhat1962cauchy}
Y.~Choquet-Bruhat, {\em The Cauchy Problem}.
\newblock New York: Wiley, 1962.

\bibitem{Pretorius:2004jg}
F.~Pretorius, ``{Numerical relativity using a generalized harmonic
  decomposition},'' \href{http://dx.doi.org/10.1088/0264-9381/22/2/014}{{\em
  Class. Quant. Grav.} {\bfseries 22} (2005) 425--452},
\href{http://arxiv.org/abs/gr-qc/0407110}{{\ttfamily arXiv:gr-qc/0407110
  [gr-qc]}}.

\bibitem{Gundlach:2005eh}
C.~Gundlach, J.~M. Martin-Garcia, G.~Calabrese, and I.~Hinder, ``{Constraint
  damping in the Z4 formulation and harmonic gauge},''
  \href{http://dx.doi.org/10.1088/0264-9381/22/17/025}{{\em Class. Quant.
  Grav.} {\bfseries 22} (2005) 3767--3774},
\href{http://arxiv.org/abs/gr-qc/0504114}{{\ttfamily arXiv:gr-qc/0504114
  [gr-qc]}}.

\bibitem{Holzegel:2011qk}
G.~Holzegel and J.~Smulevici, ``{Self-gravitating Klein-Gordon fields in
  asymptotically Anti-de-Sitter spacetimes},''
  \href{http://dx.doi.org/10.1007/s00023-011-0146-8}{{\em Annales Henri
  Poincare} {\bfseries 13} (2012) 991--1038},
\href{http://arxiv.org/abs/1103.0712}{{\ttfamily arXiv:1103.0712 [gr-qc]}}.

\bibitem{Holzegel:2013vwa}
G.~H. Holzegel and C.~M. Warnick, ``{The Einstein-Klein-Gordon-AdS system for
  general boundary conditions},''
\href{http://arxiv.org/abs/1312.5332}{{\ttfamily arXiv:1312.5332 [gr-qc]}}.

\bibitem{Holzegel:2015swa}
G.~Holzegel, J.~Luk, J.~Smulevici, and C.~Warnick, ``{Asymptotic properties of
  linear field equations in anti-de Sitter space},''
\href{http://arxiv.org/abs/1502.04965}{{\ttfamily arXiv:1502.04965 [gr-qc]}}.

\bibitem{winicour2001characteristic}
J.~Winicour, ``Characteristic evolution and matching,'' {\em arXiv preprint
  gr-qc/0102085} (2001) .

\end{thebibliography}\endgroup
\end{document}